\documentclass[11pt]{article}
\usepackage{jheppub}
\usepackage{enumerate}
\usepackage{amssymb, amsmath, bbm, dsfont}
\usepackage{graphicx}
\usepackage{color,comment}
\usepackage{soul}
\usepackage[applemac]{inputenc}
\usepackage[small]{caption}
\setcaptionwidth{14cm}

\definecolor{dm}{cmyk}{.20, 0, .30, 0}
\sethlcolor{dm}

\newcommand{\pd}{\partial}

\newcommand{\be}{\begin{equation}}
\newcommand{\ee}{\end{equation}}
\newcommand{\bea}{\begin{eqnarray}}
\newcommand{\eea}{\end{eqnarray}}
\newcommand{\bal}{\begin{align}}
\newcommand{\eal}{\end{align}}

\newcommand{\half}{\frac{1}{2}}
\newcommand{\mc}{\mathcal}

\newcommand{\im}{\mathrm{Im\;}}
\newcommand{\re}{\mathrm{Re\;}}
\newcommand{\nn}{\nonumber}

\def\bra{\langle}
\def\ket{\rangle}
\def\tr{\mathrm{tr}}
\def\ap{\alpha^{\prime}}
\def\beq{\begin{equation}}
\def\eeq{\end{equation}}
\def\mr#1{\mathrm{#1}}
\def\ov{\overline}

\newcommand{\mathscr}{\mathcal}
\newcommand{\tht}{\vartheta}

\newcommand{\beqn}{\begin{eqnarray}}
\newcommand{\eeqn}{\end{eqnarray}}

\newcommand{\thw}[2]{\vartheta[\genfrac{}{}{0pt}{}{#1}{#2} ]}
\newcommand{\tha}[3]{\vartheta[\!\begin{array}{c}{\phantom{}\vspace{-1mm}\scriptstyle#1}%
                        \\[-1.8mm]{\scriptstyle #2}\end{array}\!]( { #3} )}

\newcommand{\Ref}[1]{(\ref{#1})}

\newcommand{\dm}[1]{\marginpar[   \hfill{\hl{ \bf !}}]%
                  {{     \hl{  D  }}} {\hl{  [#1] }}}

\title{Superpotential de-sequestering  in  string models}
\author[1]{Marcus Berg,}
\author[2]{Joseph P. Conlon,}
\author[3]{David Marsh}
\author[2]{and Lukas T. Witkowski}
\affiliation[1]{Department of Physics, Karlstad University \\ 651 88 Karlstad, Sweden \\ and \\ Oskar Klein Center, Stockholm University \\ Albanova University Center \\106 91 Stockholm, Sweden}
\affiliation[2]{Rudolf Peierls Centre for Theoretical Physics,\\ University of Oxford, 1 Keble Road, Oxford OX1 3NP, UK}
\affiliation[3]{Department of Physics, \\Cornell University, Ithaca, NY 14853}

\emailAdd{marcus.berg@kau.se}
\emailAdd{j.conlon1@physics.ox.ac.uk}
\emailAdd{dm444@cornell.edu}
\emailAdd{l.witkowski1@physics.ox.ac.uk}

\abstract{
Non-perturbative superpotential cross-couplings between  visible sector matter and  Kähler moduli can lead to significant flavour-changing neutral currents in compactifications of type IIB string theory.
Here, we compute 
corrections to Yukawa couplings 
in orbifold models with chiral matter localised on D3-branes and non-perturbative effects on distant D7-branes. 
By evaluating a threshold correction to the D7-brane gauge coupling,
we determine conditions under which
the non-perturbative corrections to the Yukawa couplings appear. The
flavour structure of the induced Yukawa coupling generically
fails to be aligned with the tree-level flavour structure. 
We check our results
by also evaluating a correlation function of two D7-brane gauginos and a D3-brane Yukawa coupling. 
Finally, 
by calculating a string amplitude between $n$ hidden scalars and visible matter we show how non-vanishing vacuum expectation values of distant D7-brane scalars, if present, may correct visible Yukawa couplings with a flavour structure that differs from the tree-level flavour structure. 
}

\begin{document}
\maketitle

\section{Introduction}

Understanding  the effects of moduli stabilisation for semi-realistic models of particle physics arising from string theory is an important and challenging question in string phenomenology. 
By localising the visible sector  geometrically  in the extra dimensions, one may hope for a decoupling of  global effects and thereby a simplified framework for extracting the predictions of the model (e.g.
\cite{0005067, Verlinde:2005jr}).  A number of obstructions to such  simplifications have been discussed in the literature, and in this note we focus on one specific example of an effect induced by moduli stabilisation that can alter  the visible sector spectrum.

 {\it Sequestering} \cite{Randall:1998uk},
a strategy to surmount the famous ``supersymmetry flavour problem'', is one example
of a desirable property that is supposed to follow from localising the visible sector geometrically. 
Specifically, it was argued in \cite{Randall:1998uk} that for a five-dimensional model where only gravity propagates in the extra dimension and in which the visible and hidden sectors  are geometrically separated, the  gravity mediated soft terms vanish at tree-level \cite{Randall:1998uk}. The low-energy soft-terms were instead argued  to be flavour universal and to arise first at loop level through anomaly mediation. 

It may seem that this attractive mechanism should arise naturally in string theory models in which the visible and hidden sectors are realised as branes filling four-dimensional space-time while being geometrically separated in the internal compactification manifold. 
However, 
in \cite{Anisimov:2001zz, Anisimov:2002az} it was shown that  for a variety of string models in Type II string theory and M-theory,
 sequestering does not arise naturally. The discrepancy between these two results can be understood as a consequence of  
Kaluza-Klein (KK) modes with masses of the order of the compactification scale, and moduli which only become stabilised after compactification and which therefore have masses below the compactification scale \cite{Anisimov:2001zz, Kachru:2006em}.  These fields typically mediate interactions which spoil sequestering and re-introduce the supersymmetry flavour problem in these models.


By localising the supersymmetry breaking sector at the bottom of a \emph{warped} throat,  sequestering may still occur in large classes of string compactifications with flux \cite{Kachru:2007xp, Luty:2000ec}, consistently with the expectations from the conformal field theory (CFT) dual, known as ``conformal sequestering'' \cite{Luty:2001jh}. This type of models can be constructed in Type IIB flux compactifications \cite{Giddings:2001yu}, for example.

Furthermore, effective theories  which are not of sequestered form but satisfy
the milder criterion of  \emph{``sort-of-sequestering''}  \cite{Berg:2010ha}  still share
with fully sequestered models the favourable avoidance of the supersymmetry flavour problem at tree level. Interestingly, compactifications in the Large Volume Scenario (LVS) \cite{Balasubramanian:2005zx,Conlon:2005ki}, are of this form \cite{Berg:2010ha}, due to ``extended no-scale structure'' \cite{Cicoli:2007xp}.

However, taking the dynamics of moduli stabilisation  into account
creates another adverse effect, where even warping or ``sort-of-sequestering'' may not suffice to shield the visible sector from phenomenologically dangerous 
contributions to the soft terms.  This effect, which intimately couples the low-energy phenomenology of the localised visible sector to the details of the moduli stabilising sector, was called \emph{superpotential de-sequestering} in \cite{Berg:2010ha}.  

 This effect is particularly important  in compactification schemes in which there is a hierarchy between the gravitino mass and the scale of the soft masses, and is of particular interest in
Kachru-Kallosh-Linde-Trivedi (KKLT) type models  \cite{Giddings:2001yu,Kachru:2003aw}
 and in the Large Volume Scenario \cite{Balasubramanian:2005zx,Conlon:2005ki}, as we will now review.

\subsection*{Superpotential de-sequestering}
Non-perturbative effects in the form of Euclidean D3-branes or gaugino condensation on a stack of D7-branes wrapping some four-cycle in the internal dimensions are an essential ingredient in several celebrated moduli-stabilisation schemes in type IIB string theory. At the level of the four-dimensional effective theory, the non-perturbative effects induce a non-perturbative superpotential which depends on the volume of the corresponding four-cycle. Once supersymmetry is broken, non-vanishing F-terms can be induced for the K\"ahler moduli. Superpotential de-sequestering refers to interactions induced between the K\"ahler moduli and a localised visible sector which is geometrically separated from the four-cycle supporting the non-perturbative effects. In particular, the presence of operators of the form,
\be
W \supset {\cal O}\ e^{- T} \, ,  \label{Wdeseq}
\ee
for some visible sector operator ${\cal O}$ and for some K\"ahler modulus $T$, can give rise to  important contributions to the soft terms in certain moduli stabilisation scenarios.
We immediately
note that at first sight \Ref{Wdeseq} may seem negligible since it is
a nonperturbative contribution, but since analogous nonperturbative contributions
are also of crucial importance in stabilising the moduli
in scenarios under consideration, this argument needs careful scrutiny.
There turn out to be numerically important contributions,
and the most important contributions are those to the $B\mu$-term and the soft $A$-terms involving the Higgs field and leptons and squarks.\footnote{There are essentially no bounds on $A$-terms that do not involve the Higgs.}

In this note, we will focus specifically on superpotential operators of the form,
\be
W \supset \hat{Y}^{np}_{ijk}\ C^i C^j C^k\ e^{-T} \, , \label{Ynp}
\ee
where $\hat{Y}^{np}$ is some constant
and the $C^i$ are various charged fields. After supersymmetry breaking in which the K\"ahler modulus $T$ obtains a non-vanishing $F$-term, the induced  $A$-terms are given (in supergravity) by
\be
A_{ijk} = e^{\mathcal{K}/2}\ \bar F^m\ {\cal D}_m Y_{ijk} \, ,
\ee
in natural units, where $Y_{ijk}$ are the full, moduli-dependent superpotential Yukawa couplings. Note that this expression applies to non-canonically normalised fields, but can easily be extended to canonically normalised fields. (Even if the visible sector fields come with non-diagonal K\"ahler metric, one can still apply general coordinate transformations in field space to extend the expression to canonically normalised fields.) If one of the fields $C^i$ in equation \eqref{Ynp} is a Higgs field, soft masses for the remaining fields will be induced after electroweak symmetry breaking. These masses can be bounded by studying their effect on flavour-violating and CP-violating processes in the standard model. The consequences of the bound thus obtained depend on the moduli stabilisation scenario, and the size and structure of $\hat{Y}^{np}$. In particular, if $A \propto Y^{\rm phys}$, much weaker bounds apply than would otherwise be the case.

\begin{figure}
	\centering
		\includegraphics[width=0.40\textwidth]{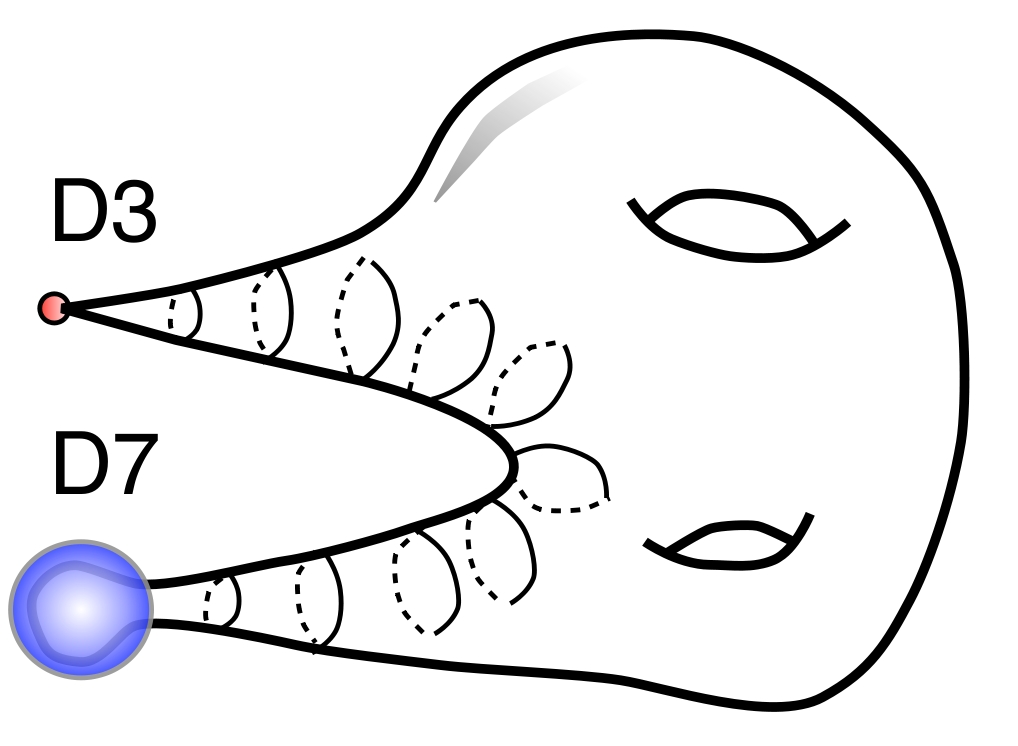}
	\caption{Calabi-Yau compactification with D3-branes at a singularity and D7-branes wrapping a small cycle. The singularity and the D7-branes share a homologous 2-cycle. This setup appears in LVS constructions.}
\label{fig:CY1}
\end{figure}

\begin{figure}
	\centering
		\includegraphics[width=0.40\textwidth]{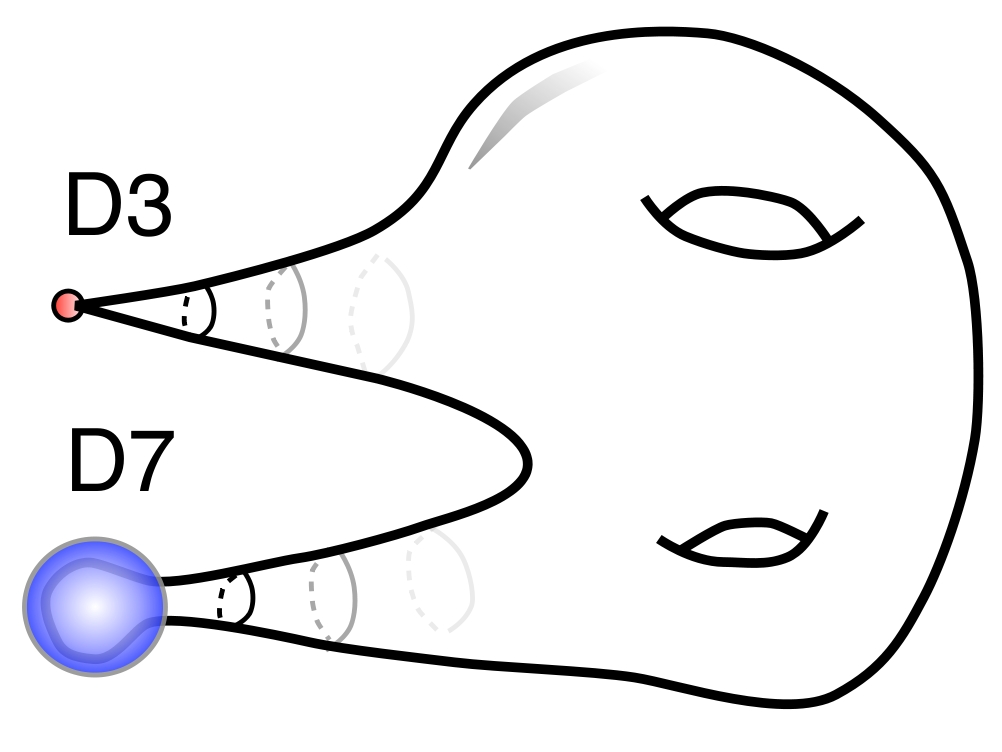}
	\caption{Calabi-Yau compactification with D3-branes at a singularity and D7-branes wrapping a small cycle. The singularity and the small cycle are geometrically separated in the bulk. This setup appears in LVS constructions. }
\label{fig:CY2}
\end{figure}

\begin{figure}
	\centering
		\includegraphics[width=0.40\textwidth]{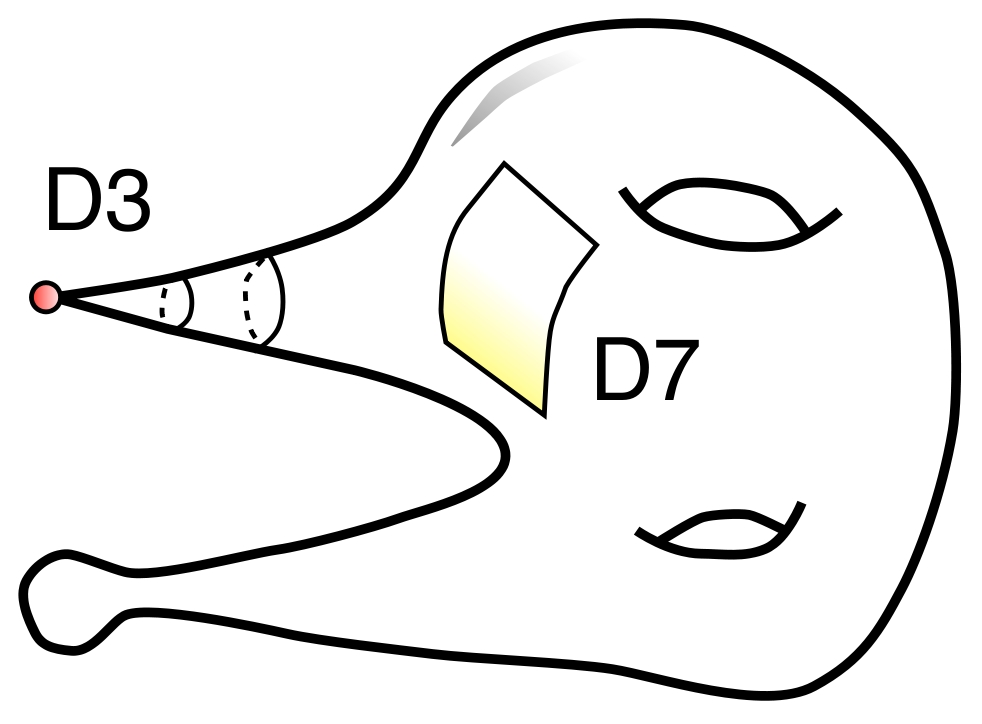}
	\caption{Calabi-Yau compactification with D3-branes at a singularity and D7-branes wrapping a bulk cycle. This setup appears in KKLT constructions.}
\label{fig:CY3}
\end{figure}

\subsection*{Computing the de-sequestering operators}
In this paper, we study de-sequestering in some simple toroidal orbifolds which serve as computable toy models for more complicated compactifications. 
We model the visible sector as arising from D3-branes at an orbifold singularity, while the D7-branes supporting the non-perturbative effects can either wrap two 2-tori of the compact space or a small 4-cycle instead. In the orbifold limit, D7-branes wrapping a small cycle become D3 branes at orbifold singularities. In our study we are motivated by previous results of flavour-violating effects due to Yukawa couplings sourced by string or D-brane instantons \cite{Abel:2006yk, Buican:2008qe, Cvetic:2009yh}. Similarly, nonperturbative effects in F-theory GUT models have been found to modify the flavour structure of the Yukawa couplings \cite{Marchesano:2009rz}.

 For non-perturbative effects arising from gaugino condensation on a stack of D7-branes, the effective superpotential  is proportional to
\be
e^{- a f_{D7}} \, ,
\ee
where $f_{D7}$ is the gauge kinetic function on the D7-branes, and $a= 2\pi/c$,  where $c$ is the dual Coxeter number
of the D7-brane gauge group, so that e.g.~$c=N$ for $SU(N)$. The dependence of the D7-brane gauge kinetic function on the visible sector fields thus induces operators of the form of equation  \eqref{Wdeseq}.
This will enable us to compute the contribution to the physical $A$-terms from superpotential de-sequestering by world-sheet techniques. Let us outline what couplings we need to compute.

In the low-energy effective field theory the quantity of interest contributes to the D7-brane gauge kinetic term of the SUSY Lagrangian:
\be
\int \textrm{d}^4 x \int \textrm{d}^2 \theta \ f_{D7}(\Phi^{i}, C^j) W_{\alpha} W^{\alpha}\ ,
\ee
where $f(\Phi^i, C^j)$ is the holomorphic gauge kinetic function depending on D3 matter fields $C^j$ and moduli $\Phi^i$, whereas $W_{\alpha}$ is the D7 field strength superfield. We can expand the gauge kinetic function in gauge-invariant combinations of the D3 matter fields $C^i$:
\be
f_{D7}(\Phi^{\alpha}, C^i)= \hat{f}(\Phi^{\alpha}) + \tilde{f}_{ij}(\Phi^{\alpha}) C^{i} C^{j} + \hat{y}_{ijk}(\Phi^{\alpha}) C^{i} C^{j} C^{k} + \ldots \ .
\ee
It is the trilinear term that will be responsible for generating superpotential terms of the form \eqref{Ynp},
so we want to determine $\hat{y}_{ijk}$. After integrating over the Grassmann coordinates $\textrm{d}^2 \theta$, the term $\hat{y}_{ijk}(\Phi^j)W_{\alpha} W^{\alpha} C^{i} C^{j} C^{k}$
will produce supermultiplet component couplings
\begin{align}  \label{components}
&\frac{1}{2}\hat{y}_{ijk} \tr( F^{\mu \nu} F_{\mu \nu}) \tr( \phi^i \phi^j \phi^k)  \qquad \textrm{and} \\
&2 \hat{y}_{ijk} \tr( \lambda \lambda) \tr(\psi^i \psi^j \phi^k) \ ,   \nonumber
\end{align}
where $A_{\mu}$ are D7 gauge bosons, $\lambda$ the corresponding gauginos and $\psi_i$ and $\phi_j$ are D3 fermionic and scalar matter fields. Thus  there are two kinds of string amplitude calculations
we could perform to obtain $\hat{y}_{ijk}$, and we will consider both. First, we compute the one-loop threshold correction to the D7-brane gauge kinetic function given by $\langle \tr( A^{\mu} A_{\mu})  \tr( \phi^i \phi^j \phi^k) \rangle$. Secondly we will calculate corrections to D3 Yukawa couplings due to gaugino condensation by evaluating $\langle \tr(\lambda \lambda) \tr(\psi^i \psi^j \phi^k) \rangle$. Both amplitudes
are computed as a cylinder diagram with two D7-brane gauge boson or gaugino vertex operators inserted at one boundary, and three visible sector matter field operators inserted at the other boundary.

\subsection*{De-sequestering results}
The string calculations will be performed for a range of orbifold models. We will find that only Yukawa couplings with particular flavour structures appear in the nonperturbative superpotential. Most importantly their flavour structure does not match that of the tree level Yukawa couplings and thus de-sequestering can introduce flavour violation.

Even though these orbifold models  are
just simple toy models, they capture the relevant features to determine whether superpotential de-sequestering can occur in these D-brane setups. We will therefore argue that our findings are generic for string backgrounds including D3/D7-branes and thus apply to popular compactification schemes with moduli stabilisation, such as
KKLT and LVS.

 While the precise expressions will be sensitive to the finer details of the model, the existence or absence of de-sequestering is governed by a few simple properties of the string theory setup. In particular, we do not need to specify the full model to determine the flavour structure of the induced operators.
 We do restrict attention to localising the visible sector on D3-branes at a supersymmetric singularity, but we consider it likely that D7-brane matter would give similar results.

 We find that the relevant string theory backgrounds fall into three classes which are displayed in figures \ref{fig:CY1}, \ref{fig:CY2} and \ref{fig:CY3}.  The setups differ in the realisation of the non-perturbative effects. We now analyse them in turn.
\begin{enumerate}[i]
\item In the first configuration shown in fig.\ \ref{fig:CY1} the D7-branes wrap a small cycle in the compact space. This is relevant in the LVS scenario, where the small cycle can be identified as a blow-up cycle. While the D3- and D7-stacks are separated in the compact space, they are connected by a homologous 2-cycle.\footnote{In orbifold models this is equivalent to the orbifold possessing a $\mathcal{N}=2$ supersymmetric sector. In this sector the stacks of branes are separated along an untwisted direction allowing for the propagation of string modes between them.} We find that this situation leads to de-sequestering: the homologous 2-cycle allows for the propagation of string modes between the stacks of branes thus inducing terms in the non-perturbative superpotential.

\item The second setup displayed in \ref{fig:CY2} is very similar to the case considered above: the D7-branes again wrap a small cycle in the compact space. This time, however, the visible stack and the non-perturbative effects do not share a homologous 2-cycle in the compact geometry.\footnote{In this case the orbifold does not display any $\mathcal{N}=2$ sectors connecting the two sectors.} As mediating closed string modes can only propagate on the appropriate cycles,
warping can help in KKLT \cite{Kachru:2007xp} and the aforementioned ``sort-of-sequestering'' can help in LVS \cite{Berg:2010ha}. We find that in this case geometric separation, together
with these two effects, appears to  suffice to avoid de-sequestering.

\item The third setup is sketched in \ref{fig:CY3} and arises in KKLT constructions where non-perturbative effects are located on bulk D7-branes wrapping a 4-cycle.\footnote{These effects are irrelevant in LVS as they are suppressed by a factor of  $\exp(- {\cal V}^{2/3})$ in the superpotential.} The effects of this setup are very similar to the first case considered above: while the two stacks of branes are separated in the compact space, they nevertheless communicate via closed string modes.\footnote{In the orbifold setting the relevant open string modes arise in the untwisted sector.} The result is again that contributions to the non-perturbative superpotential will be generated, thus leading to de-sequestering.
\end{enumerate}
The above results allow us to identify a necessary condition for de-sequestering to occur, independent of the phenomenological model: the stack of D7-branes carrying the non-perturbative effects has to either wrap a bulk 4-cycle or share a homologous 2-cycle with the stack of visible sector D3-branes.

These conditions pose further restrictions on the type of singularity at which the visible sector is realised. In case (i) above de-sequestering is absent if the singularity is too trivial as it has to share a homologous 2-cycle. On the other hand, there is no such condition in case (iii). There de-sequestering can occur for singularities as simple as $dP_0$ (which corresponds to $\mathbb{C}^3/\mathbb{Z}_3$).

Another important result is that the flavour structure of the induced Yukawa couplings does not match with the one of the tree-level Yukawa-interactions. This observation is general whenever there is de-sequestering, both in scenarios relevant to KKLT models and the LVS. Hence we confirm the validity of the effective field theory argument of \cite{Berg:2010ha}, that superpotential de-sequestering can lead to large flavour-changing neutral currents (FCNCs).

While the position moduli of D7-branes wrapping bulk cycles generically become supersymmetrically stabilised with the complex structure moduli at a relatively high scale \cite{Kachru:2007xp}, matter scalars for D7-branes wrapping small cycles may not.
We show that the Yukawa couplings of the D3-brane visible sector obtain a dependence
on fields characterising the distant D7-brane wrapping a small cycle, and the flavour structure of the resulting Yukawa couplings is different from the tree-level flavour structure.


\section{The Model: Toroidal Orbifold Backgrounds}

In this section we review some relevant background material on toroidal orbifolds for the backgrounds in which we intend to do the computation. Specifically, we will discuss different models for the visible sector and for the non-perturbative effects as well as the spectrum of local orbifold models.

\subsection{Modelling the visible and the hidden sectors}
Both the visible sector and the sector supporting the non-perturbative effects can be modeled in several different ways in  toroidal orbifolds, and here we discuss the virtues and drawbacks of some of the conceivable alternatives, which are also summarised in table \ref{table1}.
\begin{table}[h]
\begin{center}
\begin{tabular}{| c | c |c | c |} \hline
Alternative    & Visible sector  & Non-perturbative effects   & Model  \\ \hline  \hline
\#1 & Bulk D3 & Bulk D7 & not used \\ \hline
\#2 & Bulk D3 & D3 at orbifold singularity & not used \\ \hline
\#3 & Fractional D3 & Bulk D7 & ``KKLT'' \\ \hline
\#4 & Fractional D3 & D3 at orbifold singularity & ``LVS'' \\ \hline
\end{tabular}
\caption{Models of visible and hidden sectors. The unused alternatives have adjoint matter.}
\label{table1}
\end{center}
\vspace{-5mm}
\end{table}
Bulk D3-branes (i.e.~D3-branes at a smooth point) support ${\cal N}=4$ supersymmetry on the world-volume, and correspondingly the scalars of such a visible sector transform in the adjoint representation of the gauge group. This makes bulk D3-branes less than realistic as a toy model for a chiral model of particle physics and we will not discuss them any further in this work.

D3-branes at orbifold singularities give rise to  `quiver' theories, in which the scalars transform under the bifundamental representation of the gauge groups of some fractional branes \cite{Douglas:1996sw}. We will be more specific in the explicit examples discussed below.

In the various moduli stabilisation scenarios of interest, the non-perturbative effects can be realised either as gaugino condensation on a D7-brane  on a bulk cycle  or  a blow-up cycle.
The latter case is more interesting for applications to the Large Volume Scenario in which the non-perturbative effects are supported on a small (though still  above string scale) four-cycle. The former case is interesting in e.g.~the KKLT scenario. While a bulk cycle in a toroidal orbifold is realised as a D7-brane
that wraps two $T^2$'s in the internal space but is pointlike in the third $T^2$, a `small' cycle can be modeled as a D3-brane at an orbifold singularity in the toroidal computation. Let us emphasise here that the fractional D3-brane that would model a D7-brane at a small cycle should be placed at an orbifold singularity that is invariant under the orbifold action (e.g. the origin of a $\mathbb{Z}_N$ orbifold), so that it does not carry twist charge.


\begin{figure}
	\centering
		\includegraphics[width=0.75\textwidth]{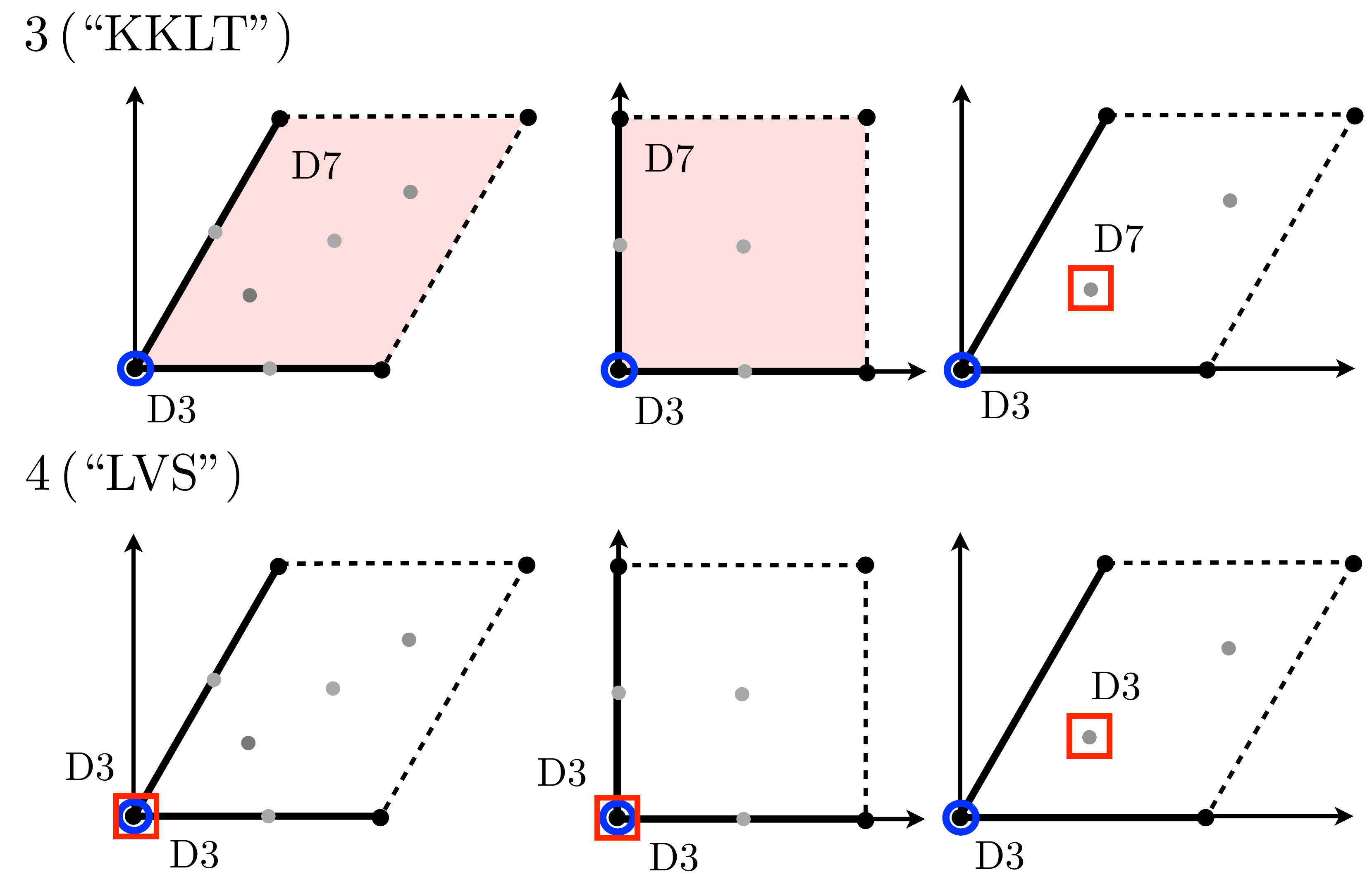}
	\caption{Brane configurations for the $\mathbb{T}^6/\mathbb{Z}_6'$ orbifold with $\theta=\frac{1}{6}(1,-3,2)$,
	for the two alternatives \#3 and \#4 of table \ref{table1}.
	(This is the AaA lattice, in the classification of \cite{Gmeiner:2007zz}.)
	 Dots indicate orbifold singularities. Stacks of D3-branes supporting visible matter are denoted by a blue circle
	 at the origin. The branes responsible for nonperturbative effects are indicated by a red box,
	 except that the bulk D7-branes wrap the first two tori of the compact space and are pointlike on the third torus.}
	\label{fig:Z6D3D37}
\end{figure}


\subsection{Local vs.~Global}
\label{localvsglobal}
 There certainly exist arrangements of
 brane stacks and orientifold planes
that ensure tadpole cancellation in global (compact) models
such as the orientifold version of $\mathbb{T}^6/\mathbb{Z}_6'$ in fig.\ \ref{fig:Z6D3D37}.
(For a review of orientifolds, see \cite{Angelantonj:2002ct}.)
We will argue that our calculation of the string one-loop renormalisation
of a certain five-point double trace operator can be performed mostly
in a local model of the singularity and
should carry over mostly unchanged to a global orientifold model. What we mean by ``mostly'' is that certain zero-mode solutions involve states that do propagate away from the singularity, but this is under control.

\begin{figure}[h]
\begin{center}
\includegraphics[width=0.6\textwidth]{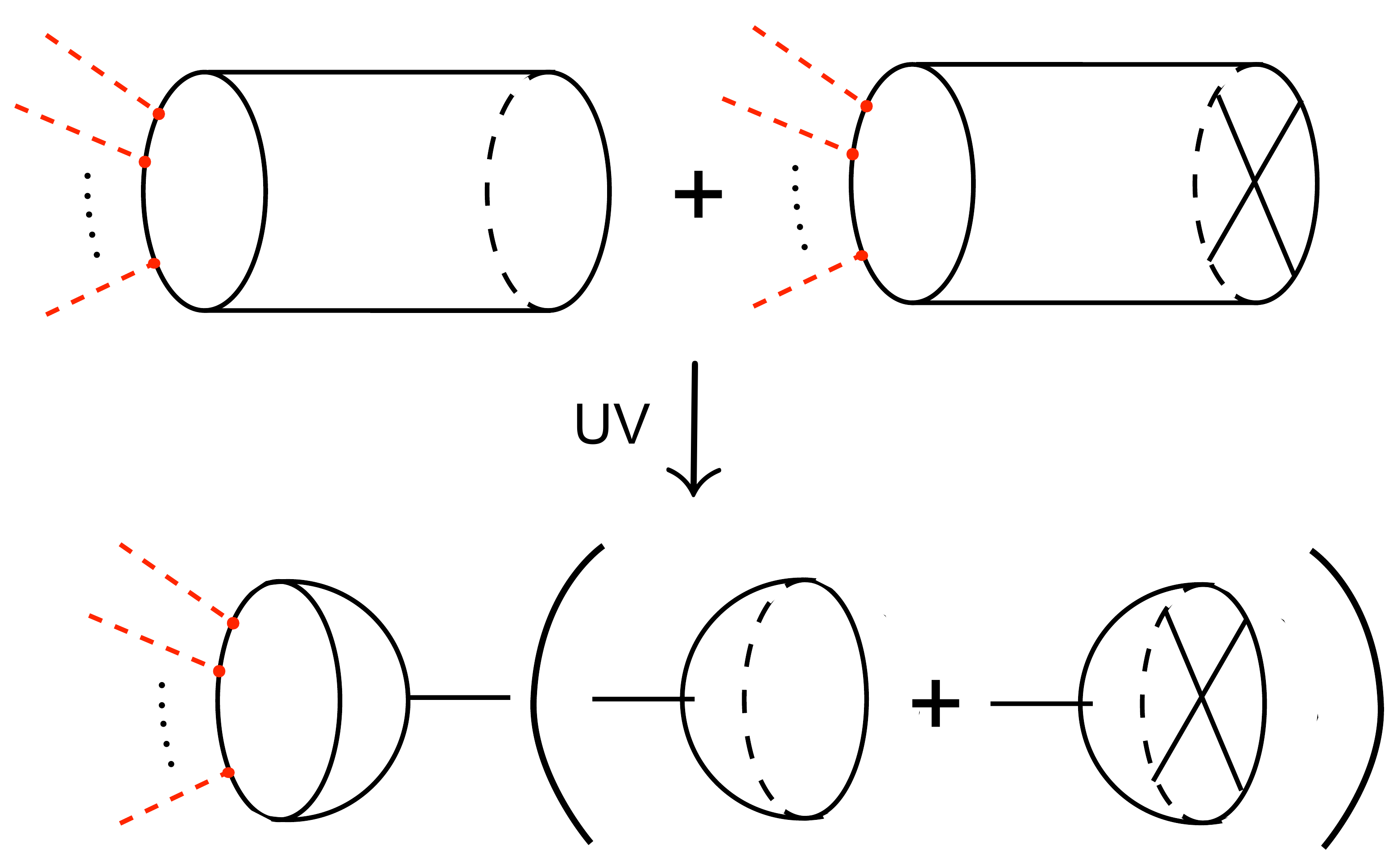}
\caption{Tadpoles as a problem
for generic planar diagrams.}
\label{tadpole}
\end{center}
\end{figure}

The statements in the previous paragraph are not obvious, but we have two simple arguments why
they are true:

a) {\it Finiteness.} As is well known,
for a generic D-brane configuration, the cylinder vacuum amplitude
 has divergences
due to long-distance propagation of massless closed string modes at zero momentum, i.e. $(1/p^2)_{p=0}$,
which can be ascribed to a tadpole diagram attached to a D-brane, i.e.\ a nonzero probability
for a zero-momentum closed string to be produced from the vacuum.
This problem is solved by imposing tadpole cancellation.
In some planar open-string amplitudes (see e.g.  \cite{Berg:2011ij})
lack of enforcement of tadpole cancellation would cause
problematic divergences, because factorisation onto closed string states
recreates  any uncancelled tadpoles on the side of the cylinder without external states inserted
(see fig.\ \ref{tadpole}).
This means that even if one tries to argue that tadpoles are
cancelled once the model is completed to a global model,
the meaning of a
calculation that suffers from this problem, but is actually performed only in the local model
where there remain uncancelled tadpoles, is unclear.

However, in the spirit of \cite{Dudas:2004nd}, certain calculations do not suffer from this problem.
In a cylinder amplitude with insertions on both sides, the situation is generically better, because
there is momentum flowing in the propagator $(1/p^2)_{p\neq 0}$, and hence no divergence.
 One expects poles from reducible diagrams, but this is perfectly physical.
  Still,  in some of our examples we are interested in a potential term, which has no derivatives, so
we want to take the external momenta on the scalar side to zero. One
could worry that the problem would reappear, but calculation shows it does not
(similar arguments recently appeared in \cite{Anastasopoulos:2011kr}).
In this limit, the amplitude behaves as (see e.g.~\Ref{polecancmom} below).
\[
{\mathcal A}\quad  \stackrel{\alpha' p^2 \rightarrow 0}{\longrightarrow} \quad  {\alpha' p^2 \over \alpha'  p^2} \sim 1
\]
From the field theory point of view, this is a nontrivial physical statement about reducible diagrams, that the vertices each supply a factor of momentum such that the overall diagram is finite in the massless and low-energy limit.
In string theory, this instead reflects something technical but essentially trivial, that
 if we begin with something finite with no momentum-dependent forefactors, then change picture (in the sense of superconformal ghost charge, see below) in a way that produces some explicit additional momenta in a given amplitude, we must also produce compensating poles of the kind above, so as to not change the final result.

b) {\it Double trace.}
As noted above, if all vertex operators had been inserted on one side
of a cylinder diagram as in fig.\ \ref{tadpole}, then there could have been a potential divergence and it would have been necessary to compute the corresponding M\"obius strip diagram.
However, the interactions we are interested in are double trace operators. This requires two boundaries on the worldsheet, and at one loop the
only topology that will allow this is the cylinder, not the M\"obius strip.
In other words, we cannot insert vertex operators on an orientifold plane (it has no dynamics of its own), so
there is no analog of the M\"obius strip amplitude for the couplings we are interested in.
Therefore, at this the order of perturbation theory, we do not need to worry about
orientifold planes: the M\"obius strip diagram cannot contribute.

\subsection{Orbifold spectrum from D-branes at singularities and Yukawa couplings}
\label{allYukawas}
We review type IIB string constructions based on $\mathbb{T}^6/\mathbb{Z}_N$ orbifolds with D3/D7 branes.
The orbifold point group $\mathbb{Z}_N$ acts on the internal space  as
\be  \label{Zaction}
Z^i \mapsto  e^{2 \pi i \theta_{i}} Z^i \; , \quad
\Psi^i \mapsto  e^{2 \pi i \theta_{i}} \Psi^i \; , \quad i=1,2,3
\ee
for complex coordinates $Z^i$ of the three $\mathbb{T}^2$'s  in $\mathbb{T}^6/\mathbb{Z}_N$,
and the corresponding
worldsheet fermions $\Psi^i$.
(For more details on our conventions, see appendix \ref{sec:cftbuildblock}.)
 The twist vector $\theta_i$ satisfies $\theta_1 + \theta_2 + \theta_3 =0$ mod $1$. For the spectrum to exhibit the field content of $\mathcal{N}=1$ supersymmetry we require that all $\theta_i \neq 0$.

The geometrical moduli of each two-torus are
\be
U = {R_2 \over R_1} e^{i\theta_U} \;, \quad  \tilde{T}_2= \sqrt{G} = R_1 R_2 \sin \theta_U \; .
\ee
where $\theta_U$ is the angle that the complex structure forms $U$ with the $x$-axis,
and $R_1$ and $R_2$ are the radii of the two one-cycles of the torus, cf.\  figure \ref{fig:Z6D3D37}.

We note that these parameters are not exactly the supergravity moduli fields,
see \cite{Lust:2004cx,Berg:2005ja,Blumenhagen:2006ci} for a detailed discussion of this, but they are the ones
that directly arise in our string theory calculation.
In particular, $\tilde{T}_2$ is {\it not} the the imaginary part of a K\"ahler modulus
corresponding to a four-cycle volume in the D3-D7 duality frame,
as it sets a two-cycle volume, which is the imaginary part of a K\"ahler modulus in the D5-D9 duality frame.
Also, for some orbifolds,
the complex structure $U$ is actually fixed --- a table can be found in \cite{Dixon:1990pc}

We will mostly focus on compactifications based on the toroidal orbifolds $\mathbb{T}^6/\mathbb{Z}_3$, $\mathbb{T}^6/\mathbb{Z}_4$, $\mathbb{T}^6/\mathbb{Z}_6$ and $\mathbb{T}^6/\mathbb{Z}_6^{\prime}$, as summarised
in table  \ref{tableYukawas}. For these models, the six-dimensional compact space can be factorised into three $\mathbb{T}^2$'s on which the individual orbifold twists act.  One can classify the action of the orbifold twists
on strings into three different sectors:
\begin{enumerate}
\item {\it Completely twisted}. the orbifold action on all subtori is non-trivial: all $\theta_i \neq 0$. The relevant physical fields exhibit $\mathcal{N}=1$ supersymmetry.
\item {\it Partially twisted}: the orbifold action leaves one $\mathbb{T}^2$ invariant, which we choose to be the third one, described by the coordinates $Z^3$ and $\bar{Z}^3$. Thus $\theta_3=0$ mod $1$ and $\theta_1+\theta_2=0$ mod $1$. These sectors appear in $\mathbb{Z}_4$, $\mathbb{Z}_6$ and $\mathbb{Z}_6^{\prime}$ orbifolds.
The relevant physical fields exhibit $\mathcal{N}=2$ supersymmetry.
\item {\it Untwisted}: the orbifold action leaves all tori invariant. These can be identified as $\mathcal{N}=4$ sectors for D3-D3 states and as $\mathcal{N}=2$ sectors for D3-D7 states.
\end{enumerate}

A first useful step in characterising the relevant aspects of the  spectrum of the global (compact) orbifold
$\mathbb{T}^6/\mathbb{Z}_N$
 is to ``zoom in'' on D3-branes at local (noncompact) $\mathbb{C}^3/\mathbb{Z}_N$  orbifold singularities, to establish which combinations of chiral superfields
give gauge-invariant Yukawa operators. Details on this can be found in \cite{0005067, Douglas:1996sw, Douglas:1997de}. The orbifold action \Ref{Zaction} on the string coordinates
is accompanied by a transformation $\gamma_{\theta}$
that acts on the Chan-Paton (CP) indices of open string states. The orbifold spectrum is found by
classifying states under this combined action into invariant and non-invariant states. The result is that the gauge group is $\prod_{i=1}^N U(n_i)$ and chiral multiplets transform as bifundamentals. For an orbifold twist $\theta=(\theta_1,\theta_2,\theta_3)= \frac{1}{N}(b_1, b_2, b_3)$ we have
\beq
\textrm{chiral multiplets}: \sum_{r=1}^3 \sum_{i=0}^{N-1} (n_i, \bar{n}_{i-b_r}) \ .
\eeq
We  label chiral multiplets as $C^r_{i, i-b_r}$ where $r$ is a label for the three two-tori of the compact space and the subscripts denote the gauge groups under which the field transforms.

An allowed Yukawa coupling corresponds to a gauge-invariant combination of three such chiral superfields of the form
\beq
\label{GaugeInvYuk}
C^r_{i, i-b_r} C^s_{i-b_r, i-b_r-b_s} C^t_{i-b_r-b_s, i} \ ,
\eeq
although an allowed coupling may not actually be present.
We summarise in table \ref{tableYukawas} the combinations $(r,s,t)$ that rise to acceptable Yukawa couplings for various orbifold singularities used in model building.
\begin{table}[h]
\begin{center}
\begin{tabular}{ | l | l | l | }
\hline
orbifold singularity & orbifold twist & allowed Yukawa couplings \\
\hline \hline
$\mathbb{C}^3/\mathbb{Z}_3$ & $\theta=\frac{1}{3}(1,1,-2)$ & all $C^r C^s C^t$ for $r,s,t =1,2,3$ \\
\hline
$\mathbb{C}^3/\mathbb{Z}_4$ & $\theta=\frac{1}{4}(1,1,-2)$ & $C^1 C^2 C^3$\\
 & & $C^1 C^1 C^3$ \\
 & & $C^2 C^2 C^3$ \\
\hline
$\mathbb{C}^3/\mathbb{Z}_6$ & $\theta=\frac{1}{6}(1,1,-2)$ & $C^1 C^2 C^3$\\
 & & $C^1 C^1 C^3$ \\
 & & $C^2 C^2 C^3$ \\
& & $C^3 C^3 C^3$ \\
\hline
$\mathbb{C}^3/\mathbb{Z}_6^{\prime}$ & $\theta=\frac{1}{6}(1,-3,2)$ & $C^1 C^2 C^3$\\
 & & $C^3 C^3 C^3$ \\
\hline
\end{tabular}
\end{center}
\caption{Gauge-invariant combinations of chiral superfields $C^r$ arising on D3-branes at
$\mathbb{C}^3/\mathbb{Z}_N$.}
\label{tableYukawas}
\end{table}
These expressions will be useful for the CFT calculations that follow: the labels $(r,s,t)$ determine the H-charges of the vertex operators to be inserted (see section \ref{sec:cftbuildblock}). A trivial observation is that the combination $C^1 C^2 C^3$ is always gauge-invariant and thus an allowed Yukawa coupling for all orbifolds.

\section{De-sequestering in string perturbation theory}
\label{sec:gaugekincorr1}

In this section we calculate the dependence of a D7 gauge coupling on chiral matter on D3 branes. This is the Lagrangian term
\be   \label{superspace}
\int d^4 x \int d^2 \theta \underbrace{Y_{\alpha \beta \gamma} \hbox{Tr}(\Phi^{\alpha} \Phi^{\beta} \Phi^{\gamma})}_{\textrm{D3 chiral matter}}
\underbrace{\hbox{Tr} (W_{\alpha} W^{\alpha})}_{\textrm{D7 gauge theory}} \; .
\ee
As stated in \Ref{components}, the superspace expansion
gives two ways to probe the existence of this term, by looking either for couplings
\be
\frac{1}{2} \int d^4 x \, \underbrace{Y_{\alpha \beta \gamma} \hbox{Tr}(\phi^{\alpha} \phi^{\beta} \phi^{\gamma})}_{\textrm{D3 chiral matter}}\hspace{1mm}\cdot\hspace{-0mm}   \underbrace{\hbox{Tr} (F_{\mu \nu} F^{\mu \nu})}_{\textrm{D7 gauge theory}}
\label{coup1}
\ee
or for couplings
\be
2 \int d^4 x\,  \underbrace{Y_{\alpha \beta \gamma} \hbox{Tr}(\psi^{\alpha} \psi^{\beta} \phi^{\gamma})}_{\textrm{D3 chiral matter}}\hspace{1mm}\cdot\hspace{-4mm}  \underbrace{\hbox{Tr} (\lambda \lambda)}_{\textrm{D7 gauge theory}}  \hspace{-3mm}.
\label{coup2}
\ee
The study of the coupling \Ref{coup2} is relegated to appendix \ref{sec:gaugekincorr2}.
In this section we will study the coupling \Ref{coup1}
of  two gauge bosons and a Yukawa-style coupling of three scalars.


\subsection{Setting up the string calculation}

\begin{figure}
	\centering
		\includegraphics[width=0.4\textwidth]{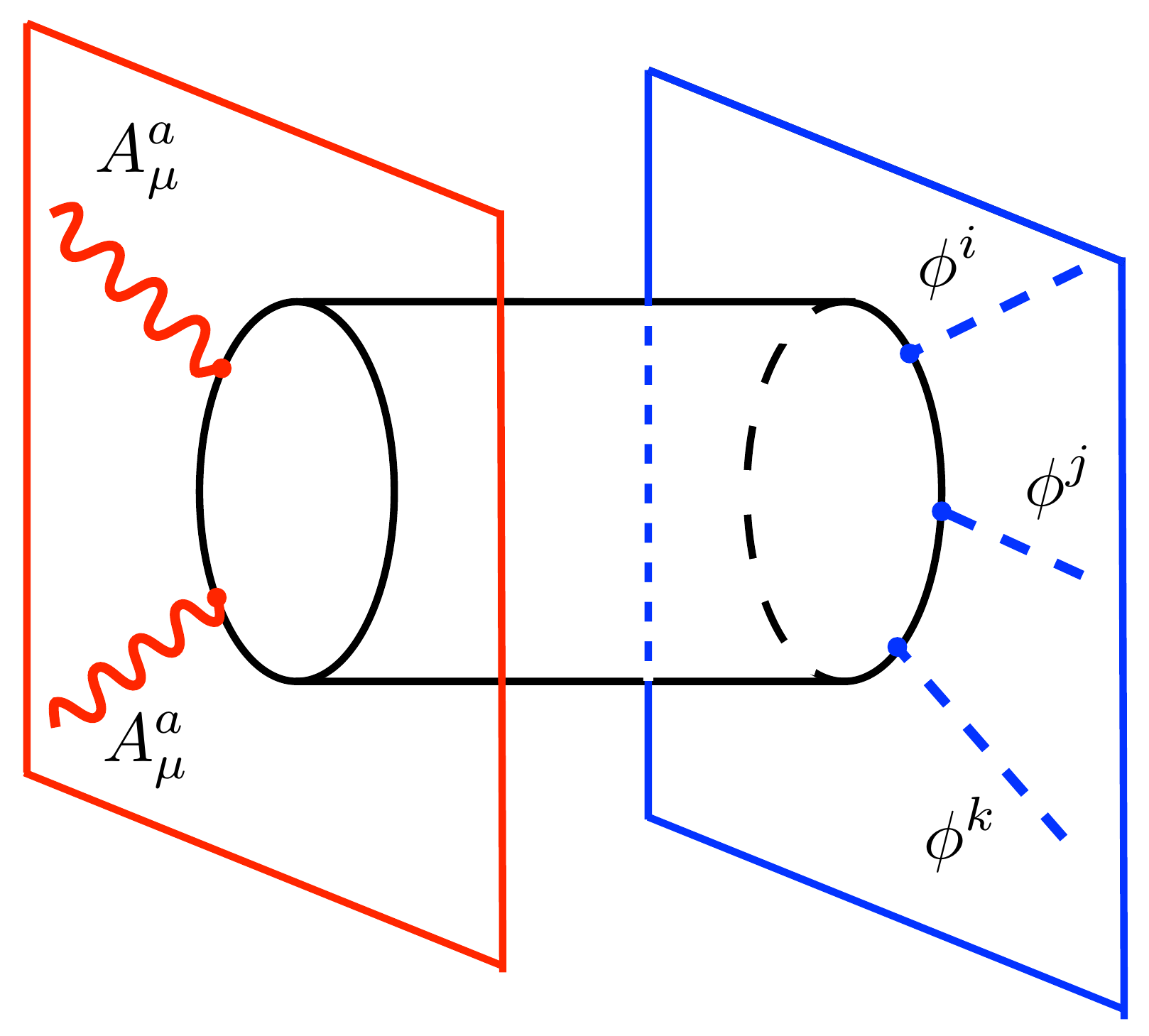}
	\caption{Cylinder amplitude that produces a threshold correction to a gauge coupling due to distant scalars. Vertex operators for gauge bosons are inserted on one boundary and the scalar operators are located on the other boundary, as in \Ref{corr}.}
		\label{fig:2A3Scalar}
\end{figure}

As we are interested in a double-trace operator, the relevant one-loop string topology is the cylinder. As shown in figure \ref{fig:2A3Scalar}, the gauge boson vertex operators are inserted on one boundary of the cylinder while the scalar operators will be located on the other boundary. In appendix \ref{sec:CFT} we review the basic ingredients of conformal field theory that are necessary to perform the computation.

We first compute this correlator for a Yukawa operator with tree-level flavour structure $C^1 C^2 C^3$, before generalising this to other flavour combinations. We begin by writing down the relevant vertex operators which we choose to be in the zero-picture for both the gauge bosons and the scalars. This will satisfy the condition that a cylinder worldsheet requires vanishing background ghost charge. The correlator is then
\be  \label{corr}
\langle   \mathcal{V}_{A_1}^0 (z_1)  \mathcal{V}_{A_2}^0 (z_2)
 \mathcal{V}_{\phi_1}^0 (z_3)  \mathcal{V}_{\phi_2}^0 (z_4)
 \mathcal{V}_{\phi_3}^0 (z_5)  \rangle
\ee
with the zero-picture vertex operators
\bal
\nn \mathcal{V}_{A_1}^0 (z_1) &= \left[\pd X^1 + i \alpha' (k_1 \cdot \psi) \psi^1 \right] e^{i k_1 \cdot X} (z_1)\\
\nn \mathcal{V}_{A_2}^0 (z_2) &= \left[\pd \bar{X}^1 + i \alpha' (k_2 \cdot \psi) \bar{\psi}^1 \right] e^{i k_2 \cdot  X} (z_2) \\
 \mathcal{V}_{\phi_1}^0 (z_3) &= \left[\pd \bar{Z}^1 + i \alpha' (k_3 \cdot \psi) \bar{\Psi}^1 \right] e^{i k_3 \cdot X} (z_3) \\
\nn \mathcal{V}_{\phi_2}^0 (z_4) &= \left[\pd \bar{Z}^2+ i \alpha' (k_4 \cdot \psi) \bar{\Psi}^2 \right] e^{i k_4 \cdot X} (z_4) \\
\nn \mathcal{V}_{\phi_3}^0 (z_5) &= \left[\pd \bar{Z}^3 + i \alpha'  (k_5 \cdot \psi) \bar{\Psi}^3 \right] e^{i k_5 \cdot X} (z_5) \ .
\end{align}
We have suppressed constant polarisations, factors of the string coupling $g_{\rm s}$ and also Chan-Paton factors,
that we will restore in section \ref{sec:discussion}.
Here, $X$ and $\psi$ are complex external coordinates, $Z^i$ and $\Psi^i$ denote
complex internal directions
for $i=1,2,3$. Note that this particular assignment for the scalars $\phi_i$ is consistent with a Yukawa coupling $C^1 C^2 C^3$.
We have chosen the gauge bosons to be polarised in the complex $X^1$ plane only (the $x^0$, $x^1$ directions in terms of real fields), thereby breaking the symmetry between $X^1$ and $X^2$,
in fact we will often  break Lorentz invariance in intermediate steps in this paper
and then assure its presence at the end of the calculation.
The bars on the vertex operators for spacetime scalars, $\partial\bar{Z}^i$ are conventional; our spacetime scalars are defined with negative orbifold charge,
as explained in appendix \ref{sec:settingup}.
Also, the momenta $k_i^{\mu}$ are external and complexified, as in eq.\ \Ref{kXcomplex} in the appendix.

We now make an observation that simplifies the calculation considerably:
as the operators above  only contain barred $\partial \bar{Z}$ and $\bar{\Psi}$ fields for the internal directions,
and no combination barred-unbarred
occurs, the (quantum)
correlators all vanish, as explained in appendix \ref{sec:intclascorr}.
The worldsheet fermion pieces of the spacetime scalar vertex operators
then cannot contribute to the result at all,
but from the scalars there will be classical contributions. The relevant terms of the scalar vertex operators are then:
\begin{align}
& \mathcal{V}_{\phi_3}^0 (z_3) = \pd \bar{Z}^1 e^{i k_3 \cdot X} (z_3) \\
& \mathcal{V}_{\phi_4}^0 (z_4) = \pd \bar{Z}^2 e^{i k_4 \cdot X} (z_4) \\
& \mathcal{V}_{\phi_5}^0 (z_5) = \pd \bar{Z}^3 e^{i k_5 \cdot X} (z_5) \ .
\end{align}
More on classical solutions is given in \ref{sec:intclascorr}.

In all, the vertex operator for a scalar from the chiral multiplet $C^i$ will just contain bosonic fields including $\pd \bar{Z}^i$. 
We continue the calculation for the case of the Yukawa coupling $C^1 C^2 C^3$ and
 then re-evaluate our results for the other cases.

In intermediate results, we will suppress overall factors and factors of $\alpha'$.

\subsection{The cylinder correlation function}

At this point our correlation function \Ref{corr} is given by:
\bea  \label{corrintermed}
&& \left\langle \left[\pd X^1 + i (k_1 \cdot \psi) \psi^1 \right](z_1) \left[\pd \bar{X}^1 + i (k_2 \cdot \psi) \bar{\psi}^1 \right](z_2) \ \pd \bar{X}^1(z_3) \pd \bar{Z}^2(z_4) \pd \bar{Z}^3(z_5) \prod_{i=1}^5 e^{i k_i \cdot X}(z_i) \right\rangle \nonumber  \\
&=& \left \langle \left[\pd X^1 + i (k_1 \cdot \psi) \psi^1 \right](z_1) \left[\pd \bar{X}^1 + i (k_2 \cdot \psi) \bar{\psi}^1 \right](z_2) \ \prod_{i=1}^5 e^{i k_i \cdot X}(z_i) \right\rangle
\langle \pd \bar{Z}^1(z_3) \pd \bar{Z}^2(z_4) \pd \bar{Z}^3(z_5) \rangle   \nonumber
\eea
The first factor is essentially a well-known one-loop two-point function of gauge bosons
(see e.g.\ \cite{Kiritsis:1997hj,Antoniadis:2002cs,Berg:2004sj}).
For completeness we sketch this calculation here.
It splits into four terms, but the two cross terms clearly vanish.
The $\partial X^1 \partial \bar{X}^1$ term actually vanishes, as we proceed to show. It consists only of the bosonic fields:
\beq
 \left\langle \pd X^1 (z_1) \pd \bar{X}^1 (z_2) \pd \bar{Z}^1(z_3) \pd \bar{Z}^2(z_4) \pd \bar{Z}^3(z_5) \prod_{i=1}^5 e^{i k_i \cdot X}(z_i) \right\rangle \ .
 \eeq
Contraction of all terms gives
\beq
\mathcal{B}(z_1, z_2, z_3, z_4, z_5) \sum_{\alpha} \langle 1 \rangle_{\alpha} =  0
\eeq
where
${\mathcal B}$ is some function of the
insertion points and the sum is over all spin structures $\alpha$.
Bosons are ignorant of the spin structure, so we were able to pull ${\mathcal B}$ out of the sum,
then all that remains is the partition function which  vanishes in a supersymmetric model.
The amplitude therefore reduces to the fourth term in \Ref{corrintermed}:
\beq
\left\langle (k_1 \cdot \psi) \psi^1 (z_1) (k_2 \cdot \psi) \bar{\psi}^1 (z_2) \right\rangle \left\langle  \prod_{i=1}^5 e^{i k_i \cdot X}(z_i) \right\rangle \left\langle \pd \bar{Z}^1(z_3) \pd \bar{Z}^2(z_4) \pd \bar{Z}^3(z_5) \right\rangle \ .
\eeq
which splits into correlation functions involving the  fermions, the momentum exponentials and the internal bosonic fields.
As the fermion correlator is universal to all cases we evaluate it next.

\subsection{Fermion part of gauge-boson two-point-function}

The computation of the fermion correlator is straightforward,
see for example \cite{10075145}.
 The basic correlator of fermionic fields on the cylinder
 is given in \Ref{wsfermion}.
Then:
\beq
\left\langle (k_1 \cdot \psi) \psi^1 (z_1) (k_2 \cdot \psi) \bar{\psi}^1 (z_2) \right\rangle= (k_1 - a_1) \cdot (k_2 - a_2) S_\alpha^2 (z_1-z_2) \langle 1 \rangle_\alpha
\eeq
The factor $(k_1 - a_1) \cdot (k_2 - a_2)$ encodes the kinematics of the gauge-boson kinetic term: $F^{\mu \nu}F_{\mu \nu}$, which we can formally extract, so
in the rest of the calculation we suppress this factor. To sum over spin structures, we reexpress \cite{Kiritsis:1997hj,Antoniadis:2002cs,0607224}
\beq  \label{szegosquare}
S_\alpha^2 (z_1-z_2) = \wp (z_1-z_2) - e_{\alpha -1},
\eeq
where $e_{\alpha-1} =  -4 \pi i \pd_\tau \ln \left( {\vartheta_\alpha(0, \tau)/ \eta(\tau)} \right)$, see e.g. \cite{0607224}. Here $\wp$ is the Weierstrass function
\be \label{weierstrass}
{\wp}(z,\tau) = -\partial^2_{z} \log \tht_1(z,\tau) +4\pi i \partial_{\tau} \log \eta(\tau) \, .
\ee
For comments on these relations, see appendix \ref{sec:fermcorr}.

As the Weierstrass $\wp$-function
and the $ \pd_\tau \ln  \eta(\tau)$ piece of $e_{\alpha-1}$ are both
 independent of
spin structure $\alpha$, the summation over spin structures for these terms
produces the supersymmetric partition function, which vanishes.
Therefore this correlator only receives contributions from \beq
-4 \pi i  \frac{\pd_\tau \vartheta_\alpha(0, \tau)}{\vartheta_\alpha(0, \tau)}  \ .
\eeq
Now as $\pd_\tau \vartheta_\alpha=1/(4\pi i) \cdot \pd_z^2 \vartheta_\alpha$ we can write the contributions of the fermionic parts of gauge bosons to the overall amplitude as
\beq
-\frac{\vartheta_\alpha^{\prime \prime}(0, \tau)}{\vartheta_\alpha(0, \tau)} \langle 1 \rangle_\alpha,
\eeq
where primes denote derivatives with respect to $z$. We see that the contribution is independent of the worldsheet positions of the gauge bosons, a general phenomenon
in orbifold sectors with enhanced supersymmetry. Later, we will combine this result with the bosonic correlators and the fermionic partition function. Next, we consider another ingredient of the calculation --- the classical solutions over winding modes.

\subsection{Correlator of internal bosonic fields --- sums over winding modes}
\label{winding}

The part of the overall correlation function that is sensitive to the flavour structure of the Yukawa coupling is the correlator of internal bosonic fields. A potential Yukawa coupling of the form $C^1 C^2 C^3$ is probed by a correlator
\beq
\left\langle \pd \bar{Z}^1(z_3) \pd \bar{Z}^2(z_4) \pd \bar{Z}^3(z_5) \right\rangle \ .
\eeq
As we discuss in detail in appendix \ref{sec:intclascorr}, this
has a classical part which is found as  a sum over classical solutions
\beq
\langle ( \pd \bar{Z}^i )^3 \rangle =\sum_{\substack{\textrm{classical} \\ \textrm{solutions}}} (\pd \bar{Z}_\textrm{cl}^i)^3 \ \langle 1 \rangle \ e^{-S_{\textrm{cl}}} \ .
\eeq
Non-trivial classical backgrounds correspond
to winding modes of stretched strings, so we need to examine the appearance of such modes in our setup.
We will use the classification in section \ref{allYukawas} to distinguish the various
actions of the orbifold on the (super-)string coordinates.

\subsection*{D3-D3 models}

We begin by examining the case where both gauge bosons and chiral matter are located on stacks of D3-branes.
We generate chiral matter by placing a D3 brane at an orbifold singularity at $(0,0,0)$,
and place another stack supporting the gauge bosons away from the origin at $(0,0,r_c)$
where we choose the two stacks of branes only to be separated in the third 2-torus by a distance $r_c=r_1 + \bar{U} r_2$. (For our convention about this, see \ref{sec:intclascorr}.) This setup is general enough to examine the desired physical effects.

Winding solutions for open strings exist for directions with Dirichlet-Dirichlet (DD) boundary conditions. In principle, strings can stretch in any of compact dimensions for this setup.
 However the contribution of such modes depends on the orbifold twists (see section \ref{allYukawas}):
\begin{enumerate}
\item Completely twisted directions do not allow for winding modes as they are projected out. A classical solution cannot be given for any combination of $\langle \pd \bar{Z}^r \pd \bar{Z}^s \pd \bar{Z}^t \rangle$ and thus completely twisted sectors do not contribute to the overall result.
\item In partially twisted sectors, winding is only allowed along directions on the untwisted 2-torus,
which we have chosen to be the third, and only combinations of fields polarised in this 2-torus can contribute. Thus only three-scalar-operators of the form $\langle \pd \bar{Z}^3 \pd \bar{Z}^3 \pd \bar{Z}^3 \rangle$ give a non-vanishing classical solution whereas  other configurations do not contribute.
\item Untwisted sectors can in principle support winding modes on all of the subtori. However, this is not actually relevant as the contributions of the untwisted sectors are zero due to the vanishing of the quantum correlator for $\mathcal{N}=4$ supersymmetry. This will be different for D3-D7 models.
\end{enumerate}
To summarise, for D3-D3 models the only contribution to the amplitude arises from Yukawa couplings of the form $C^r C^r C^r$ where $r$ labels a 2-torus that is left untwisted in a $\mathcal{N}=2$ orbifold sector. Gauge-invariant operators of this form only arise in  the following orbifolds: $\mathbb{Z}_6$ with $\theta=\frac{1}{6}(1,1,-2)$ and $\mathbb{Z}_6^{\prime}$ with $\theta=\frac{1}{6}(1,-3,2)$. The relevant superfield term
there is $C^3 C^3 C^3$ and the classical solution is then given by (see  \ref{sec:intclascorr}):
\begin{align}
\label{clsolD3D3}
\nn \langle \pd \bar{Z}^3 \pd \bar{Z}^3 \pd \bar{Z}^3 \rangle= & \ \sum_{m,n}
\alpha'^{3/2} c^3 (\bar{r}_c + m + n{U})^3 e^{-t L_{mn}^2}
 \\
 &=\sum_{m,n}  \bar{r}_{mn}^3
 e^{-t L_{mn}^2} \\
&= \left(-{\sqrt{\alpha'} \over \pi ct }\frac{\partial}{\partial {r}_c}\right)^3 \mathcal{Z}(t)
\end{align}
where we defined
\be  \label{defrtilde}
{r}_{mn} = \alpha'^{1/2} c\, (r_c + m + n\bar{U})
\ee
and we have from \Ref{classact} that
\be   \label{Lmndef}
 L_{nm}^2 =  \pi c^2   |r_c + n+m\bar{U}|^2  = { \pi \over \alpha'} |r_{mn}|^2
\ee
 and
where $\mathcal{Z}(t)$ is the partition function over winding modes, as given in equation \eqref{Zt}.

\subsection*{D3-D7 models}

In this case the gauge fields are located on a stack of D7-branes whereas chiral superfields are supported on a stack of D3-branes. We let the D7-branes wrap the first two subtori and be located at $r_c$ of the third 2-torus, with D3-branes at the origin.
We first note that D3-D7 winding modes are only allowed on the third torus, as on the first two subtori the Neumann-Dirichlet boundary conditions
prohibit such modes.
Thus the classical solution automatically vanishes for any combination of fields except $\langle \pd \bar{Z}^3 \pd \bar{Z}^3 \pd \bar{Z}^3 \rangle$.

As before, winding states are only allowed on untwisted tori. We therefore restricted to partially twisted and untwisted sectors of the orbifold as defined above. For D3/D7 systems the untwisted sectors only preserve $\mathcal{N}=2$ supersymmetry, and so the  correlator does not vanish
and such sectors can give non-zero results. This gives an effect in more models than for the D3-D3 case where only $\mathbb{Z}_6$ and $\mathbb{Z}_6^{\prime}$ models were interesting --- here we also have non-zero correlators for $\langle \pd \bar{Z}^3 \pd \bar{Z}^3 \pd \bar{Z}^3 \rangle$ in the untwisted sector of $\mathbb{Z}_3$ orbifolds.

In an orbifold we also need to include images of the D7-branes if they are not at orbifold fixed points.
 For each stack of D7-branes we add $N-1$ identical stacks at the image loci of the original stack. Thus, for D7-branes located at $z_3=r_c$ on the third 2-torus we include images at $z_3= e^{2 \pi i \theta_3 j } r_c$ for $j=1, \ldots N-1$.
 To obtain a result invariant under the orbifold, we also include strings stretching between the D3-branes and all the image D7-branes.
   In our present calculation this leads to the following modification: each stack of D7-branes will give rise to winding solutions stretching from the D3-stack to the D7-stack. The corresponding classical partition function $\mathcal{Z}(t)$ is the same for strings stretching between the D3-branes and any of the stacks of D7-branes. This is due to the fact that the lattice used to construct the torus is invariant under the orbifold action. However, as $r_c \rightarrow e^{2 \pi i \theta_3 j } r_c$ for image branes the internal correlator $\langle \pd \bar{Z}^3 \pd \bar{Z}^3 \pd \bar{Z}^3 \rangle$ is modified accordingly:
\beq
\label{joeade}
 \left(-{\sqrt{\alpha'} \over \pi ct }\frac{\partial}{\partial {r}_c}\right)^3 \mathcal{Z}(t)
 \rightarrow
  \left(-{\sqrt{\alpha'} \over \pi ct }e^{-2 \pi i \theta_3 j } \frac{\partial}{\partial {r}_c}\right)^3 \mathcal{Z}(t)
 \eeq
for $j=1, \ldots N-1$. However, the only orbifold models that allow for a non-vanishing correlator $\langle \pd \bar{Z}^r \pd \bar{Z}^r \pd \bar{Z}^r \rangle$ are generated by the point groups $\mathbb{Z}_3$, $\mathbb{Z}_6$ or $\mathbb{Z}_6^{\prime}$ for which $\theta_r=\pm \frac{1}{3}$ (see table \ref{tableYukawas}).
Correspondingly, all images contribute the same result:
\beq
 \left(-{\sqrt{\alpha'} \over \pi ct }\frac{\partial}{\partial {r}_c}\right)^3 \mathcal{Z}(t) \rightarrow e^{\pm 2 \pi i j } \left(-{\sqrt{\alpha'} \over \pi ct }\frac{\partial}{\partial {r}_c}\right)^3 \mathcal{Z}(t)= \left(-{\sqrt{\alpha'} \over \pi ct }\frac{\partial}{\partial {r}_c}\right)^3 \mathcal{Z}(t)  .
\eeq

\subsection{Completing the calculation}

In this section we combine the results of the previous sections. We also need to include the correlator over momentum exponentials which contributes
\beq
\left\langle \prod_{i=1}^5 e^{i k_i \cdot X}(z_i) \right\rangle = \prod_{i < j} e^{-k_i \cdot k_j \mathcal{G}(z_i-z_j)}
\eeq
where $\mathcal{G}(z_i-z_j)$ is the Green's function with NN boundary conditions. The amplitude is then integrated over worldsheet positions and the modular parameter of the cylinder $t$.

\subsection*{D3-D3 models}
In \S\ref{winding} we established that the only non-zero contribution arises in partially twisted sectors of the orbifold. To be specific we take $\theta_3=0$ mod $1$ and $\theta_1+\theta_2=0$ mod $1$. Collating the previous results, the complete amplitude is
\begin{align}
\nn \mathcal{A} \propto & \int \frac{\textrm{d}t}{t} \int \textrm{d}z_1 \textrm{d}z_2 \textrm{d}z_3 \textrm{d}z_4 \textrm{d}z_5 \frac{1}{(2 \pi^2 t)^2} \left(\prod_{i < j} e^{-k_i \cdot k_j \mathcal{G}(z_i-z_j)} \right) \\
\nn & \sum_{\substack{\alpha, \beta \\ = 0, 1}} \frac{\eta_{\alpha \beta}}{2} (-1)^{\alpha} \frac{{\vartheta}_{\alpha \beta}^{\prime \prime}(0)}{\eta^3} \frac{{\vartheta}_{\alpha \beta}(0)}{\eta^3} \left( \prod_{k=1}^2 (-2 \sin \pi \theta_k)\frac{\vartheta_{\alpha \beta}(\theta_k)}{\vartheta_{1}(\theta_k)}\right) \\
&  \left(-{\sqrt{\alpha'} \over \pi ct }\frac{\partial}{\partial {r}_c}\right)^3 \mathcal{Z}(t) \ ,
\end{align}
where $\mathcal{Z}(t)$ is the partition function given in equation \eqref{Zt}.
We can apply the Riemann identity \eqref{D3D3partial} to collapse the sum over spin structures to a constant.\footnote{If we repeated the analysis for the untwisted orbifold sector, the sum over spin-structures would be identically zero due to \eqref{D3D3untw}. Hence untwisted sectors do not contribute in the D3-D3 case.} Further, in the partially twisted sectors of the relevant orbifolds ($\theta^3$ sector of both $\mathbb{Z}_6$ and $\mathbb{Z}_6^{\prime}$) we have $\theta_1=\theta_2=1/2$. Thus we are left with:
\begin{align}
\label{result1D3D3}
\mathcal{A} \propto & \int \frac{\textrm{d}t}{t} \int \textrm{d}z_1 \textrm{d}z_2 \textrm{d}z_3 \textrm{d}z_4 \textrm{d}z_5 \frac{ 1}{(2 \pi^2 t)^2} \left(\prod_{i < j} e^{-k_i \cdot k_j \mathcal{G}(z_i-z_j)} \right) \left(-{\sqrt{\alpha'} \over \pi ct }\frac{\partial}{\partial {r}_c}\right)^3 \mathcal{Z}(t) \ .
\end{align}
Before analysing this expression we present the result for the D3-D7 model as it will be identical.

\subsection*{D3-D7 models}
For the case of D3-D7 models the untwisted sector only preserves $\mathcal{N}=2$ (in contrast to $\mathcal{N}=4$ for D3-D3) and we thus expect a non-zero result. We combine our previous results with the appropriate partition function to obtain
\begin{align}
\nn \mathcal{A} \propto & \int \frac{\textrm{d}t}{t} \int \textrm{d}z_1 \textrm{d}z_2 \textrm{d}z_3 \textrm{d}z_4 \textrm{d}z_5 \frac{1}{(2 \pi^2 t)^2} \left(\prod_{i < j} e^{-k_i \cdot k_j \mathcal{G}(z_i-z_j)} \right) \\
\nn & \sum_{\substack{\alpha, \beta \\ = 0, 1}} \frac{\eta_{\alpha \beta}}{2} \frac{{\vartheta}_{\alpha \beta}^{\prime \prime}(0)}{\eta^3} \frac{{\vartheta}_{\alpha \beta}(0)}{\eta^3} {\left(\frac{\thw{ 1/2 -\alpha/2}{ \beta/2} (0)}{\thw{0} {1/2} (0)} \right)}^2 \\
&\left(-{\sqrt{\alpha'} \over \pi ct }\frac{\partial}{\partial {r}_c}\right)^3 \mathcal{Z}(t) \ ,
\end{align}
with $\mathcal{Z}(t)$ again given by \eqref{Zt}.
As above we get
\begin{align}
\label{result2D3D7}
\mathcal{A} \propto & \int \frac{\textrm{d}t}{t} \int \textrm{d}z_1 \textrm{d}z_2 \textrm{d}z_3 \textrm{d}z_4 \textrm{d}z_5 \frac{1}{(2 \pi^2 t)^2} \left(\prod_{i < j} e^{-k_i \cdot k_j \mathcal{G}(z_i-z_j)} \right) \left(-{\sqrt{\alpha'} \over \pi ct }\frac{\partial}{\partial {r}_c}\right)^3 \mathcal{Z}(t) \ ,
\end{align}
which is identical to \eqref{result1D3D3} and hence we discuss both results together.

\subsection{Result --- evaluating the integral over $t$}
\label{disc1}

We now perform the integrals in equations \eqref{result1D3D3} and \eqref{result2D3D7}. The only momentum-dependent term is the correlator over momentum exponentials. This is also the only term depending on the worldsheet positions.
As we are only interested in the amplitude at vanishing momenta $k_i$, the integrand can be considerably simplified. After the spin structure sum has collapsed to a number, there is no possible source of pinching singularities that could make the exponential of bosonic propagators  $e^{-2 \alpha' k_i\cdot k_j{\mathcal G}(z_i-z_j)}$ contribute. The limit $\alpha' k_i \rightarrow 0$ can then be taken without subtleties, removing any dependence of the amplitude on world-sheet positions. Subsequently, the integral over world-sheet positions can be performed giving a contribution $\int \prod_{i=1}^5 \textrm{d}z_i \propto t^5$.

Then, ignoring other overall factors, we want to compute the following integral over the world-sheet modulus $t$, where we now include the winding modes explicitly:
\be  \label{startint}
{\cal A} \sim \int_0^{\infty} \frac{dt}{t} \frac{1}{t^2} t^5 \sum_{m,n} \bar{r}_{mn}^3  \exp\left( -  { \pi t \over \alpha'} |r_{mn}|^2 \right) \, ,
\ee
where ${r}_{mn}$ was defined in \Ref{defrtilde}
and $L_{mn}$ is given in \Ref{Lmndef}. The $t$ integral is elementary and produces
\be  \label{intintermediate}
{\cal A} \sim \sum_{m,n} \bar{ r}_{mn}^3 \frac{2}{ ( \pi  / \alpha')^3  |r_{mn}|^6} = {2 \over \pi^3} \left( \frac{2\alpha' U_2}{ \tilde{T}_2}
\right)^{3/2} \ \sum_{m,n} \frac{1}{ (r_c + m + n\bar{U})
^3} \, .
\ee
The double sum over $m$ and $n$ is of the form
\be  \label{wp}
\sum_{m,n}{2 \over (z-(m+n\bar{U}))^3} =-\wp'(z,\bar{U}) ,
\ee
where we identified the first derivative $\wp'(z)$ of the Weierstrass $\wp(z)$ function.
More explicitly, our  first answer for the amplitude is
\be
{\cal A} \sim {1 \over \pi^3} \left( \frac{2\alpha' U_2}{ \tilde{T}_2}
\right)^{3/2} \partial_{r_c} \wp \left(r_c, \bar{U}\right) \, . \label{eq:answ}
\ee
We can arrive at the result \eqref{eq:answ} from \Ref{startint} by a different method.
 In the integrand, we restore the triple derivative from
 above and instead of \Ref{startint}, we recast  the sum over winding modes as a generalised (Riemann) theta function \eqref{doubletheta}:
\be
\mathcal{A} = \mathcal{C} {\left(\frac{U_2}{\tilde{T}
_2}\right)}^{3/2} \int_0^{\infty} \frac{\textrm{d}t}{t} \frac{1}{t^2} t^5\left(-{\sqrt{\alpha'} \over \pi ct }\frac{\partial}{\partial {r}_c}\right)^3\tha{\vec{R} }{ \vec 0 } {\vec 0 , itG/2} \ ,
\ee
where $G$ is the metric on the torus wrapped \eqref{torusmetric}, $\vec{R}=(r_1,r_2)$ is a real 2-component vector with $r_c=r_1 + \bar{U} r_2$ and $\mathcal{C}$ is a numerical constant which we leave undetermined. We now want to move the derivatives with respect to the complex separation $r_c$ outside of the integral over $t$. Unlike
in the previous method, the integral now apparently becomes divergent, so
we  introduce a UV cutoff $\Lambda$  in this intermediate step. Fortunately, we know from the previous method
that the result is finite, so this is no cause for concern. The remaining integral
is now the same as that which occurs in usual gauge coupling renormalisation (i.e.\ without the three scalar insertions we have here), so we can perform the integral
as in \cite{0404087} to find:
\be
\label{AASSSresult}
\mathcal{A} = -\mathcal{C} {\left(\frac{U_2}{\tilde{T}
_2}\right)}^{3/2} \frac{\partial^3}{{\partial r_c}^3} \int_{1/ \Lambda^2}^{\infty} \frac{\textrm{d}t}{t} \tha{ \vec{R}}{ \vec 0}{\vec 0 , itG/2} = -\mathcal{C} {\left(\frac{U_2}{\tilde{T}
_2}\right)}^{3/2} \frac{\partial^3}{{\partial r_c}^3} \left(\Lambda^2 \sqrt{G} - {\cal G}(0, r_c)
 \right) \ ,
\ee
with the scalar boson propagator on the torus\footnote{see e.g.\
section 7.2 of \cite{Polchinski1}. Note
that we give two arguments for ${\mathcal G}(x,y)$,
because although here the transverse space is flat
and we have translational invariance, in general one might be interested
in background gravitational and $p$-form fields
in which case the Green's function is not a function only of the difference $x-y$.}
\be   \label{bosprop}
{\cal G}(0, r_c) = -\ln {\left|\frac{\vartheta_1(r_c,\bar{U})}{\eta(\bar{U})} \right|}^2 - \frac{2 \pi {\textrm{Im}(r_c)}^2}{U_2} \, ,
\ee
i.e.\ ${\cal G}(0, r_c) $ is  the solution of the Laplace equation on the torus with a delta function source and a neutralising background charge.
Happily, as expected from our previous calculation, the cutoff $\Lambda$ disappears from the final result in virtue of the differentiation. Further, we see that the last term of the above expression is proportional to $r_c^2$ and will be annihilated by the three derivatives. We find \footnote{The sum and product
expressions for the theta functions only converge for $U_2<0$ in this case,
which is unconventional. To recover the $U_2>0$ convention, we could switch
the convention of negative orbifold charge for the scalars, so those
vertex operators become unbarred, and simultaneously switch the definition
of $r_c$ to $\bar{r}_c$.}
\be     \label{Aresult}
\mathcal{A} = -\mathcal{C} {\left(\frac{U_2}{\tilde{T}
_2}\right)}^{3/2} \frac{\partial^3}{{\partial r_c}^3} \ln \vartheta_1 \left(r_c,\bar{U} \right)  =   \mathcal{C} {\left(\frac{U_2}{\tilde{T}
_2}\right)}^{3/2} \frac{\partial^3}{{\partial r_c}^3}{\cal G}(0, r_c) \ .
\ee
To establish that this expression is in fact equivalent to \Ref{eq:answ},
we merely need to note \eqref{weierstrass}.

We have performed the integral over the cylinder worldsheet modulus $t$ in two different ways,
and we would like to highlight the differences between the two calculations.
If there had been a lower power of $t$ than two
or a lower power of $\bar r_{mn}$ than three in \Ref{startint} in \Ref{intintermediate}, the integral would have failed to converge, just like the integral did in the second method when we peeled off the three derivatives.
 These powers
originally came from  from three holomorphic vertex operators for the three spacetime scalars
in the Yukawa-like coupling,
and this is an example of the good behaviour of this amplitude alluded
to in section \ref{localvsglobal}. In the analogous
calculation of the  gauge coupling correction in   \cite{Berg:2004sj},  without the Yukawa-style triple scalar insertions, those divergences appeared in each separate diagram due to tadpoles in the open-string UV limit, and could only be cancelled between diagrams by enforcing tadpole cancellation. We see that for the particular calculation of interest in this paper, this is not a problem.

\subsubsection{Discussion}
\label{sec:discussion}
The above results give the dependence of the gauge kinetic function on a stack of D3/D7-branes due to
chiral matter on another stack of D3-branes.  As discussed in section \ref{superspace}, the term in the superspace action we are
interested in is
\be  \label{fWW}
\int d^4 x\,  d^2 \theta f_a(\Phi) \hbox{Tr}(W_{\alpha} W^{\alpha}).
\ee
First, we will see if we can understand the scaling behaviour obtained in \Ref{Aresult}.
For this purpose, let us set the two torus radii equal: $\tilde{T}_2 = R_1 R_2 \sin \theta_U$ and $R_1 \sim R_2 \sim R$.
We find
\be
\label{sugraX}
\mathcal{A} \sim  {\left(\frac{U_2}{\tilde{T}
_2}\right)}^{3/2} \frac{\partial^3}{{\partial r_c}^3} \ln \vartheta_1(r_c,\bar{U}) \, \sim \, \tilde{T}_2^{-3/2} \sim R^{-3}\; .
\ee
We can in fact understand the $R^{-3}$ scaling from an analysis in four-dimensional effective supergravity. We have computed the superfield component coupling\footnote{Note that this is Weyl invariant,
so it is not affected by the Weyl rescaling  we need to go to Einstein frame.}
\be  \label{intphi3FF}
{1 \over 2}\hat{y}_{ijk} \int \!d^4 x\, \sqrt{-g}~{\rm Tr}\left( \phi_1 \phi_2 \phi_3\right) g^{\mu \nu} g^{\lambda \rho} {\rm Tr} \left( F_{\mu \lambda}  F_{\nu \rho} \right) \; .
\ee
in the string effective action.
In terms of superfields, we are looking at a term $\Phi^1 \Phi^2 \Phi^3$ in the gauge kinetic function $f_a(\Phi^i)$.
The interaction \Ref{intphi3FF} comes from the term \Ref{fWW} in the superspace action, once the matter fields are canonically normalised.
The K\"ahler metric for matter on D3 branes scales as $K_{i \bar{i}} \sim \mc{V}^{-2/3} \sim R^{-4}$. Performing the canonical normalisation and restricting to flat space,
we obtain
\be
\int \! d^4 x \,  \frac{1}{R^3 M_s^3}~{\rm Tr}( \phi_1 \phi_2 \phi_3) ~{\rm Tr}(F_{\mu \nu} F^{\mu \nu}) \; ,
\ee
where $M_s \sim {M_P}/{\sqrt{\mc{V}}}$. The $R^{-3}$ scaling of (\ref{sugraX}) is then consistent with supergravity expectations.

Our second comment is that as noted in the introduction, \Ref{fWW} generates two distinct Lagrangian terms,
$$
\int d^4 x ~{\rm Tr}(\phi_1 \phi_2 \phi_3)
~{\rm Tr}( F_{\mu \nu} F^{\mu \nu})\quad  \hbox{ and } \quad \int d^4 x
~{\rm Tr}(\psi_1 \psi_2 \phi_3)~{\rm Tr}( \lambda_\alpha \lambda^{\alpha}).
$$
where $\lambda$ is the gaugino.
In this section we have studied the former coupling. In Appendix \ref{sec:gaugekincorr2} we give the details of a  consistency check that shows that the the latter coupling perfectly agrees with these results, as it must.

\begin{figure}
	\centering
		\includegraphics[width=0.40\textwidth]{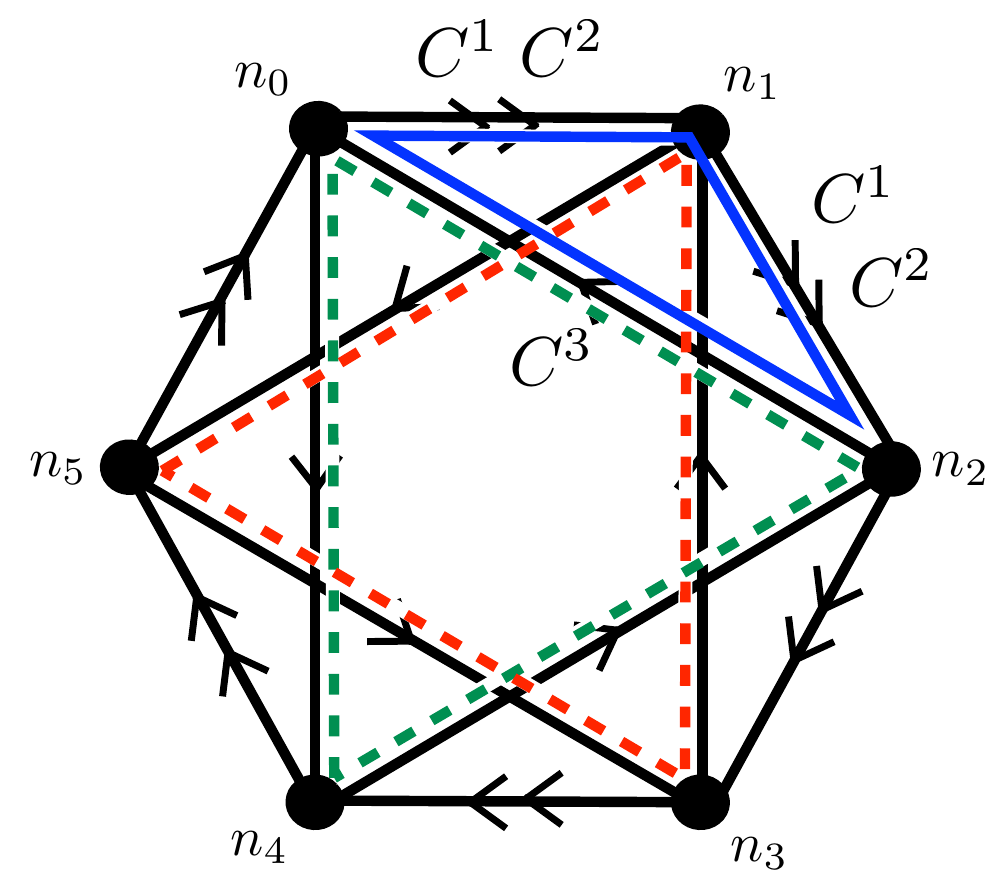}
	\caption{Quiver theory on the worldvolume of D3-branes at $\mathbb{C}^3/\mathbb{Z}_6$ orbifold singularities. There are three matter superfields for each label $C^r$ with $r=1,2,3$. The fields $C^1$ and $C^2$ form the outer matter lines while the field $C^3$ is responsible for the inner matter lines.
	The blue solid triangle is the original $C^1C^2C^3$ operator,
	the red and green dashed triangles are two examples of $C^3C^3C^3$ operators
	generated by de-sequestering.}
\label{fig:Z6quiver}
\end{figure}

We found that the only non-zero contribution can arise from Yukawa couplings with flavour structure $C_{(l)}^3 C_{(m)}^3 C_{(n)}^3$, where the sub-index enumerates three fields originating from the third torus forming a triangle in the quiver diagram (for example, see figure \ref{fig:Z6quiver}). Below, for completeness, we list all supersymmetric orbifolds and denote as ``relevant A-term coupling'' those operators
 that receive a non-vanishing contribution in our calculation. \\ \\
\begin{tabular}{|c|c|c|c|c|}\hline
Orbifold & Twist vector $\theta$ & Partially twisted sectors & \multicolumn{2}{|c|}{Relevant A-term coupling}  \\ \cline{4-5}
group  & & & D3-D3 & D3-D7 \\ \hline \hline
$\mathds{Z}_3$ & $\frac{1}{3} (1,1,-2)$ & & & $(C^3)^3$ \\ \hline
$\mathds{Z}_4$ & $\frac{1}{4} (1,1,-2)$ & $Z^2$ under $\theta^2$ & & \\ \hline
$\mathds{Z}_6$ & $\frac{1}{6} (1,1,-2)$ & $Z^3$ under $\theta^3$ & $(C^3)^3$ & $(C^3)^3$ \\ \hline
$\mathds{Z}_6'$ & $\frac{1}{6} (1,-3, 2)$ & $Z^2$ under $\theta^2$ and $\theta^4$, $Z^3$ under $\theta^3$ &$(C^3)^3$ & $(C^3)^3$  \\ \hline
$\mathds{Z}_7$ & $\frac{1}{7} (1, 2,-3)$ &  & &  \\ \hline
$\mathds{Z}_8$ & $\frac{1}{8} (1,3,-4)$ & $Z^3$ under $\theta^2$ & &  \\ \hline
$\mathds{Z}_8'$ & $\frac{1}{8} (1, -3, 2)$ & $Z^3$ under $\theta^4$ & &  \\ \hline
$\mathds{Z}_{12}$ & $\frac{1}{12} (1,-5, 4)$ & $Z^3$ under $\theta^3$ &$(C^3)^3$ & $(C^3)^3$  \\ \hline
$\mathds{Z}_{12}'$ & $\frac{1}{12} (1, 5, -6)$ & $Z^3$ under $\theta^n$ for $n=2, 4, 6, 8, 10$. & & \\ \hline
\end{tabular} \\ \\

The entries in the table take into account the fact that de-sequestering is mediated by winding states in {\it partially twisted} sectors in the case of D3-D3 models (``LVS''), and in {\it untwisted} sectors in the case of D3-D7 models (``KKLT''). Most importantly, the flavour structure of the induced non-perturbative superpotential terms does not line up with the flavour structure of the tree-level Yukawas. The tree-level superpotential $W = \epsilon_{r,s,t} \textrm{Tr}(C^r C^s C^t)$ only contains the field combination $C^1C^2C^3$ and permutations. Hence, de-sequestering is indeed a possible source of flavour violation.

We will now make another observation about a detailed difference between the ``KKLT'' and ``LVS'' realisations
of the nonperturbative effects in the orbifold,
by reinstating the traces over Chan-Paton-factors. To be explicit, we will examine the example of the $\mathbb{Z}_6$ orbifold with twist vector $\theta= \frac{1}{6}(1,1,-2)$. We focus on the Chan-Paton factors on the matter D3-brane. The action of the orbifold on the Chan-Paton factors is captured by an orbifold twist matrix $\gamma_{\theta}$ which transforms in the regular representation. For $\omega=e^{-\frac{2 \pi i}{6}}$ we can write
\be
\gamma_{\theta_{D3}} = \textrm{diag}(\mathbb{I}_{n_0}, \omega \mathbb{I}_{n_1}, \omega^2 \mathbb{I}_{n_2}, \omega^3 \mathbb{I}_{n_3}, \omega^4 \mathbb{I}_{n_4}, \omega^5 \mathbb{I}_{n_5}) \ .
\ee
The partially twisted $\mathcal{N}=2$ sector is generated by $\theta^3$ and thus is accompanied by a twist matrix
\be
\gamma_{\theta_{D3}}^3 = \textrm{diag}(\mathbb{I}_{n_0}, - \mathbb{I}_{n_1}, \mathbb{I}_{n_2}, - \mathbb{I}_{n_3},  \mathbb{I}_{n_4}, - \mathbb{I}_{n_5}) \ .
\ee
We see that the trace over Chan-Paton factors on D3-branes produces a factor $\textrm{Tr}(C^3 C^3 C^3)$ in untwisted sectors and $\textrm{Tr}(C^3 C^3 C^3 \gamma_{\theta_{D3}}^3)$ in the
partially twisted sector.
Thus, reinstating the Chan-Paton traces in front of our previously calculated amplitudes they read:
\begin{align}
\textrm{untwisted:} \qquad & \textrm{Tr}(C^3 C^3 C^3) \textrm{Tr}(AA) \times \mathcal{A}(r_c, U) \\
\textrm{partially twisted:} \qquad & \textrm{Tr}(C^3 C^3 C^3\gamma_{\theta_{D3}}^3) \textrm{Tr}(AA\gamma_{\theta_{D7}}^3) \times \mathcal{A}(r_c, U) \; .
\end{align}
In the following we denote the Chan-Paton factors for the visible matter fields as $C_{i,j}^3$ where the labels $i,j$ indicate that the field transforms in the bifundamental representation $(n_i, \bar{n}_j)$. The allowed combinations can be read off the quiver diagram shown in figure \ref{fig:Z6quiver}. Given the form of $\gamma_{\theta_{D3}}$ above we find that
\begin{align}
\textrm{Tr}(C_{0,2}^3 C_{2,4}^3 C_{4,0}^3)&=\hphantom{-} \textrm{Tr}(C_{0,2}^3 C_{2,4}^3 C_{4,0}^3 \ \gamma_{\theta_{D3}}^3) \ , \\
\textrm{Tr}(C_{1,3}^3 C_{3,5}^3 C_{5,1}^3)&=- \textrm{Tr}(C_{1,3}^3 C_{3,5}^3 C_{5,1}^3 \ \gamma_{\theta_{D3}}^3) \ .
\end{align}
It is the appearance of the minus sign in the second line that we want to emphasise. The traces on the left appear in models with bulk D7-branes, while the traces on the right are found in de-sequestering calculations with D7-branes wrapping a small cycle. Thus we observe that, depending on the way of accommodating the non-perturbative effects, one can generate couplings with opposite signs:
\be
\hat{Y}^{np}(\textrm{bulk D7})= - \hat{Y}^{np}(\textrm{D7 on small cycle}) \ .
\ee
Thus the details of superpotential de-sequestering depend
on the particular realisation of nonperturbative stabilisation. This
observation seems like it could have some phenomenological relevance but we will
not explore it further in this paper.

\subsubsection{Connections to the closed string calculation}
\label{sec:closed}
The above result was obtained from an analysis of a one-loop open string amplitude. We can reinterpret this scattering as an exchange of a closed string in the tree channel of the cylinder amplitude. To make contact with this  closed string picture, we reexpress our brane setup in supergravity as follows. The two stacks of branes are separated by a distance $r_c$ which we can understand as the position modulus of the D3-branes supporting the chiral matter.
This matter can source closed string fields which propagate through the geometry and couple to the distant brane, thereby affecting its gauge coupling.
In \cite{0607050}, the correction to the D7 gauge coupling due to D3-brane position moduli was calculated by backreacting a D3-brane at a smooth point. 
Since the D3-brane sources the warp factor, the volume of the four-cycle wrapped by the D7-brane depends  on the D3-brane position. In our notation their result (eq. (23) on p.12) becomes:
\bea
\label{closedresult}
\delta \left(\frac{8 \pi^2}{g^2} \right) &=& T_3~\delta V_{\Sigma_4} = T_3~V_{\Sigma_4}~{\cal G}(0, r_c) = \nonumber \\
 &=& \frac{\pi}{U_2} {\left[\textrm{Im} \left( r_c \right) \right]}^2 - \frac{1}{2} \ln {\left| \vartheta_1 \left( r_c  ,U \right) \right|}^2 \ ,
\eea
where $V_{\Sigma_4}$ denotes the warped volume of the four-cycle wrapped by the D7-branes. The equation $\delta V_{\Sigma_4} = V_{\Sigma_4}~{\cal G}(0, r_c) $ follows since the perturbation in the warp factor is the Green's function of the Laplace problem, and the six-dimensional problem simplifies on this background to just the torus transverse to the wrapped D7-brane.  Note that here $r_c$ denotes the position modulus of the D3-brane, and there are no chiral (or adjoint) matter \emph{fluctuations} on the D3-brane worldvolume.

On the contrary, our calculation involves not the D3-brane positions but the chiral matter on their worldvolume. Our result \eqref{AASSSresult} is closely related to equation \eqref{closedresult} above; more precisely, our expression is given by three  derivatives with respect to the position modulus of the above result! 
Of course, we arrived at this result in an open string approach, and we did not attempt to rederive it directly in the closed string picture.

Heuristically, the three derivatives of the Green's function appear as the leading order coefficient of the Taylor expansion of \eqref{closedresult},
\be
\delta \left(\frac{8 \pi^2}{g^2} \right) = T_3~\Big( {\cal G}(0, r_c+\delta r_c) -  {\cal G}(0, r_c) \Big) = T_3~(\delta r_c)^3 \partial_{r_c}^3  {\cal G}(0, r_c) + \ldots \label{eq:naive}
\ee
where the ellipsis denotes higher order terms. The cubic term appears (by assumption) as the first gauge invariant fluctuation of the D3-brane matter around the fixed point and gives the coupling between the operator $C^3 C^3C^3$  and the volume swept by the wrapped D7-brane.

Here however, we will not make  the relation between the operators of the D3-brane gauge theory and the spatial fluctuation $\delta r_c$ around the orbifold point more precise, but we note
  that,  somewhat surprisingly, the most naive argument appears to give the correct result.
 Perhaps the very special nature of the
 states that are exchanged (corresponding to partially twisted and untwisted open string states)
 could serve to explain this simplicity.
 We will not delve deeper into this exciting topic in this paper.

\subsection{Cross-couplings due to distant branes}
It is somewhat counterintuitive
that branes can have non-negligible interactions
 even at relatively large separation.
(This is not particularly new to this paper, but has appeared
in the literature many times, e.g.~\cite{Antoniadis:1998ax,0404087,Baumann:2006cd}.)
Here
we summarise a few arguments that might provide some intuition.

\begin{figure}
\begin{center}
\includegraphics[width=0.3\textwidth]{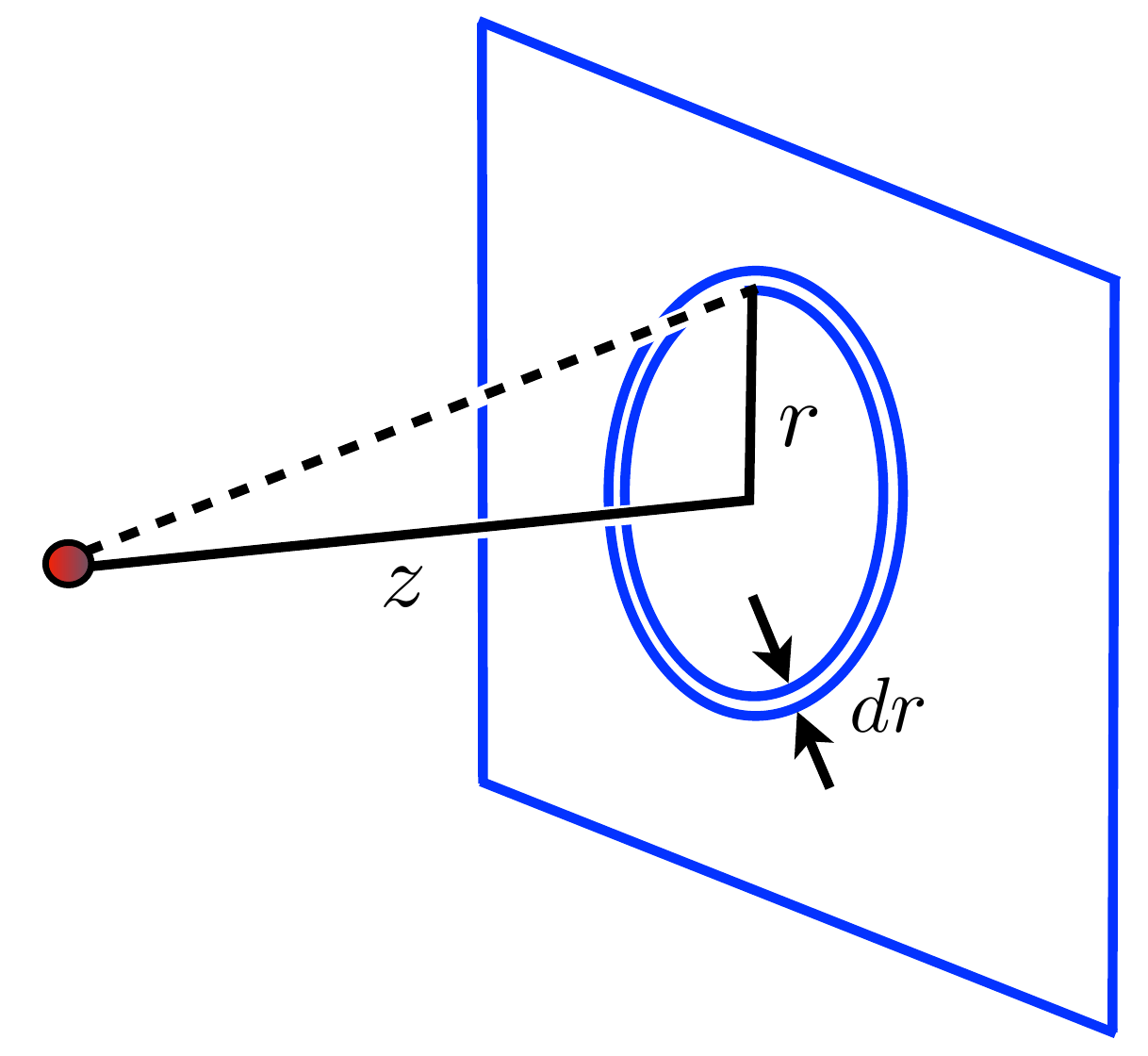}
\caption{D3-D7 noncompact toy model.
The influence of a point source on
an area element (blue circle) of thickness $dr$, where the source is located a distance $z$ from an infinite
sheet.}
\label{noncomp}
\end{center}
\end{figure}
If the perturbation due to a D3-brane
on a volume element at the position of the D7-brane is $\delta g \sim 1/R^4$ where $R=\sqrt{r^2+z^2}$ (see fig.\ \ref{noncomp}), then the total correction from integrating over volume elements
affected by this perturbation is
\[
\int r^3 dr {1 \over ((r^2+z^2)^{1/2})^4}\; \stackrel{r\rightarrow \infty}{\longrightarrow}\;  \log r \;  .
\]
The picture here is that the D7-brane is practically ``infinite'' from the point of view of the D3-brane, and therefore it  ``sees more'' of the D7-brane as it moves away, so this motion fails  to suppress the cross-interaction.
 (Of course the actual D7 is not really infinite.)
It may also be useful to note that we are computing corrections to gauge couplings, not a potential energy. If it had been the latter, one could have argued
that the logarithm would be differentiated when we compute the physical force which should therefore fall off, but this is not
the case for the gauge coupling correction, which is itself a physical quantity.

A related argument, also in effectively noncompact
(though actually only large) extra dimensions, was given in \cite{Antoniadis:1998ax}. These authors
argued that a tadpole on a D-brane corresponds to propagation of a closed string state in
the dimensions  transverse to the brane, hence for two transverse dimensions one
obtains a logarithmic correction to gauge couplings,
and they used this to argue for a gauge hierarchy with logarithmic running
due to two extra dimensions opening up at a certain scale.
The physics is different, but the logic is somewhat similar to our geometric argument in section \ref{sec:closed} that
these effects do not decouple because the partially twisted (D3-D3) or untwisted (D3-D7) open strings
correspond to closed string states that only propagate
along homologous 2-cycles. It requires some further work to make this precise
in for example a  Calabi-Yau compactification. We comment on this in the conclusions.

Another argument, unrelated to the two arguments above, is about towers of winding states in compact spaces. We see that the zero-mode sum in \Ref{Zintcl}
is an infinite sum over a tower of winding states. (Note that they are not massive string states in the sense of oscillator excitations, but rather towers  on top of a single light open string state, with masses controlled
by the size of the torus rather than just the string scale.)
As we see in \Ref{Zt}, if the torus is itself not particularly degenerate, this infinite sum over exponentially suppressed contributions  is a perfectly finite function of separation.

Finally, the appearance of non-perturbative corrections to the Yukawa couplings 
from the cycles on which D7-branes have undergone gaugino condensation is furthermore motivated already in four dimensional supersymmetric field theory. In $SU(N)$ Yang-Mills theory with $N>2$, UV cut-off $\Lambda_{UV}$, and with one massive flavour $\varphi$ with mass $m \ll \Lambda_{UV}$ transforming in the fundamental representation of $SU(N)$, gaugino condensation gives rise to a non-perturbative Affleck-Dine-Seiberg (ADS) superpotential of the form
\be
W_{UV} = \Lambda^3_{UV} \, .
\ee
The effective potential valid for energies below $\Lambda \ll m$ can be found by integrating out $\varphi$, which then gives rise to the superpotential,
\be
W_{\Lambda} = \Lambda_{UV}^{3-1/N}~m^{1/N} \, . \label{eq:ADSIR}
\ee
If the mass of $\varphi$ depends on other fields which may be uncharged under the $SU(N)$ gauge group, then the non-perturbative superpotential of \eqref{eq:ADSIR} induces a non-trivial potential for these fields.

It was pointed out in \cite{McAllister:2005mq} that if  the $SU(N)$ gauge theory with gaugino condensation is realised as the world-volume theory on a stack of $N$ D7-branes, then the presence of a distant D3-branes gives rise to massive $37$-strings stretching between the D3-brane and the stack of D7-branes, and
the effective superpotential for energies below the mass of the $37$-string is given by \eqref{eq:ADSIR}, where $m$ denotes the mass of the $37$-string. This mass depends linearly on the distance between the D3-brane and the D7-brane, and therefore on the D3-brane position modulus.

The result of the corresponding open string computation for a D3-brane at a singularity presented here suggests that a similar interpretation may be possible also in this case: equation \eqref{result2D3D7} shows that the coupling between a Yukawa operator on the D3-brane and the distant D7-brane is proportional to three derivatives of the partition function --- and thereby on the masses of the strings stretching between the D3-brane and the D7-brane --- with respect to the brane separation. This form of the coupling suggests that the mass $m$ of equation \eqref{eq:ADSIR} may depend on the operators on the D3-brane as through a Taylor expansion in the fluctuations of the brane separation. This dependence would explain why geometric separation is not sufficient to decouple the localised visible sector from the distant non-perturbative effects.




\section{Scalar corrections to Yukawa couplings}
\label{sec:corrscalar}
\subsection{Motivation}
Visible sector Yukawa couplings may also depend on hidden sector scalars. In particular in \cite{Berg:2010ha}, it was noticed that the position  of a hidden sector D3-brane can affect the $F$-term of the volume modulus, which may significantly affect the overall magnitude of the induced soft terms in the visible sector.

In this section, we consider the effects of non-vanishing vacuum expectation values of hidden sector  scalars arising from distant D7-branes wrapped on either bulk or collapsed four-cycles. While position moduli of D7-branes wrapping bulk cycles in flux compactifications generically obtain masses at the order of the flux scale, this does not have to be the case for branes wrapping small cycles. In this section, we simply compute some of the consequences of non-vanishing vevs for D7-brane scalars, and we do not address the model building required to ensure the presence of such vevs.

For non-vanishing correlators of the form,
\be
\langle \tr(\tilde{\phi} \tilde{\phi} \ldots \tilde{\phi}) \tr(\psi \psi \phi) \rangle \, ,
\ee
a vev for the hidden sector scalars $\tilde \phi$ may affect the visible sector Yukawa couplings. If the flavour structure of these couplings differs from the tree-level flavour structure, then these couplings may  allow for a Froggatt-Nielsen  mechanism for branes at singularities where a small vev for $\tilde{\phi}$ represents the expansion parameter for the CKM matrix.

Furthermore, non-vanishing couplings of this form signal the presence of certain operators in either the Kähler potential or the superpotential of the effective four-dimensional supergravity. For instance, the dependence of a  Yukawa coupling on two hidden sector fields is probed by  $\langle \tr(\tilde \phi \tilde \phi) \tr(\psi \psi \phi) \rangle$. For a visible sector with a diagonalised Kähler metric, $K_{\alpha \bar \beta} = {\rm diag}(K_{\alpha})$, the physical Yukawa coupling  is given by \cite{Kaplunovsky:1993rd},
\be
\label{Yphyshid}
Y_{\alpha \beta \gamma}^{\textrm{phys}}= e^{\mathcal{K}/2} \frac{Y_{\alpha \beta \gamma}^{\textrm{hol}}}{\sqrt{K_{\alpha}K_{\beta} K_{\gamma}}} \ .
\ee
By expanding this expression to quadratic order in the hidden sector fields, we note that it receives contributions from operators of the form,
\be
K \supset \tilde Z~\tilde C \tilde C + \sum_{k} c_{k}~C_k C^*_k~\tilde C \tilde C \, , \label{eq:Kops}
 \ee
 and
 \be
 W \supset w_{klm}C^k C^l C^m~\tilde C~\tilde C \, . \label{eq:Wops}
 \ee


A vev for $\tilde \phi$ would therefore supersymmetrically induce small changes in the Yukawa couplings and in the normalisation of the kinetic term. While the Kähler potential operators are not relevant for supersymmetry breaking, the superpotential operator induces a contribution to the holomorphic $A$-term of the form,
\be
A_{klm} \supset w_{klm} \tilde \phi~\tilde F_{\tilde \phi} \, .
\ee
The corresponding operator in the Lagrangian is dimension six, and if one of the fields $\phi$ is a Higgs field which obtains a vev $v$ after electroweak symmetry breaking, then this operator is given by,
\be
\Big|\frac{ \hat A_{kl H}}{M_{Pl}^2} v \tilde \phi~\tilde F_{\tilde \phi} \Big| <  \Big|\hat A_{kl H}~ \frac{\tilde \phi}{M_{Pl}} \Big|~\frac{v}{M_{Pl}}~m_{3/2} \ll |\hat A_{klH}| m_{3/2} \, ,
\ee
where $\hat A$ is dimensionless and we have used that $|\tilde F_{\tilde \phi}| < m_{3/2}$ in a spacetime with a small cosmological constant. This bound implies that in compactifications in which $m_{3/2} \approx m_{soft}$, these operators are hardly relevant, and this type of \emph{perturbative superpotential de-sequestering} does not appear to give rise to any new bounds on hidden sector matter due to flavour constraints.  In compactifications with a large hierarchy between $m_{soft}$ and $m_{3/2}$ these operators are enhanced, however we know of no example in which they are important.

Moreover, let us revert the assignment of visible and hidden sector operators and regard the visible sector dependence of the \emph{hidden} sector Yukawa coupling. These operators can be computed as
\be
\langle \tr(\phi \phi) \tr(\tilde \psi \tilde \psi \tilde \phi) \rangle \, ,
\ee
and they can arise as a consequence of any of the operators \eqref{eq:Kops} or \eqref{eq:Wops}, with the tilded fields un-tilded and with the un-tilded fields tilded.

The presence of a holomorphic  ``$Z$-term'' in the Kähler potential can be important for Higgs phenomenology, as this coefficient affects the normalised  $\mu$-term \cite{Kaplunovsky:1993rd}:
\be
\hat{\mu} = \left(e^{\hat{K}/2} \mu + m_{3/2} Z - F^{\bar{m}} \pd_{\bar{m}} Z\right) {\left(\tilde{K}_{H_1} \tilde{K}_{H_2}\right)}^{-1/2} \ .
\ee
Here $\hat K$ denotes the zeroth order term in a Taylor expansion of the Kähler potential in the visible sector fields, and $\mu$ denotes the superpotential mass term.

Corrections due to more than two scalars also induce changes to the Yukawa couplings and we will examine them by also calculating the correlator $\langle \tr(\tilde{\phi} \tilde{\phi} \ldots \tilde{\phi}) \tr(\psi \psi \phi) \rangle$.

\subsection{Summary of results}
Let us briefly summarise the results of the detailed calculation in Appendix \ref{2scalar}.

We begin with corrections due to two hidden scalars. Interestingly, the induced contributions to Yukawa couplings are exactly aligned with the tree-level flavour structure. The only non-vanishing amplitude includes the superfields $\langle \tr(\tilde{C}^3 \tilde{C}^3) \tr(C^1 C^2 C^3) \rangle$ and arises in $\mathbb{Z}_4$ and $\mathbb{Z}_6^{\prime}$ orbifolds with $\theta=\frac{1}{4}(1,1,-2)$ and $\theta=\frac{1}{6}(1,2,-3)$ respectively. For both D3-D3 and the D3-D7 setups the calculation yields
\begin{align}
\nn \mathcal{A} = \mathcal{C} \ \frac{U_2}{\tilde{T}_2} \ \frac{\partial^2}{{\partial r_c}^2} \int_{1/ \Lambda^2}^{\infty} \frac{\textrm{d}t}{t} \mathcal{Z}(t) &=  \mathcal{C} \ \frac{U_2}{\tilde{T}_2} \ \frac{\partial^2}{{\partial r_c}^2} \left(\frac{\Lambda^2}{\tilde{T}_2} - \ln {\left|\frac{\vartheta_1(r_c,\bar{U})}{\eta(\bar{U})} \right|}^2 - \frac{2 \pi{\textrm{Im}(r_c)}^2}{U_2} \right) \\
&= - \mathcal{C} \ \frac{U_2}{\tilde{T}_2} \  \frac{\partial^2}{{\partial r_c}^2}{\mathcal G}(0,r_c) \ ,
\end{align}
with the boson propagator on the torus from \ref{bosprop}. Appendix \ref{2scalar} contains a detailed derivation of this result.
We determine the volume dependence for $R_1\sim R_2 \sim R$ as
\be
\mathcal{A} = - \mathcal{C} \ \frac{U_2}{\tilde{T}_2}  {\mathcal G}''(0,r_c) \sim \frac{1}{\tilde{T}_2} \sim \frac{1}{R^2} \ .
\ee

As discussed above, this result  does not only provide us with the correction to Yukawa couplings due to two hidden scalars, but by swapping the definition of visible and hidden sectors it may also indicate  the presence of a visible sector Higgs ``$Z$-term'' of the form  $Z_{33} C^3 C^3$. 

The findings are different for corrections due to three scalars. The new contributions to Yukawa couplings do not have to display the flavour structure observed at tree-level. For example, the $\mathbb{Z}_6^{\prime}$ orbifold with $\theta=\frac{1}{6}(1,2,-3)$ allows for non-zero correlators
\begin{align}
\nn & \langle \tr(\tilde{C}^2 \tilde{C}^2 \tilde{C}^2) \tr(C^1 C^2 C^3) \rangle \ \textrm{and} \\ \nn & \langle \tr(\tilde{C}^1 \tilde{C}^2 \tilde{C}^3) \tr(C^2 C^2 C^2) \rangle \ ,
\end{align}
thus inducing a new Yukawa coupling $C^2 C^2 C^2$. The amplitude evaluates to
\begin{equation}
\mathcal{A} = - \mathcal{C} \ \frac{U_2^{3/2}}{\tilde{T}_2^{3/2}} \ \frac{\partial^3}{{\partial r_c}^3}{\mathcal G}(0,r_c) \ ,
\end{equation}
which scales as $\tilde{T}_2^{-3/2} \sim R^{-3}$.

In general, we find that for $n$ hidden scalar insertions the annulus amplitude yields
\begin{equation}
\mathcal{A} = - \mathcal{C} \ \frac{U_2^{n/2}}{\tilde{T}_2^{n/2}} \ \frac{\partial^n}{{\partial r_c}^n}{\mathcal G}(0,r_c) \sim R^{-n} \ .
\end{equation}

\subsection*{Corrections with non-tree flavour structure from two scalars}
Last, we comment on the vanishing of the correlators $\langle \tr(\tilde{C}^3 \tilde{C}^3) \tr(C^1 C^1 C^3) \rangle$ in $\mathbb{Z}_4$ orbifolds and $\langle \tr(\tilde{C}^3 \tilde{C}^3) \tr(C^2 C^2 C^2) \rangle$ in $\mathbb{Z}_6^{\prime}$ models. In the latter case the H-charges of the vertex operators in the canonical picture for the five complex directions of spacetime take the form:
\bal
\nn \tilde{\phi}_1^{-1}(z_1) \ = & \ \frac12 \left(0,0,0,0,--\right) \\
\nn \tilde{\phi}_2^{-1}(z_2) \ = & \ \frac12 \left(0,0,0,0,--\right) \\
 \psi_1^{-\half}(z_3) \ = & \ \frac12 \left(+,-,+,-,+\right) \\
\nn \psi_2^{-\half}(z_4) \ = & \ \frac12 \left(-,+,+,-,+\right) \\
\nn \phi^{-1}(z_5) \ = & \ \frac12 \left(0,0,0,--,0\right) \ ,
\end{align}
where $\pm$ denotes $\pm1/2$ while $--$ denotes $-1$, and we refer the reader to Appendix \ref{sec:CFT} for an explanation of our conventions. 
To arrive at a consistent ghost background charge of $0$ for the cylinder, we need to picture change four of the vertex operators. It is not hard to see that all attempts of picture-changing fail to give a configuration where all the H-charges cancel. The scalars $\tilde{\phi}_1$ and $\tilde{\phi}_2$ can only be picture-changed externally or along the third internal direction. This restriction makes it impossible to cancel the H-charges along the first two internal directions using the remaining two vertex operators. Hence, the whole amplitude is bound to vanish. We will give a more systematic explanation for this failure in the next section. The correlator $\langle \tr(\tilde{C}^3 \tilde{C}^3) \tr(C^1 C^1 C^3) \rangle$ evaluates to zero for the same reasons. We conclude that only Yukawa couplings with tree-level flavour structure receive corrections due to two scalars.

\subsection{Corrections from $n$ scalars}
\label{nscalars}
The above analysis can be straightforwardly generalised to the case of $n$ hidden sector open string scalars, as we will here first demonstrate for the correlator $\langle \tr(\tilde{C}^3 \ldots \tilde{C}^3) \tr(C^1 C^2 C^3) \rangle$ before considering the general case. 
 The H-charges of the canonical picture vertex operators are given by,
\bal
\nn \tilde{\phi}_1^{-1}(z_1) \ = & \ \frac12 \left(0,0,0,0,--\right) \\
\nn \tilde{\phi}_2^{-1}(z_2) \ = & \ \frac12 \left(0,0,0,0,--\right) \\
\nn \qquad \vdots \\
 \tilde{\phi}_n^{-1}(z_n) \ = & \ \frac12 \left(0,0,0,0,--\right) \\
\nn \psi_1^{-\half}(w_1) \ = & \ \frac12 \left(+,-,-,+,+\right) \\
\nn \psi_2^{-\half}(w_2) \ = & \ \frac12 \left(-,+,+,-,+\right) \\
\nn \phi^{-1}(w_3) \ = & \ \frac12 \left(0,0,0,0,--\right) \ ,
\end{align}
and the total ghost charge is $n+2$. For the above setup the H-charges along the external and the first two internal directions vanish. Thus we only need to ensure that we cancel the H-charges along the third internal direction to arrive at a consistent amplitude. Given the setup above, the overall H-charge along the third internal direction is $-n$. The process of picture-changing one vertex operator can raise or lower the H-charge by one unit. In particular, we note that scalar operators can only be picture-changed externally or in one internal direction: internal picture-changing on the $i$-th torus lifts the H-charge $(--) \rightarrow 0$ and introduces the worldsheet field $\pd \bar{Z}^i$. Fermion vertex operators can be picture-changed in all directions, both externally and internally. However, the internal H-charge modification $+ \rightarrow -$ introduces a factor of $\pd Z^i$ while the process $- \rightarrow +$ yields a factor of $\pd \bar{Z}^i$. We can thus now formulate a strategy for successful picture-changing: we need to picture-change $n+2$ of the $n+3$ vertex operators and raise the H-charge along the third internal direction by $n$ units. This can be achieved by picture-changing $n$ vertex operators internally with $e^{\phi} \Psi^3 \pd \bar{Z}^3$. The procedure will introduce a factor of $(\pd \bar{Z}^3)^n$ into the amplitude which will contribute its classical correlator in terms of winding states. After that we are still left with two vertex operators which have to be picture-changed. As the overall H-charge is already cancelled we need to picture-change one operator with $e^{\phi} \Psi^i \pd \bar{Z}^i$ while the other one is contracted with $e^{\phi} \bar{\Psi}^j \pd Z^j$. There are several potential contributions to the amplitude:
\begin{enumerate}
\item \textbf{Purely internal picture-changing:} \newline
We can choose to also modify the remaining two vertex operators internally. However, in this case we will be left with a configuration as in \ref{purelyint}: the only non-zero H-charges appear in the vertex operators for the fermions. Thus, the spinorial correlation function becomes a two-fermion amplitude which we found to vanish in \ref{purelyint} ($\propto \vartheta_1(0)=0$). hence, purely internal picture-changing does not contribute.
\item \textbf{Picture-change scalars internally, picture-change Yukawas externally:} \newline
Here we choose to picture-change the remaining two vertex operators externally. Internal picture-changing is performed on the $n$ hidden scalars while external picture-changing is applied to one fermion and the Higgs. This way of picture-changing generalises the procedure of \S\ref{IntExt}.
\item \textbf{Mixed picture-changing:} \newline
In this case we picture-change the visible Higgs and $n-1$ hidden scalars internally while one hidden scalar and one visible fermion are modified externally. Compare with \S\ref{mixedpc}.
\end{enumerate}
The procedure outlined above is a straightforward generalisation of the two-scalar calculation presented before. To examine  the possibility of Yukawa couplings with a different flavour structure,  we now consider the more general correlator involving any combination of superfields $\langle \tr(\tilde{C}^a \tilde{C}^b \ldots \tilde{C}^m) \tr(C^r C^s C^t) \rangle$. Nevertheless, the basic strategy for picture-changing remains unaltered: out of the $n+3$ vertex operators we still need to change $n+2$ to arrive at the correct ghost charge. The overall H-charge is still $-n$; however, now it can be distributed over all three internal directions. Thus, we arrive at the following strategy for picture-changing: we again need to picture-change $n$ vertex operators internally; the remaining two will be modified in the external directions as purely internal picture-changing leads to vanishing results (see \ref{purelyint}). It is at this point that we can use the condition coming from the classical solution over winding states to constrain the possible configurations. Every internal picture-changing operator introduces a factor $\pd \bar{Z}^i$ into the amplitude. These fields can only contribute to the amplitude via their classical correlator. Given our brane setup (e.g.~see figure \ref{fig:Z6D3D37}) these winding state solutions only exist along one of the three two-tori and vanish for the remaining directions. We continue by analysing a case where winding solutions are only allowed on the third two-torus; the classical correlators of $\pd \bar{Z}^1$ and $\pd \bar{Z}^2$ hence vanish in our models. To arrive at a non-zero result we need to obtain a classical correlator of the form $\langle (\pd \bar{Z}^3)^n \rangle$. This can only occur if the initial H-charge of $-n$ was located entirely along the third internal direction. Hence we have arrived at a condition which severely constrains the possible arrangements of superfields $\langle \tr(\tilde{C}^a \tilde{C}^b \ldots \tilde{C}^m) \tr(C^r C^s C^t) \rangle$. One possible configuration is shown by the vertex operators above. It corresponds to corrections to Yukawa-couplings with tree-level flavour structure $C^1 C^2 C^3$. Having completed our analysis of H-charge cancellation and the restrictions imposed by the winding states we thus conclude that there is one other  configuration that satisfies the conditions laid out before:\footnote{The configurations $\langle \tr(\tilde{C}^1 (\tilde{C}^3)^{n-1}) T\  \tr(C^2 C^3 C^3) \rangle$ and $\langle \tr(\tilde{C}^2 (\tilde{C}^3)^{n-1}) \  \tr(C^1 C^3 C^3) \rangle$ also satisfy the criteria of H-charge cancellation and the existence of a non-zero classical winding solution. Nevertheless, these correlators vanish for small momenta $k_i \rightarrow 0$. External picture-changing introduces a momentum prefactor $k_ik_j$. To arrive at a finite result this prefactor needs to be cancelled by a momentum pole. In contrast to the correlators in the main text the two amplitudes above do not generate a momentum pole. This pole usually appears when the worldsheet positions collide in the expression $\left(\frac{\vartheta_1(z_i - z_j)}{\vartheta_1^{\prime}(0)} \right)^{-1+k_i k_j}$. For the given amplitudes above $z_i$ and $z_j$ are localised on different cylinder boundaries and hence the limit $z_i \rightarrow z_j$ never occurs.}
\be
\langle \tr(\tilde{C}^1 \tilde{C}^2 (\tilde{C}^3)^{n-2}) \  \tr(C^3 C^3 C^3) \rangle \ .
\ee
Most importantly, this combination of superfields constitutes a Yukawa coupling with the non-tree flavour structure $C^3 C^3 C^3$. We can write down the corresponding vertex operators:
\bal
\nn \tilde{\phi}_1^{-1}(z_1) \ = & \ \frac12 \left(0,0,--,0,0\right) \\
\nn \tilde{\phi}_2^{-1}(z_2) \ = & \ \frac12 \left(0,0,0,--,0\right) \\
\nn \tilde{\phi}_3^{-1}(z_3) \ = & \ \frac12 \left(0,0,0,0,--\right) \\
\nn \qquad \vdots \\
\nn \tilde{\phi}_n^{-1}(z_n) \ = & \ \frac12 \left(0,0,0,0,--\right) \\
 \psi_1^{-\half}(w_1) \ = & \ \frac12 \left(+,-,+,+,-\right) \\
\nn \psi_2^{-\half}(w_2) \ = & \ \frac12 \left(-,+,+,+,-\right) \\
\nn \phi^{-1}(w_3) \ = & \ \frac12 \left(0,0,0,0,--\right) \ .
\end{align}
In this case picture-changing is considerably simple: there is only one non-vanishing combination. We picture-change the first two scalars $\phi_1$ and $\phi_2$ externally while all other vertex operators except for one are picture-changed internally along the third direction. We pick up the classical correlator $\langle (\pd \bar{Z}^3)^n \rangle$ as expected.

Last, we will apply the restrictions provided by gauge invariance to single out the physically relevant correlation functions. For $n=2$ only superfield combinations of the form $(\tilde{C}^i \tilde{C}^i)$ are gauge-invariant in orbifolds with $\mathbb{Z}_2$ element. For $n=3$ the gauge-invariant combinations of hidden superfields are Yukawa-like terms and are summarised in table \ref{tableYukawas}. Hence, for $n=2$ and $n=3$ we will list all gauge-invariant combinations for the $\mathbb{Z}_3$, $\mathbb{Z}_4$, $\mathbb{Z}_6$ and $\mathbb{Z}_6^{\prime}$ orbifolds which lead to non-zero correlation functions: \\

\begin{tabular}{ | l | l | l | }
\hline
Toroidal orbifold $\mathbb{T}^6/\mathbb{Z}_N$  & $n=2$ scalars & $n=3$ scalars\\
\hline \hline
$\mathbb{Z}_3$: $\theta=\frac{1}{3}(1,1,-2)$ &   & $\tr(\tilde{C}^3 \tilde{C}^3 \tilde{C}^3) \  \tr(C^1 C^2 C^3)$ \\
 & & $\tr(\tilde{C}^1 \tilde{C}^2 \tilde{C}^3) \  \tr(C^3 C^3 C^3)$ \\
\hline
$\mathbb{Z}_4$: $\theta=\frac{1}{4}(1,1,-2)$ & $\tr(\tilde{C}^3 \tilde{C}^3) \  \tr(C^1 C^2 C^3)$ & \\
\hline
$\mathbb{Z}_6$: $\theta=\frac{1}{6}(1,1,-2)$ & & $\tr(\tilde{C}^3 \tilde{C}^3 \tilde{C}^3) \  \tr(C^1 C^2 C^3)$\\
& & $\tr(\tilde{C}^1 \tilde{C}^2 \tilde{C}^3) \  \tr(C^3 C^3 C^3)$ \\
\hline
$\mathbb{Z}_6^{\prime}$: $\theta=\frac{1}{6}(1,2,-3)$ & $\tr(\tilde{C}^3 \tilde{C}^3) \  \tr(C^1 C^2 C^3)$ & $\tr(\tilde{C}^2 \tilde{C}^2 \tilde{C}^2) \  \tr(C^1 C^2 C^3)$\\
& & $\tr(\tilde{C}^1 \tilde{C}^2 \tilde{C}^3) \  \tr(C^2 C^2 C^2)$ \\
\hline
\end{tabular}  \\ \\

Thus we confirm our previous finding that two-scalar-operators only correct Yukawa couplings with tree-level flavour structure. More crucially, we find that Yukawa operators with different flavour structure can be generated by three scalar insertions. This is one of the main results of this paper: we demonstrate that corrections due to hidden scalars can induce new Yukawa couplings and thus possible new flavour violations.

To complete this section we still need to evaluate the correlator $\langle (\phi)^n \ (\psi \psi \phi) \rangle$ explicitly. We will do so for the superfield combination $\langle \tr(\tilde{C}^1 \tilde{C}^2 (\tilde{C}^3)^{n-2}) \  \tr(C^3 C^3 C^3) \rangle$, but the result will be general for any correction to Yukawa couplings due to $n$ distant scalars. As described above, we picture-changed the hidden fields $\tilde{C}^1$ and $\tilde{C}^2$ externally which leaves us with a momentum prefactor $k_1 k_2$. We can evaluate the fermionic and ghost correlator using the standard CFT results. After applying the Riemann identity \eqref{Riemann} we are left with (for D3-D3 models)
\be
2 {\vartheta_1^{\prime}}^3 \ \frac{\vartheta_1(-z_1+z_2 + \theta_1) \vartheta_1(\theta_2) \vartheta_1(z_2-w_1 + \theta_3)}{\vartheta_1(z_1-z_2) \vartheta(z_2-w_1)}
\ee
Further we need to consider the classical correlator $\langle (\pd \bar{Z}^3)^n \rangle$. It can be rewritten in terms of a derivative according to \eqref{windingderivative}:
\be
\langle (\pd \bar{Z}^3)^n \rangle = \left(-{\sqrt{\alpha'} \over \pi ct }\frac{\partial}{\partial {r}_c}\right)^n \mathcal{Z}(t) \ .
\ee
The above winding solution is only valid in a partially twisted sector and thus we set $\theta_3=0$. We multiply by the appropriate internal bosonic partition function and the relevant terms from the momentum correlators we arrive at
\begin{align}
\nn \mathcal{A} \propto (k_1k_2) \int_0^{\infty} & \frac{\textrm{d}t}{t} \int \prod_{i=1}^{n} \textrm{d}z_i \prod_{i=1}^{3} \textrm{d}w_i \frac{1}{{(2 \pi^2 t)}^2} \ \prod_{i=1}^2 (-2 \sin \pi \theta_i) \\
& {\left(\frac{\vartheta_1(z_1-z_2)}{\vartheta_1^{\prime}(0)} \right)}^{-1+k_1k_2} \frac{\vartheta_1(-z_1 + z_2+\theta_1)}{\vartheta_1(\theta_1)} \langle (\pd \bar{Z}^3)^n \rangle\ .
\end{align}
The integrals over worldsheet positions $z_i$ and $w_i$ can also be performed without difficulty. The integration over $z_1-z_2$ generates a momentum pole $1/k_1k_2$ (the mechanism is exactly the same as in the case of two-scalar-corrections) while the remaining $n+2$ integrations just contribute ${(it/2)}^{n+2}$ in the $k_i \rightarrow 0$ limit. However, at the same time the classical correlator contains $n$ factors of $\frac{1}{t} \frac{\pd}{\pd r_c}$. In the end the result has the following $t$-dependence: $\frac{\textrm{d}t}{t} \frac{1}{t^2} t^{n+2} \frac{1}{t^n} \mathcal{Z}(t)=\frac{\textrm{d}t}{t} \mathcal{Z}(t)$. We thus obtain the result:
\be
\mathcal{A} = \mathcal{C} \frac{\partial^n}{{\partial r_c}^n} \int_{0}^{\infty} \frac{\textrm{d}t}{t} \mathcal{Z}(t) \sim \frac{1}{R^n} \ .
\ee

For three-scalar-corrections we evaluate the result explicitly. As in previous chapters we evaluate the integral as in \cite{0404087}. After introducing a UV-cutoff and rewriting the sum over winding modes in terms of a genus-two theta function \eqref{Zt} we can obtain:
\begin{align}
\nn \mathcal{A} = \mathcal{C} \frac{\partial^3}{{\partial r_c}^3} \int_{1/ \Lambda^2}^{\infty} \frac{\textrm{d}t}{t} \mathcal{Z}(t) &={\left(\frac{U_2}{\tilde{T}_2}\right)}^{3/2}  \mathcal{C} \frac{\partial^3}{{\partial r_c}^3} \left(\frac{\Lambda^2}{\tilde{T}_2} - \ln {\left|\frac{\vartheta_1({r_c},\bar{U})}{\eta(\bar{U})} \right|}^2 - \frac{{\textrm{Im}(r_c)}^2}{2 \pi U_2} \right) \\
&= - \mathcal{C} {\left(\frac{U_2}{\tilde{T}_2}\right)}^{3/2}{\mathcal G}'''(0,r_c) \ ,
\end{align}
again with ${\mathcal G}$ from \Ref{bosprop}. The scaling of this is
\be
\mathcal{A} = - \mathcal{C} {\left(\frac{U_2}{\tilde{T}_2}\right)}^{3/2}{\mathcal G}'''(0,r_c) \sim \frac{1}{\tilde{T}_2^{3/2}} \sim \frac{1}{R^3} \ .
\ee
We end this section by emphasising that three-scalar-corrections not only contribute to Yukawa couplings with tree level flavour structure $C^1 C^2 C^3$, but also generate Yukawa couplings with flavour structure $C^r C^r C^r$, thus providing new possible sources of flavour-changing neutral currents.

\section{Conclusion}
\label{sec:conclusion}
In this work we have investigated certain non-perturbative corrections to the  Yukawa couplings in type IIB  orbifold models in which the visible sector is realised as a stack of D3-branes on an orbifold singularity. For non-perturbative effects arising from gaugino condensation on a stack of distant D7-branes, these non-perturbative superpotential operators can be computed as one-loop threshold corrections to the gauge kinetic function of the D7-branes. We have calculated the corresponding open string cylinder amplitude both in the case of D7-branes wrapping a large `bulk' cycle in the internal dimensions, and in the case of D7-branes wrapping a small cycle, which in the orbifold limit corresponds to a stack of D3-branes at an orbifold singularity.

It was shown in  \cite{Berg:2010ha} that if the induced operators carries  a non-trivial flavour structure, constraints on flavour-changing  neutral currents may severely constrain the moduli stabilisation sector. In the models we have considered in
this work, we have verified that  the superpotential de-sequestering operators indeed carry a non-trivial flavour structure and we have extracted a set of conditions under which superpotential de-sequestering can be expected to occur also for more general models.

Furthermore, we have shown that the presence of hidden sector scalars may also induce corrections to the Yukawa couplings with non-trivial flavour structure. The induced operators may allow for a Froggat-Nielsen mechanism for flavour, and we note that one of the computed correlation functions is correlated with a contribution to the physical Higgs $\mu$-term proportional to $m_{3/2}$.  

In the first out of three calculations supporting  our conclusions we determined a threshold correction to the gauge kinetic function of D3/D7-branes due to distant chiral scalars: $\langle \tr(A_{\mu} A^{\mu}) \tr(\phi^i \phi^j \phi^k) \rangle$. We noted that the scalar insertion in the amplitude translates into a classical correlator over the fields ${\langle \pd \bar{Z}^i \pd \bar{Z}^j \pd \bar{Z}^k\rangle}$. In D3/D7 orbifold models a non-zero classical solution exists in terms of winding states around the compact space. For D3/D7 orbifold constructions non-zero winding solutions only exist in partially twisted or untwisted sectors for $i=j=k$, thus constraining the flavour structure of the induced Yukawa-couplings. For our particular setups based on $\mathbb{T}^6/\mathbb{Z}_3$, $\mathbb{T}^6/\mathbb{Z}_6$ and $\mathbb{T}^6/\mathbb{Z}_6^{\prime}$ we find:
\be
\label{deseq}
\langle \tr(A_{\mu} A^{\mu}) \tr(\phi^i \phi^i \phi^i) \rangle = - \mathcal{C} \frac{U_2^{3/2}}{\tilde{T}_2^{3/2}}\frac{\partial^3}{{\partial r_c}^3} \ln \vartheta_1\left(r_c,\bar{U}\right) =  - \mathcal{C} \frac{U_2^{3/2}}{\tilde{T}_2^{3/2}}~\frac{\partial^3}{\partial r_c^3}{\mathcal G}(0,r_c) \ ,
\ee
where $r_c$ is the dimensionless distance between the stacks of branes supporting the gauge theory and the chiral matter, $U$ is the complex structure modulus of the two-torus wrapped,
and ${\mathcal G}$ is the boson propagator on the torus. The existence of this threshold correction thus implies the generation of a non-perturbative superpotential of the form
\be
W \supset \hat{Y}^{np}_{iii}\ C^i C^i C^i\ e^{-T} \, ,
\ee
where $C^i$ are chiral superfields. Crucially, we observe that the flavour structure of the non-perturbative Yukawa couplings fails to align with the tree-level flavour structure,  $W_{\rm tree}=\epsilon_{rst} C^r C^s C^t$. 
We conclude that $\hat{Y}^{np}_{ijk} \neq c Y^{\rm phys}_{ijk}$ which establishes superpotential de-sequestering in local orbifold models.

This result is corroborated by the second computation of this paper in which we evaluated a cylinder correlator between two D3/D7 gaugini and D3 chiral matter: $\langle \tr(\lambda \lambda) \tr(\psi^i \psi^j \phi^k) \rangle$. This correlator is  related by four-dimensional supersymmetry to the first amplitude and  we confirm 
that it reproduces the same result, i.e.~equation \eqref{deseq}.

 Finally, we studied corrections to visible sector Yukawa couplings due to hidden scalars on a distant brane stack. The string calculation consisted of a cylinder correlator with $n$ hidden scalar insertions and a Yukawa operator, $\langle (\tilde{\phi})^n \ (\psi \psi \phi)\rangle$, which we evaluated for arbitrary $n$. This is possible as the bosonic correlator in the internal directions decouples from the remaining correlation function, only contributing its classical correlator over winding modes to the result. Again, the classical solution plays a crucial role: it determines the flavour structure of the induced Yukawa couplings. In summary we find that tree-level Yukawa couplings are corrected due to hidden scalar matter and, further, we also establish that Yukawa couplings with different flavour structure can be induced: $\hat{Y}^{\rm phys} \neq c Y^{\rm tree}$. Overall, we find that corrections to Yukawa couplings due to $n$ scalars take the form
\be
\label{nresult}
\mathcal{A} = \mathcal{C} {\left(\frac{U_2}{\tilde{T}_2} \right)}^{n/2}~\frac{\partial^n}{\partial r_c^n}{\mathcal G}(0,r_c) \, .
\ee

While our computations are performed in the open string channel, we expect that a similar analysis should  be possible by adopting a geometrical closed string approach. From this perspective the volume of the four-cycle wrapped by the D7-branes --- and thereby the D7-brane gauge coupling --- will be modified both by the presence of D3-branes and by the matter fluctuations on their worldvolume. 
Performing this analysis explicitly could offer new interesting insights.

Last, we note that our results were obtained in local orbifold models. While these setups offer calculability they are merely toy-versions of more realistic string models. It will be an interesting, but formidable task to extend the study of superpotential de-sequestering to more realistic string compactifications. Rewriting our results in a closed string picture could offer insights into possible generalisations to more realistic compact spaces.
In particular, we note that in more realistic models
than orbifolds, such as the del Pezzo singularities
considered in e.g.~\cite{Conlon:2008wa, Dolan:2011qu}, it is  natural to try to understand the flavour structure as geometric
symmetries of the singularity. Our statement is essentially that
the exchanged closed fields, in particular ``partial blowup'' modes that resolve a co-dimension two singularity (corresponding
to partially twisted open strings),
break these geometric symmetries  \cite{Conlon:2011jq, Maharana:2011wx}.
Of course, global effects always break such local symmetries (Calabi-Yau compactifications have no global isometries), but for example in LVS, such effects are very small, whereas we have indicated
(as made somewhat more explicit in  \cite{Berg:2010ha}) that these effects can be fairly big.

Nevertheless, our results are a strong indication that superpotential de-sequestering is an effect not uncommon to string theory models.

\acknowledgments
We thank Marco Bill\'o, Mark Goodsell,  Michael Haack, Ben Heidenreich and Michael Klaput for helpful discussions. We are grateful to Liam McAllister for comments on the draft of this paper. LTW is supported by the Science and Technology Facilities Council (STFC). JC thanks the Royal Society for a University Research Fellowship.
DM gratefully acknowledges support from the Gålö foundation. MB is supported by the Swedish Research Council (Vetenskapsr{\aa}det),
and gratefully acknowledges support for this work by the Swedish Foundation for International Cooperation in Research and Higher Education.
We all thank the Isaac Newton Institute in Cambridge for a stimulating atmosphere during the completion of this project.

\appendix

\section{CFT Building Blocks}
\label{sec:cftbuildblock}
\label{sec:CFT}

We will use standard CFT methods in string perturbation theory in this paper.
Here we give a reasonably self-contained toolbox collection for this purpose.
Useful references are \cite{Polchinski1, Polchinski2, Friedan:1985ge, Atick:1986rs, Atick:1986ns, Atick:1987gy, Abel:2004ue, Abel:2005qn, 0607224, Benakli:2008ub, Schlotterer:2010mw}.

\subsection{Worldsheet geometry and coordinates}

All the amplitudes evaluated in this paper are cylinder (annulus) amplitudes and so all correlators refer to this worldsheet topology, except where otherwise noted. We  begin with a brief description of this geometry. The cylinder has a single real modulus $t$ and is parameterised by a complex coordinate $z$. The boundary circles are located  at $\re{z}=0$ and $\re{z}=1/2$ and are parameterised by $0\leq \im{z}\leq {t}/{2}$. The long cylinder limit is given by $t\rightarrow 0$ and corresponds to UV in the open string channel and IR in the closed string channel. The long strip limit is $t \rightarrow \infty$ and  corresponds to IR in the open string channel and UV in the closed string channel. There is a single conformal Killing vector corresponding to translations parallel to the boundary.

The target space coordinates are the real worldsheet bosons $X^{M}\left(z,\bar{z}\right)$ where $M=0,...,9$. We split them as $X^{M}=\left\{X^{\mu},X^m\right\}$ with $\mu=0,..,3$ denoting external directions and $m=4,..,9$ denoting internal directions. It will be useful to group the directions into complex pairs and we define \cite{Lust:2004cx,Berg:2005ja,Blumenhagen:2006ci}
\be
Z^{i} =c( X^{2i+2} + \bar{U} X^{2i+3}) \;,  \label{cplxcoord}
\ee
where $i=1,2,3$,  and the normalisation
\be
c = \sqrt{\tilde{T}_2 \over 2 U_2}  \label{normc}
\ee
is fixed so that
\be  \label{Znorm}
\langle \partial Z^i(z_1) \partial \bar{Z}^j(z_2)\rangle = {\delta^{ij} \over (z_1-z_2)^2}
\ee
in the torus metric
\be  \label{torusmetric}
G_{ij} = {\tilde{T}_2  \over  U_2} \left( \begin{array}{cc} 1 \; \; & U_1 \\ U_1 \; \; & |U|^2 \end{array} \right)  \; .
\ee
Note that with an off-diagonal piece in $G_{ij}$, the real coordinates
have nonvanishing contractions in different directions along the same torus (for example $X^4$ and $X^5$), but
the $Z^i$ are precisely such that they do not, as in \Ref{Znorm}.
As an example, let us compute the tree-level two-point functions
\bea
\langle \partial Z^1  \partial Z^1 \rangle_{D^2} &=&
{1 \over (z-w)^2} {c^2 \over \tilde{T}_2 U_2}  \left(  \langle \partial X^4 \partial X^4 \rangle
+ 2\bar{U} \langle \partial X^4 \partial X^5 \rangle + \bar{U}^2 \langle \partial X^5 \partial X^5 \rangle  \right) \\
&=& 0 \nonumber
\eea
using  the explicit form of $\langle \partial X^m \partial X^n \rangle_{D^2} =G^{mn}/(z-w)^2$
for the real coordinates.
It  is an important fact in the main text that the correlator
of two un-barred (or two barred) fields  $\partial Z^3 \partial Z^3$ vanishes.
(This is the ``quantum'' tree-level correlator. Later, we will also talk about classical contributions to  the one-loop amplitude,
but at tree-level, there is no 1-cycle on the worldsheet that can be wrapped in spacetime.)
We can also compute the unbarred-barred correlator
\bea
\langle \partial Z^1  \partial \bar{Z}^1 \rangle_{D^2} &=&
{1 \over (z-w)^2} {c^2 \over \tilde{T}_2 U_2}  \left(  \langle \partial X^4 \partial X^4 \rangle
+ 2U_1 \langle \partial X^4 \partial X^5 \rangle + |U|^2 \langle \partial X^5 \partial X^5 \rangle  \right) \\
&=& {1 \over (z-w)^2} {c^2 \cdot 2U_2^2 \over \tilde{T}_2 U_2}  \nonumber
\eea
from which we see
that given the torus metric \Ref{torusmetric}, the normalisation \Ref{normc} for $c$
is precisely what produces the simple normalisation \Ref{Znorm}
for the tree-level two-point function.

 To save space we occasionally drop  indices on the coordinates unless needed and denote
\be
Z = X^1 + \bar{U} X^2 \;.
\ee

There are two basic boundary conditions that can be imposed at each end of the cylinder
\be  \label{NeuBC}
\hbox{Neumann: } \partial_n X (z, \bar{z}) \equiv \half(\partial + \bar{\partial}) X(z, \bar{z}) = 0,
\ee
\be \label{DirBC}
\hbox{Dirichlet: } \partial_t X (z, \bar{z}) \equiv \half(\partial - \bar{\partial}) X(z, \bar{z}) = 0.
\ee
We have also defined the normal and tangential derivatives. In principle we can consider different boundary conditions at each end of the annulus but since we are only studying D3 branes we restrict either to NN or DD boundary conditions. Also, we occasionally denote the coordinate dependence $X(z)$ without implying holomorphic properties.

The cylinder can be obtained from the torus by quotienting under the identification $z \to 1 - \bar{z}$, with boundaries at $z = 1 - \bar{z}$. This is useful for relating bosonic ($X(z,\bar{z})$) correlators on the torus to those on the cylinder. The method of images
can then be used to obtain the cylinder correlators by starting with torus correlators and adding an image field at $1- \bar{z}$ for any field at $z$. The sign of the image correlator is positive for Neumann boundary conditions and negative for Dirichlet boundary conditions. The torus modular parameter $\tau$ is related to the cylinder modulus by $\tau={it}/{2}$.

\subsection{Vertex operators}
\label{sec:vertexop}

In this section we briefly summarise the expressions for the vertex operators. We  calculate  cylinder amplitudes for which the ghost charge should be zero, or equivalently, the sum of all the vertex operator `pictures' should vanish.
The positions of the vertex operators will be integrated.

The bosonic vertex operator for a four-dimensional scalar $\phi$ is given in the $(-1)$ picture as
\be
{\cal V}_{\phi}^{(-1)} \left(z\right) = t^a e^{-\phi} \psi^i e^{ik \cdot x} \left(z\right)\;.
\ee
Here $z$ denotes the point on the worldsheet at which the vertex operator is inserted.  The scalar
Chan-Paton wavefunction is denoted $t^a$ and the field $\phi$ is the ghost from bosonising the $(\beta, \gamma)$ CFT.
The complexified spinorial field $\psi^i$ can be bosonised in terms of free fields $H_i$ so that
\be
\label{psiboson}
\psi^i(z) = e^{iH_i(z)}\;.
\ee
Here $i$ labels the complex direction.
Note that this bosonisation is purely local as the $\psi^i$ correlators depend on the spin structure and so cannot be
globally bosonised. However these
amplitudes (which we give in \S\ref{sec:fermcorr} below) are fixed uniquely in terms of this local bosonisation.
For economy of notation we typically suppress the CP index and wavefunction so that for a four-dimensional scalar
we have the $(-1)$-picture vertex operator
\be
\label{scalar-1}
{\cal V}_{\phi}^{-1}\left(z\right) = e^{-\phi} \psi^i e^{ik \cdot x} \left(z\right)\;.
\ee
The four-dimensional gauge field vertex operator is given by
\be
\label{gauge-1}
{\cal V}_{A}^{-1} \left(z\right) = A^a e^{-\phi} \epsilon_{\mu} \psi^{\mu} e^{ik \cdot x} \left(z\right)\;.
\ee
Here again $\psi^{\mu}$ can be bosonised with H-charge of $\pm 1$ and $\epsilon_{\mu}$ is the polarisation vector of the gauge boson which satisfies $\epsilon \cdot k = 0$.

The spacetime fermion vertex operator in the $-\half$ picture is given by
\be
{\cal V}_{\psi}^{-\half} \left(z\right) = e^{-\frac{\phi}{2}} S_{10} e^{ik \cdot x} \left(z\right)\;.
\ee
and if we add a Chan-Paton factor it looks like a gaugino
\be
{\cal V}_{\lambda}^{-\half} \left(z\right) = \lambda^a e^{-\frac{\phi}{2}} S_{10} e^{ik \cdot x} \left(z\right)\;.
\ee
Here $S_{10}$ is the ten-dimensional spin field which can be locally bosonised to
\be
\label{spinbos}
S_{10} = \prod_{i=1}^5 e^{iq^iH_i} \;,
\ee
where the H-charges $q_i$ are given by the spin $\pm \half$ of the complex direction components of the spinor.

Without loss of generality for the fermions we impose the external spinors $S_1 = |+-\rangle$ and $S_2 = |-+\rangle$. This restriction is equivalent
to boosting to a frame with $k_1 = (k, k, 0, 0)$ and $k_2 = (k, -k, 0, 0)$. To see this, note that the physical state condition
$(k \cdot \Gamma) | \psi \rangle = 0$ gives
$$
(k \cdot \Gamma) | \psi \rangle = (k_0 \Gamma^0 \pm k_1 \Gamma^1) | \psi \rangle = -k_1 \Gamma^0(\Gamma^0 \Gamma^1 \mp 1) | \psi \rangle
= - 2k_1 \Gamma^0 (S_0 \mp 1/2) | \psi \rangle = 0,
$$
and so $S_0 = \pm 1/2$ for $k_1 = \pm k_0$. The GSO conditions then require $S_1 = |+-\rangle$ and $S_2 = |-+\rangle$.
We also note that for the complex momenta the physical state conditions imply
\bal
k_3^{1+} = k_3^{2+} = k_3^{2-} &= 0 \label{k3condition} \\
k_4^{1-} = k_4^{2-} = k_4^{2+} &= 0 \label{k4condition} \ .
\end{align}

To bring the amplitude into the appropriate zero ghost charge picture we can change pictures following the
prescription of Friedan, Martinec and Shenker \cite{Friedan:1985ge} using
\be
{\cal V}^{i+1}\left(z\right) = \underset{{z\rightarrow w}}{\mr{lim}} e^{\phi(z, \bar{z})} \mr{T_F} \left(z\right) {\cal V}^{i}\left(w\right) \;,
\ee
where we have the picture changing operator
\be
\mr{T_F}\left(w\right) = \frac12 \left( \psi_i \partial \overline{X}^i\left(w\right)  + \overline{\psi}_i\partial X^i\left(w\right)  \right) \;.
\ee
In practice the picture changing is evaluated using the operator product expansions (OPE)
\bea
e^{iaH\left(w\right)}e^{ibH\left(z\right)} &=& \left(w-z\right)^{ab}e^{i\left(a+b\right)H\left(z\right)} + ... \;, \label{opespin}\\
e^{ia\phi\left(w\right)}e^{ib\phi\left(z\right)} &=& \left(w-z\right)^{-ab}e^{i\left(a+b\right)\phi\left(z\right)} + ... \;, \label{opeghost} \\
\partial X\left(w\right)e^{ikX(z)} &=& -\frac{i\alpha'}{2}k^+\left(w-z\right)^{-1} e^{ikX(z)} + ... \;, \label{opeboson} \\
\partial \overline{X}\left(w\right)e^{ikX(z)} &=& -\frac{i\alpha'}{2}k^-\left(w-z\right)^{-1} e^{ikX(z)} + ... \;, \label{opecmplxboson}
\eea
where the ellipses denote less divergent terms. Recall that the $H_i$ are free fields and so
only OPEs with the same direction are non-vanishing.
We have also introduced  complex momenta $k^{\pm}=k^1\pm ik^2$ and defined
\be
\label{kXcomplex}
kX\left(z\right) \equiv \half \left(k^+ \cdot \overline{X}\left(z\right) + k^- \cdot X\left(z\right)\right)\;,
\ee
so that in complex notation we can write
\be
k\cdot X\left(z\right) = k_i X^i\left(z\right) \;.
\ee

For the $(-1)$-picture scalar vertex operator shown in \eqref{scalar-1} above, this procedure leads to
the zero-picture vertex operator
\be
\label{scalar0}
{\cal V}_{\phi}^{0} \left(z\right) = \left[\pd Z^{i} + i (k \cdot \psi)\Psi^{i} \right]e^{ik \cdot X} \left(z\right)\;.
\ee

\subsection{Bosonic Correlators}
\label{sec:boscorr}

We first evaluate the bosonic correlators, namely those involving the worldsheet bosons $X(z,\bar{z})$.
 Since the bosons are free worldsheet fields, for a correlator to be non-vanishing it must involve the same complex directions. Therefore such a correlator can be labeled by the associated direction: correlators involving $Z^{1,2,3}$ are labeled internal, while $X^{1,2}$ (also complexified) are external.

 Internal correlators are either twisted or untwisted. Twisted correlators involve directions in which an orbifold twist acts
 whereas untwisted directions have no orbifold action. We will not encounter twisted correlators in the calculations that follow. In addition to the quantum contribution untwisted correlators may also have a classical contribution coming the zero mode solutions. For the case of D3 branes studied  here, this is associated to winding modes in the compact space.

For external modes with Neumann boundary conditions, the classical contribution instead comes from momentum modes. The quantum correlator is the same as for internal directions except with Neumann boundary conditions. We now proceed to calculate the correlators according to the preceding classification.

\subsubsection{Internal untwisted quantum correlators}

The quantum bosonic correlator on the cylinder can be derived from that on the covering torus (denoted by a subscript ${\cal T}$) which reads
\be
\langle Z(z) \overline{Z}(w) \rangle_{\mr{{\cal T},qu}} = -\alpha' \log |\vartheta_1 (z-w)|^2 + \frac{2\pi\alpha'}{\im{\tau}} \left(\im (z-w)\right)^2\;.
\label{torusuntwicorr}
\ee
Here $\tau$ is the torus modular parameter.
For comparison with say \cite{Polchinski1} note that $Z$ here is a complexified coordinate.
As only correlators involving the same directions are non-vanishing
\be
\langle Z(z) Z(w) \rangle_{\mr{{\cal T},qu}} = 0 \;,
\ee
as the two real directions give equal contributions of opposite sign.
From (\ref{torusuntwicorr}) one can obtain correlators on the cylinder (denoted by a subscript ${\cal A}$) through use
of the method of images.
\begin{align}
\nn \langle Z(z) \overline{Z}(w) \rangle_{\mr{{\cal A}}} =  \half & \left[ \langle Z(z) \overline{Z}(w) \rangle_{\mr{{\cal T}}} \, \pm \, \langle Z(1-\bar{z}) \overline{Z}(w) \rangle_{\mr{{\cal T}}}\right. \\
& \left. \pm \langle Z(z) \overline{Z}(1 - \bar{w})\rangle_{\mr{{\cal T}}} + \langle Z(1-\bar{z}) \overline{Z}(1-\bar{w}) \rangle_{\mr{{\cal T}}} \right],
\end{align}

where the plus sign applies for Neumann boundary conditions and the minus sign applies for Dirichlet boundary conditions. We can write the Neumann and Dirichlet correlator explicitly as
\bea
\label{neumannxx}
\bra Z (z) \ov{Z} (w) \ket_{\mr{{\cal A},qu}}^{\mr{N}} &=&  -\alpha' \left(
\log \left|\vartheta_1 (z - w) \right|^2 + \log \left| {\vartheta_1 (\ov{z} + w)}\right|^2\right) + \frac{8\pi\alpha'}{t} \left(\im (z-w)\right)^2 \;, \\
\label{dirichletxx}
\bra Z (z) \ov{Z} (w) \ket_{\mr{{\cal A},qu}}^{\mr{D}} &=&  -\alpha' \left(
\log \left| \vartheta_1 (z - w) \right|^2 - \log \left| {\vartheta_1 (\ov{z} + w)}\right|^2 \right)\;.
\eea
Here we have used the relation $\tau=\frac{it}{2}$ for the modular parameters of the cylinder and the covering torus.
The Dirichlet correlator has no zero mode since the string center of mass is fixed, whereas for Neumann boundary conditions the string can take any position.
Note also that when restricted to the boundary the Dirichlet correlator vanishes
\be
\left.\bra Z (z) \ov{Z} (w) \ket_{\mr{{\cal A},qu}}^{\mr{D}}\right|_{\mr{Boundary}} = 0\;.
\ee

Vertex operator computations with the bosonic fields normally involve not the bare fields but rather their derivatives.
For operators on the boundary, under Neumann boundary conditions the vertex operators involve tangential derivatives $\partial_t Z$ whereas for Dirichlet boundary conditions vertex operators involve normal derivatives $\partial_n Z$. The relevant boundary correlators are
\bea
\label{Neumanntt}
\bra \partial_t Z(z) \partial_t \ov{Z}(w) \ket_{\mr{{\cal A},qu}}^{N} &=&
-\frac{\alpha'}{2} \left( \partial_z \partial_w \log \vartheta_1 (z-w)+ \mr{c.c.} \right) + \frac{4 \pi \alpha'}{t} \;, \\
\label{Dirictt}
\bra \partial_n Z(z) \partial_n \ov{Z}(w) \ket_{\mr{{\cal A},qu}}^{D} &=&
-\frac{\alpha'}{2} \left( \partial_z \partial_w \log \vartheta_1 (z-w) + \mr{c.c.} \right) \;.
\eea

\subsubsection{Internal classical correlators}
\label{sec:intclascorr}

In our following analysis we will encounter untwisted Dirichlet correlators of the general form $\bra \partial_n Z(z) \partial_n Z(w) \ket^{D}_{\mr{{\cal A}}}$ which do not receive quantum contributions, and purely come from the winding modes along the compact internal directions.

We begin with the classical action for the worldsheet cylinder which is given by
\be
S = \frac{1}{2\pi\alpha'}\int \! d^2 \sigma \,  G_{ij} \partial X^i \bar{\partial}X^j =
\frac{2\pi}{\alpha'}\int \! d^2 z\,  (\partial {\bar Z}^i \bar{\partial} Z^i + {\partial} Z^i \bar{\partial}\bar{ Z}^i) \;.
\ee
where $d^2\sigma$ are the coordinates of e.g.\  \cite{Polchinski1},
which we rescale as $d^2\sigma= (2\pi)^2 d^2 z$,
so that Re $z$ goes between $0$ and $1/2$.
Also, we used the complex coordinates $Z$ from \Ref{cplxcoord} above.
Note
that there is no explicit metric in the $Z$ coordinates.
A classical solution is \cite{Berg:2007wt}
\be  \label{classol}
Z_{\rm cl}(z,\bar{z}) = \sqrt{\alpha'} c\left[ r_c +  ( n+ m\bar{U}) \right](z+\bar{z})
\ee
where we defined $c=\sqrt{\tilde{T}_2/(2U_2)}$ in \Ref{normc} above,
and $r_c = r_1 + \bar{U}r_2$ with $r_1,r_2$ real.
This embeds $z=0$ (the D3-brane side of the cylinder) into the origin of this torus,
and $z=1/2$ (the D7-brane side of the cylinder) will be at $r_c$, or lattice shifts from $r_c$.
Also, $Z_{\rm class}$ has the right periodicities\footnote{There are no hidden moduli
dependences in the embedding periodicities.}
\bea
X_{\rm cl}^4 &\rightarrow&  X^4 + n \sqrt{\alpha'} \\
X_{\rm cl}^5 &\rightarrow&  X^5 + m \sqrt{\alpha'}
\eea
as $z\rightarrow z + 1/2$. The classical action of the solution \Ref{classol}
is
\be  \label{classact}
S =\frac{2\pi }{\alpha'}\int \! d^2 z\,  2\alpha' c^2|r_c + n+m\bar{U}|^2
= \pi c^2 t  \, |r_c + n+m\bar{U}|^2 =: t L_{nm}^2
\ee
where $L_{mn}^2$ is the rescaled length-squared of the winding string.
We note for later use that
\be \label{noteforlater}
{\partial L_{mn}^2  \over \partial \bar{r}_c} = \pi c^2 (r_c + n+m\bar{U}) \; .
\ee
From \Ref{classact}, the classical part of the partition function associated to the internal directions is given by the sum over the winding modes
\be  \label{Zintcl}
{\cal Z}_{\mr{int,cl}} (t) = \sum_{n,m} e^{-tL_{nm}^2} \;.
\ee
In the main text
we encounter correlation functions with insertions of the internal fields $\pd Z^i(z)$:
\[
\langle \partial_n  Z^i \rangle
\]
where $\partial_n$ is given in \Ref{NeuBC} above.
In the functional integral, the correlation function is given by the functional integral  over fluctuations around the classical solution weighted by the action. Denoting the fluctuations by $\eta^I$ we have:
\beq
\langle \pd Z^i \rangle =\sum_{\substack{\textrm{classical} \\ \textrm{solutions}}}  \int \textrm{d} \left[ \eta^I \right] \left(\pd Z_\textrm{cl}^i + \eta^I \right) e^{-S[\pd Z_\textrm{cl}^i + \eta^I]} \ .
\eeq
The action splits into the classical action and the action of pure fluctuations
(see for example \cite{Polchinski1}, Ch.8): $S[\pd Z_\textrm{cl}^I + \eta^I]= S_{\textrm{cl}} + S[\eta^I]$.
We have
\begin{align}
\langle \pd Z^i \rangle &=\sum_{\substack{\textrm{classical} \\ \textrm{solutions}}}   \left( \pd Z_\textrm{cl}^i \int \textrm{d} \left[ \eta^I \right] e^{-S[\eta^I]}+ \int \textrm{d} \left[ \eta^I \right] \eta^I e^{-S[\eta^I]} \right) e^{-S_{\textrm{cl}}} \\
&=\sum_{\substack{\textrm{classical} \\ \textrm{solutions}}}  \pd Z_\textrm{cl}^i \ \langle 1 \rangle \ e^{-S_{\textrm{cl}}}
\end{align}
where we used that the one-point-function of a quantum fluctuation vanishes,
and where $\langle 1 \rangle$ denotes the vacuum quantum partition function.
Here the sum over all classical background solutions is equivalent to the sum over all winding modes. We
find from \Ref{classol} and \Ref{classact} that
\be
\langle \pd Z^i \rangle= \langle 1 \rangle \sum_{m,n}  \sqrt{\alpha'} c\left( r_c +  ( n+ m\bar{U}) \right)\ e^{-tL_{mn}^2}
\ee
Using \Ref{noteforlater}, we can rewrite this more compactly as
\be
\langle \pd Z^i \rangle= -{\sqrt{\alpha'} \over \pi ct }\frac{\partial}{\partial \bar{r}_c} \sum_{m,n} e^{-tL_{mn}^2} \langle 1 \rangle
\ee
We will actually use the expression for the barred fields
\be
\langle \pd \bar{Z}^i \rangle= -{\sqrt{\alpha'} \over \pi ct }\frac{\partial}{\partial {r}_c} \sum_{m,n} e^{-tL_{mn}^2} \langle 1 \rangle
\ee
Going back to the un-barred fields, we note that the one-point function
is essentially the derivative of the partition function with respect to the position of the D3 brane,
\be
\langle \pd Z^i \rangle=  -{\sqrt{\alpha'} \over \pi ct }\frac{\partial}{\partial \bar{r}_c} \mathcal{Z}(t)
\ee
with the partition function $\mathcal{Z}(t) =\sum \int e^{-S}$.

This calculation for the insertion of a single $\langle \partial_n Z^i \rangle$ extends directly to the case of multiple insertions:
instead of a single derivative we now have (for $n$ insertions of $\partial_n Z^i$)
\be
\label{windingderivative}
\langle {(\pd Z^i)}^n \rangle=\left(-{\sqrt{\alpha'} \over \pi ct }\frac{\partial}{\partial \bar{r}_c}  \right)^n \mathcal{Z} (t) \ .
\ee
The classical partition function, i.e. the sum over winding modes (with the quantum partition function $\langle 1 \rangle$ suppressed), can be conveniently rewritten in terms of a Riemann theta function given in the appendix \eqref{doubletheta}:
\be
\label{Zt}
\mathcal{Z}(t) = \sum_{m,n} e^{- \pi c^2 t  \, |r_c + n+m\bar{U}|^2} = \tha{\vec R}{\vec 0}{\vec 0, it {G}/2} \ ,
\ee
where
\be
\vec{R} =\left( \begin{array}{c} r_1 \\ r_2 \end{array} \right)
\ee
in $r_c = r_1 + \bar{U}r_2$
and $G$ is the torus metric \Ref{torusmetric}. This expression will be useful for evaluating integrals over the cylinder modulus $t$. We also note that the modular S transformation of the genus two theta function
\be
\tha{ \vec{R} }{ \vec 0 }{\vec 0 , itG} = \sqrt{G}^{-1} t^{-1}\tha{ \vec{0} }{ \vec 0 }{\vec{R} ,it^{-1}G^{-1}}
\ee
is identical to a Poisson resummation of the sum over winding modes.

\subsubsection{Momentum exponential correlators and pole structures}
\label{sec:momenexpopole}

We also encounter correlators involving exponentials $e^{ikX}$. These are most easily calculated using real coordinates $x^{M}$ and momenta $k^M$. The relevant correlator
\be
\bra \prod_i e^{ik_i \cdot x\left(z,\bar{z}\right)} \ket \;,
\ee
is evaluated by contracting the scalars using the real forms\footnote{These are simply related to the complex versions by a factor of $\half$.} of the cylinder correlators (\ref{neumannxx}) and (\ref{dirichletxx}).
In general this is given by
\be
\prod_{i<j} e^{- k_i \cdot k_j \mc{G}(z_i - z_j)},
\ee
where $\mc{G}(z_i - z_j)$ is the bosonic correlator.

However, we also provide more explicit expressions in the case we
only require
the Neumann correlator in the limit $z_i\rightarrow z_j$, where we can drop the zero mode piece of (\ref{neumannxx}).  This is given by
\be
\bra \prod_i e^{ik_i \cdot x\left(z_i,\bar{z}_i\right)} \ket^{N}_{{\cal A}} = \prod_{i<j} \left| \frac{\vartheta_1\left(z_{ij}\right)}{\vartheta_1'(0)} \right|^{\alpha' k_i k_j} \;. \label{expcorrreal}
\ee
We may also write (\ref{expcorrreal}) in complex co-ordinates and momenta as
\be
\bra \prod_i e^{ik_iX\left(z_j\right)} \ket^{N}_{{\cal A}} = \prod_{i<j} \left| \frac{\vartheta_1\left(z_{ij}\right)}{\vartheta_1'(0)} \right|^{\frac{\alpha'}{2}\left(k^+_i k^-_j + k^-_i k^+_j \right)} \;. \label{momexpcorrcom}
\ee
where we recall that the complex notation $k_iX\left(z\right)$ is defined in (\ref{kXcomplex}).

Another correlator that we require is
\be
\bra \partial X(w) \prod_i e^{ik_iX\left(z_j\right)} \ket^{N}_{{\cal A}} = -i\alpha' \prod_{i<j} k_j^+ \frac{\vartheta'_1\left(w-z_j\right)}{\vartheta_1\left(w-z_j\right)} \left| \frac{\vartheta_1\left(z_{ij}\right)}{\vartheta_1'(0)} \right|^{\frac{\alpha'}{2}\left(k^+_i k^-_j + k^-_i k^+_j \right)} \;, \label{momexpdercorr}
\ee
which can be deduced by acting on (\ref{expcorrreal}) with a derivative.

At this point we discuss a principle which greatly simplifies our calculations. The important point is that since we are probing
non-derivative terms in the action we do not need to know the full amplitude but rather only its zero momentum limit $k_i\rightarrow 0$. Given this it seems naively that bosonic correlators such as (\ref{momexpdercorr}) vanish. However it is also possible to generate a pole in the amplitude which when combined with the correlator (\ref{expcorrreal}) can generate inverse powers of momenta that cancel against the momenta in the amplitude leaving a result that is non-vanishing in the zero momentum limit. To see this consider the amplitude factor
\be
\label{polecancmom}
{\cal A} \supset \underset{{k_1\cdot k_2\rightarrow 0}}{\mr{lim}} \left[ \left(k_1 \cdot k_2\right) \int dz_1 \left|\frac{\vartheta_1\left(z_1-z_2\right)}{\vartheta_1'\left(0\right)}\right|^{k_1\cdot k_2} \left(\frac{\vartheta'\left(0\right)}{\vartheta_1\left(z_1-z_2\right)}\right)\right] = \frac{\left(k_1 \cdot k_2\right)}{\left(k_1 \cdot k_2\right)} = 1\;,
\ee
where we have used
\be
\frac{\vartheta_1\left(z\right)}{\vartheta_1'\left(0\right)} = z + {\cal O}\left(z^3\right) \;.
\ee
The pole at $z_1=z_2$ has canceled the vanishing momentum prefactor. In practice this means that
evaluating certain amplitudes can simply amount to analysing their pole structure.

\subsection{Fermionic and Ghost Correlators}
\label{sec:fermcorr}
The correlator of worldsheet fermions in the even spin structures is
\beq  \label{wsfermion}
S_\alpha (z_1-z_2)= \frac{\vartheta_\alpha(z_1-z_2) \vartheta_1^{\prime}(0)}{\vartheta_1(z_1-z_2) \vartheta_\alpha(0)} \ .
\eeq
In the main text we used
\beq  \label{szegosquare2}
S_\alpha^2 (z_1-z_2) = \wp (z_1-z_2) - e_{\alpha -1},
\eeq
and
\be \label{weierstrass2}
{\wp}(z,\tau) = -\partial^2_{z} \log \tht_1(z,\tau) +4\pi i \partial_{\tau} \log \eta(\tau) \, .
\ee
These relations simply follow from the fact that the Weierstrass function
is the unique elliptic (doubly periodic and meromorphic) function
with a double pole at the origin. Therefore
both $S_{\alpha}^2$ and $\partial^2_{\nu} \log \tht_1(\nu,\tau)$
must be expressible in $\wp$.  (Note
that the quasiperiodicity of $\tht_1$ disappears when taking the logarithm
and two derivatives.) The $z$-independent shifts, the second
terms in each of \Ref{szegosquare2} and \Ref{weierstrass2}, adjust
the zeros.

The amplitudes also involve correlators of spin fields, which after bosonisation as in (\ref{spinbos}) correspond
 to correlators of $H$ fields.
This includes the case of the $\psi$ correlators which are spin fields with $\pm 1$ H charge.
The correlators depend on the spin structure, denoted by indices $\left(\alpha\beta\right)=\left\{\left(00\right),\left(10\right),\left(01\right),\left(11\right)\right\}$, and read
\be
\label{hchargecorr}
\bra \prod_i e^{ia_iH\left(z_i\right)} \ket = K_{\alpha\beta} \left[\prod_{i<j} \left(\frac{\vartheta_1\left(z_{ij}\right)}{\vartheta'_1\left(0\right)}\right)^{a_ia_j}  \right]\vartheta_{\alpha\beta}\left(\sum_i a_i z_i + \theta_I\right) \;,
\ee
where $\theta_I$ is the orbifold twist in torus $I$. The constants $K_{\alpha\beta}$ are determined for each amplitude by the factorisation limit. This amounts to taking the limit $z_i \rightarrow z_j$ for all $i,j$ so that the amplitude factorises to the field theory amplitude times the string partition function. The spin structure is then matched to that of the partition function.
Note that using (\ref{opespin}) we deduce that only correlators where the total $H$-charge is zero are non-vanishing. This is known as $H$-charge conservation. These correlators were derived by Atick and Sen by considering their OPEs with the stress tensor, giving a set
of differential equations that can be solved to obtain the correlator. The details can be found in
\cite{Atick:1986ns, Atick:1986rs, Abel:2004ue}.

The ghost correlators can be found by the same method \cite{Atick:1986ns, Atick:1986rs}. The resulting correlators are very similar to the fermionic
correlators except with signs and powers reversed,
\be
\label{ghostchargecorr}
\bra \prod_i e^{ia_i\phi\left(z_i\right)} \ket = K_{\alpha\beta} \left[\prod_{i<j} \left(\frac{\vartheta_1\left(z_{ij}\right)}{\vartheta'_1\left(0\right)}\right)^{-a_ia_j}  \right]\vartheta_{\alpha\beta}^{-1}\left(-\sum_i a_i z_i \right) \;.
\ee
Again, the factors $K_{\alpha\beta}$ are determined by factorisation onto the partition function limit.

\subsection{Partition functions}
\label{sec:partfuncpre}

In the $2,3,4$ spin structures - those involving $\vartheta_{00}, \vartheta_{01},$ and  $\vartheta_{10}$ - the partition functions for the non-compact dimensions are given as follows
\begin{align}
\mathrm{Bosonic} : \quad & \frac{1}{\eta^4 (it)} \frac{1}{(4\pi^2 \ap t)^2}, \nonumber \\
\mathrm{Fermionic} :\quad& \bigg(\frac{\vartheta_\nu (0)}{\eta (it)} \bigg)^2, \nonumber \\
bc\ \mathrm{ghosts} :\quad& \eta^2 (it), \nonumber \\
\beta \gamma \ \mathrm{ghosts} :\quad& \frac{\eta(it)}{\vartheta_\nu (0)}, \nonumber \\
\mathrm{Total} :\quad& \frac{\vartheta_\nu (0)}{\eta^3 (it)}  \frac{1}{(4\pi^2 \ap t)^2}.
\end{align}
For the 1 spin structure, which involves $\vartheta_{11}$, the above expressions must be changed, and they become
\begin{align}
\mathrm{Bosonic} :\quad& \frac{1}{\eta^4 (it)} \frac{1}{(4\pi^2 \ap t)^2}, \nonumber \\
\mathrm{Fermionic} :\quad& \bigg(\eta^4 (it) \bigg)^2, \nonumber \\
bc\ \mathrm{ghosts} :\quad& \eta^2 (it), \nonumber \\
\beta \gamma \ \mathrm{ghosts} :\quad& \frac{1}{\eta^2 (it)}, \nonumber \\
\mathrm{Total} :\quad& \frac{1}{(4\pi^2 \ap t)^2}.
\end{align}
which assumes that the zero modes in the fermionic sector are saturated.
If this is not the case that the partition function vanishes due to integrating over the fermionic zero modes.
Note that we require no additional insertions for the $\beta \gamma$ ghosts; their zero modes must be explicitly excluded.
In practice however the effect of the fermionic and ghost partition functions are already incorporated into the
correlators (\ref{hchargecorr}) and (\ref{ghostchargecorr}).

The internal bosonic partition functions depend on the boundary conditions for the coordinate field $X$. In D3-D3 models all internal directions exhibit DD boundary conditions. In D3-D7 constructions we also have mixed ND boundary conditions in two of the three complex dimensions. The bosonic partition function for one compact torus $I$  with twist $\theta_I \ne 0$ is\footnote{For the partition function derivation see for example \cite{Conlon:2009xf,Conlon:2009kt}.}
\beq
\mathcal{Z}_{I} =   \mathcal{Z}(t) \times \left\{ \begin{array}{cc} \frac{-2\sin \pi \theta_I}{\vartheta_1 (\theta_I)} & \textrm{DD directions} \\
\frac{1}{\vartheta_4 (\theta_I)} & \textrm{ND directions} \end{array} \right. ,
\eeq
while for an untwisted torus of area $\tilde{T}_2$ and complex structure $U= U_1 + i U_2$ it is
\beq
\mathcal{Z}_{I} =   \mathcal{Z}(t) \times \left\{ \begin{array}{cc} \frac{1}{\eta^3 (it/2)} & \textrm{DD directions} \\
\frac{1}{\vartheta_4 (0)} & \textrm{ND directions} \end{array} \right. ,
\eeq
where $\mathcal{Z}(t)$ is the partition function over classical winding solutions as described in \S\ref{sec:intclascorr}. If winding strings do not contribute the classical partition function just provides a factor $\mathcal{Z}(t)=1$.

\section{Theta functions and identities}
\label{identities}

The standard notation for the Jacobi theta functions is:
\begin{equation}
\tha{ a }{ b }{z, \tau}= \sum_{n = -\infty}^{\infty} \mathrm{exp} \bigg[ \pi i (n+a)^2 \tau + 2 \pi i (n+a)(z+b) \bigg] \ .
\end{equation}
The following notation is also used: $\vartheta_{\alpha \beta}(z) \equiv \tha{ \alpha/2 }{ \beta/2}{z}$, and
\begin{eqnarray}
\vartheta_1 \equiv \vartheta_{11}, & \vartheta_2 \equiv \vartheta_{10} \ ,  \nonumber \\
\vartheta_3 \equiv \vartheta_{00},  & \vartheta_4 \equiv \vartheta_{01}.
\end{eqnarray}
Expansions of the functions for $q=e^{\pi i \tau}$ are
\begin{eqnarray}
\vartheta_{00} (z , \tau) & = \vartheta_3 =& 1 + 2\sum_{n=1}^{\infty} q^{n^2} \cos 2\pi n z \nonumber \\
\vartheta_{01} (z , \tau) & = \vartheta_4 =& 1 + 2\sum_{n=1}^{\infty} (-1)^n q^{n^2} \cos 2\pi n z \nonumber \\
\vartheta_{10} (z , \tau) & = \vartheta_2 =& 2q^{1/4}\sum_{n=0}^{\infty} q^{n(n+1)} \cos \pi (2n+1) z \nonumber \\
\vartheta_{11} (z , \tau) & = \pm \vartheta_1 =& 2q^{1/4}\sum_{n=0}^{\infty} (-1)^n q^{n(n+1)} \sin \pi (2n+1) z \nonumber \\
\end{eqnarray}
The Dedekind $\eta$ function is defined as
\begin{eqnarray}
\eta (\tau) &=& q^{1/12} \prod_{m=1}^{\infty} (1-q^{2m}) \\
&=& \left[ \frac{\vartheta_1^{\prime} (0 , \tau)}{-2\pi} \right]^{1/3}.
\end{eqnarray}

It will be useful to reexpress the sum over winding modes in terms of the following generalised
(Riemann) theta-functions:
\begin{equation}
\label{doubletheta}
\tha{ \vec{\alpha} }{ \vec\beta}{\vec \nu , G} = \sum_{\vec n\in \mathbb{Z}^N}
 e^{i\pi(\vec n+\vec \alpha)^{\rm T} G (\vec n+\vec \alpha)}
e^{2\pi i(\vec \nu+\vec \beta)^{\rm T}(\vec n+\vec\alpha)}  \ ,
\end{equation}
where $G$ is an $N \times N$ matrix with $\textrm{Im}(G)>0$. The case $N=1$ gives the usual Jacobi theta functions while the case $N=2$ will be used to reexpress the sum over winding modes on a two-torus.

\subsection*{Riemann summation formula}
We can simplify expressions involving theta functions using the generalised Riemann summation formula:
\begin{align}
\label{Riemann}
&\sum_{\alpha, \beta} {(-1)}^{\alpha +\beta+ \alpha \beta} \prod_{i=1}^4 \vartheta \left[\begin{array}{c} \alpha/2 +c_i \\ \beta/2 +d_i \end{array}\right] \left(z_i, \tau \right) =\\
\nn &= 2 \vartheta \left[\begin{array}{c} 1/2 \\ 1/2 \end{array}\right] \left(\sum_i \frac{z_i}{2}, \tau \right) \vartheta \left[\begin{array}{c} 1/2+c_2 \\ 1/2+d_2 \end{array}\right] \left(\sum_i \frac{z_1+z_2-z_3-z_4}{2}, \tau \right) \\
\nn & \times \vartheta \left[\begin{array}{c} 1/2+c_3 \\ 1/2+d_3 \end{array}\right] \left(\sum_i \frac{z_1-z_2+z_3-z_4}{2}, \tau \right) \vartheta \left[\begin{array}{c} 1/2+c_4 \\ 1/2+d_4 \end{array}\right] \left(\sum_i \frac{z_1-z_2-z_3+z_4}{2}, \tau \right) \ .
\end{align}
We also have the five-theta identity \cite{Atick:1986rs}:
\begin{align}
\label{5Riemann}
\nn & \sum_{\alpha, \beta} \eta_{\alpha \beta} \ \vartheta_{\alpha \beta}(z_1) \vartheta_{\alpha \beta}(z_2) \vartheta_{\alpha \beta}(z_3) \vartheta_{\alpha \beta}(z_4) \vartheta_{\alpha \beta}(z_5) \vartheta_{\alpha \beta}^{-1}(z_1+z_2+z_3+z_4+z_5) \\
= \ & -2 \ \vartheta_1 (z_1+z_2+z_3+z_4) \vartheta_1 (z_2+z_3+z_4+z_5) \vartheta_1 (z_1+z_3+z_4+z_5) \\
& \times \ \vartheta_1 (z_1+z_2+z_4+z_5) \vartheta_1 (z_1+z_2+z_3+z_5) \vartheta_1^{-1}(2[z_1+z_2+z_3+z_4+z_5])
\end{align}

\subsection*{Identities for D3-D3 models}

In particular, we can use the above relation to derive the following identities that arise in the context of D3-D3 calculations. For untwisted sectors we find
\begin{equation}
\label{D3D3untw}
\sum_{\alpha, \beta} \eta_{\alpha \beta} \ {\vartheta}_{\alpha \beta}^{\prime \prime}(0) \prod_{i=1}^{3} \vartheta_{\alpha \beta}(0) = 0
\end{equation}
where $\eta_{\alpha \beta} = {(-1)}^{\alpha + \beta - 2 \alpha \beta}$ and derivatives are w.r.t.~$z$. In the context of fully twisted sectors we can substitute
\begin{equation}
\sum_{\alpha, \beta} \eta_{\alpha \beta} \frac{{\vartheta}_{\alpha \beta}^{\prime \prime}(0)}{\eta^3} \prod_{i=1}^{3} \frac{\vartheta_{\alpha \beta}(\theta_i)}{\vartheta_{1}(\theta_i)} = -2 \pi \sum_{i=1}^3 \frac{\vartheta_{1}^{\prime}(\theta_i)}{\vartheta_{1}(\theta_i)} \ .
\end{equation}
We will rely on the following identity to simplify results for partially twisted sectors of the orbifold:
\begin{equation}
\label{D3D3partial}
\sum_{\alpha, \beta} \eta_{\alpha \beta} {(-1)}^{\alpha} \frac{{\vartheta}_{\alpha \beta}^{\prime \prime}(0)}{\eta^3} \frac{{\vartheta}_{\alpha \beta}(0)}{\eta^3} \frac{\vartheta_{\alpha \beta}(\theta_1)}{\vartheta_{1}(\theta_1)} \frac{\vartheta_{\alpha \beta}(\theta_2)}{\vartheta_{1}(\theta_2)} = -4\pi^2
\end{equation}
where $\theta_1 + \theta_2 =1$ mod 2.

\subsection*{Identities for D3-D7 models}
Expressions arising in the context of D3-D7 models can be simplified as follows. For untwisted sectors we find:
\begin{equation}
\label{D3D7untw}
\sum_{\alpha, \beta} \eta_{\alpha \beta} \frac{{\vartheta}_{\alpha \beta}^{\prime \prime}(0)}{\eta^3} \frac{{\vartheta}_{\alpha \beta}(0)}{\eta^3} {\left(\frac{\thw{1/2 -\alpha/2}{\beta/2} (0)}{\thw{0}{ 1/2}(0)} \right)}^2 = 4\pi^2 \ .
\end{equation}
Correspondingly, for partially twisted sectors $\theta_3=0$ (mod $1$) we can use:
\begin{equation}
\sum_{\alpha, \beta} \eta_{\alpha \beta} \frac{{\vartheta}_{\alpha \beta}^{\prime \prime}(0)}{\eta^3} \frac{{\vartheta}_{\alpha \beta}(0)}{\eta^3} \left(\frac{\thw{1/2 -\alpha/2}{\beta/2} (\theta_1)}{\thw{ 0}{ 1/2 } (\theta_1)} \right) \left(\frac{\thw{1/2 -\alpha/2 }{ \beta/2}(\theta_2)}{\thw{0 }{ 1/2 }(\theta_2)} \right) = 4\pi^2 \ .
\end{equation}

\section{Yukawa couplings and gaugino condensation}
\label{sec:gaugekincorr2}

In this section we perform a complementary calculation to the determination of the holomorphic $A$-terms by the computation of a threshold correction to the D7-brane Yukawas. It involves computing the cylinder correlation function $\langle \tr(\lambda \lambda) \tr(\psi \psi \phi) \rangle$. In principle it should give an identical result and the fact it does we take as strong confirmation of our results.


To compute the cylinder amplitude $\langle \tr(\lambda \lambda) \tr(\psi \psi \phi) \rangle$, we insert
gaugino and Yukawa vertex operators on opposite boundaries of the cylinder. As before, the gaugini can arise from a stack of either D3- or D7-branes.

\subsection{Setting up the calculation}
\label{sec:settingup}
We follow the general strategy in section \ref{sec-strategy}.
For $\langle (\lambda \lambda) (\psi \psi \phi) \rangle$, the canonical picture H-charges are
\bal
\nn \lambda_1^{-\half}(z_1) \ = & \ \frac12 \left(+,-,-,-,-\right) \\
\nn \lambda_2^{-\half}(z_2) \ = & \ \frac12 \left(-,+,-,-,-\right) \\
 \psi_1^{-\half}(z_3) \ = & \ \frac12 \left(+,-,-,+,+\right) \\
\nn \psi_2^{-\half}(z_4) \ = & \ \frac12 \left(-,+,+,-,+\right) \\
\nn \phi^{-1}(z_5) \ = & \ \frac12 \left(0,0,0,0,--\right) \ .
\end{align}
Under an orbifold twist the vertex operators transform as $\psi_1 \rightarrow e^{-2 \pi i \theta_1} \psi_1$, $\psi_2 \rightarrow e^{-2 \pi i \theta_2} \psi_2$ and $\phi \rightarrow e^{-2 \pi i \theta_3} \phi$ and thus the Yukawa couplings above come from the superfield term $C^1 C^2 C^3$. Later we also study Yukawa couplings with general flavour structure $C^r C^s C^t$. The gaugino vertex operators are invariant under orbifold twists as expected: $\lambda \rightarrow e^{-\pi i (\theta_1+\theta_2+\theta_3)} \lambda = \lambda$. Picture changing\footnote{When picture-changing fermionic vertex operators as described in the main text difficulties can arise due to the appearance of terms singular in $(z-w)$. To avoid these issues picture-changing should be performed by inserting one picture-changing-operator (PCO) $e^{\phi(z)} T_F(z)$ into the amplitude for each vertex operator to be modified. After the amplitude has been calculated with the PCO insertions we can safely take the limit where the PCO positions approach the insertion points of the vertex operators. These problems with the fermionic vertex operators only originate when picture-changing in the external directions due to OPEs with the momentum exponentials. Thus we do not encounter any difficulties in the given situation and the cavalier way of picture-changing gives the correct result.}, we arrive at the following H-charge configuration:
\bal
\label{HchargeGY}
\nn \lambda_1^{-\half}(z_1) \ = & \ \frac12 \left(+,-,-,-,-\right) \\
\nn \lambda_2^{-\half}(z_2) \ = & \ \frac12 \left(-,+,-,-,-\right) \\
 \psi_1^{+\half}(z_3) \ = & \ \frac12 \left(+,-,+,+,+\right) \\
\nn \psi_2^{+\half}(z_4) \ = & \ \frac12 \left(-,+,+,+,+\right) \\
\nn \phi^{0}(z_5) \ = & \ \frac12 \left(0,\ 0, \ 0, \ 0, \ 0\right)
\end{align}
and the vertex operators for the Yukawa coupling are given by
\begin{align}
\mathcal{V}_{\psi_1}^0 (z_3) &= \pd \bar{Z}^1 e^{i q_3 \cdot H} e^{i k_3 \cdot X} (z_3) \\
\mathcal{V}_{\psi_2}^0 (z_4) &= \pd \bar{Z}^2  e^{i q_4 \cdot H} e^{i k_4 \cdot X} (z_4) \\
\mathcal{V}_{\phi}^0 (z_5) &= \pd \bar{Z}^3  e^{i q_5 \cdot H} e^{i k_5 \cdot X} (z_5)
\end{align}
where $q_3$, $q_4$ and $q_5$ are given by the H-charges displayed above. Again, we find that a Yukawa-coupling of the chiral superfields $C^1 C^2 C^3$ leads to the appearance of the internal bosonic fields $\pd \bar{Z}^1\pd \bar{Z}^2\pd \bar{Z}^3$. We also can check this explicitly for a Yukawa coupling arising from $C^3 C^3 C^3$. In this case we have to start with the following assignment of H-charges in the canonical picture
\bal
\nn \lambda_1^{-\half}(z_1) \ = & \ \frac12 \left(+,-,-,-,-\right) \\
\nn \lambda_2^{-\half}(z_2) \ = & \ \frac12 \left(-,+,-,-,-\right) \\
 \psi_1^{-\half}(z_3) \ = & \ \frac12 \left(+,-,+,+,-\right) \\
\nn \psi_2^{-\half}(z_4) \ = & \ \frac12 \left(-,+,+,+,-\right) \\
\nn \phi^{-1}(z_5) \ = & \ \frac12 \left(0,0,0,0,--\right) \ .
\end{align}
which exhibit the correct behaviour under orbifold twists. Again, picture-changing has to occur in the internal directions (and, in fact, only on the third 2-torus). The resulting H-charges are the same that we obtained in the previous case \eqref{HchargeGY} as it is the only configuration that allows for a full cancellation of H-charge. Further, we find that the vertex operators for the Yukawa coupling pick up the bosonic field combination $\pd \bar{Z}^3\pd \bar{Z}^3\pd \bar{Z}^3$.

\subsection{Classical solution - winding modes}

The classical correlator from $\langle \partial \bar{Z}^r \partial \bar{Z}^s \partial \bar{Z}^t \rangle$ is the same as in \S\ref{winding}:
the only non-zero contribution can arise from a Yukawa-coupling of the form $C^r C^r C^r$. For $r=3$ this is
\begin{align}
\langle \pd \bar{Z}^3 \pd \bar{Z}^3 \pd \bar{Z}^3 \rangle= \ &\sum_{m,n} 8 {\left(\frac{\tilde{T}_2}{2U_2} \right)}^{3/2} {\left(m + n{U} +{\bar{r}_c} \right)}^3 \ e^{-\frac{\pi \tilde{T}_2 t}{\alpha^{\prime} 2 U_2} {\left|m + n\bar{U} +{r_c} \right|}^2} \\
= \ & \left( -{\sqrt{\alpha'} \over \pi ct }\frac{\partial}{\partial r_c} \right)^3 \mathcal{Z}(t)
\end{align}
where $\mathcal{Z}(t)$ is the partition function over winding modes, provided the third two-torus is unaffected by the orbifold twist.

\subsection{Fermionic and ghost amplitude}

The correlator of the spinorial and ghost degrees of freedom is independent of the structure of the Yukawa coupling and thus we can calculate it independently for all cases we want to consider later. The configuration of ghost and H-charges is summarised in \eqref{HchargeGY}.

\subsection*{D3-D3 models}
In this case the spin-structure dependent terms of the correlators over fermionic and ghost fields contribute
\begin{align}
&\frac{\vartheta_{\alpha \beta} \left(\frac{z_1-z_2+z_3-z_4}{2} \right)\vartheta_{\alpha \beta} \left(\frac{-z_1+z_2-z_3+z_4}{2} \right)}{\vartheta_{\alpha \beta} \left(\frac{z_1+z_2-z_3-z_4}{2} \right)}  \\
\nn &\vartheta_{\alpha \beta} \left(\frac{-z_1-z_2+z_3+z_4}{2} + \theta_1 \right) \vartheta_{\alpha \beta} \left(\frac{-z_1-z_2+z_3+z_4}{2} + \theta_2 \right) \vartheta_{\alpha \beta} \left(\frac{-z_1-z_2+z_3+z_4}{2} + \theta_3 \right) \ .
\end{align}
Jacobi theta-functions are symmetric or antisymmetric under a change of sign of the argument $z$. Thus the above expression is left invariant if we swap the sign of the argument for two of the above functions. It will be helpful to modify the argument of the second theta function in the numerator and the theta function in the denominator accordingly:
\begin{align}
&\frac{\vartheta_{\alpha \beta} \left(\frac{z_1-z_2+z_3-z_4}{2} \right)\vartheta_{\alpha \beta} \left(\frac{z_1-z_2+z_3-z_4}{2} \right)}{\vartheta_{\alpha \beta} \left(\frac{-z_1-z_2+z_3+z_4}{2} \right)}  \\
\nn &\vartheta_{\alpha \beta} \left(\frac{-z_1-z_2+z_3+z_4}{2} + \theta_1 \right) \vartheta_{\alpha \beta} \left(\frac{-z_1-z_2+z_3+z_4}{2} + \theta_2 \right) \vartheta_{\alpha \beta} \left(\frac{-z_1-z_2+z_3+z_4}{2} + \theta_3 \right) \ .
\end{align}
 The classical amplitudes vanishes unless $\theta_3 = 0$ (no twisting on the third torus). We can then simplify the result by summing over spin-structures and employing a Riemann identity \eqref{Riemann}:
\begin{align}
\sum_{\alpha \beta} \eta_{\alpha \beta} \ & \vartheta_{\alpha \beta} \left(\frac{z_1-z_2+z_3-z_4}{2} \right) \vartheta_{\alpha \beta} \left(\frac{z_1-z_2+z_3-z_4}{2} \right) \\
\nn & \times \ \vartheta_{\alpha \beta} \left(\frac{-z_1-z_2+z_3+z_4}{2} + \theta_1 \right) \vartheta_{\alpha \beta} \left(\frac{-z_1-z_2+z_3+z_4}{2} + \theta_2 \right) \\
& =- 2 \vartheta_1(z_1-z_4) \vartheta_1(z_2-z_3) \vartheta_1(\theta_1) \vartheta_1(\theta_2)
\end{align}
The spin-structure-independent terms for the fermions and ghosts evaluate to:
\beq
\label{spinindep}
\frac{\vartheta_1^{\prime}(0)}{\vartheta_1(z_1-z_4)} \frac{\vartheta_1^{\prime}(0)}{\vartheta_1(z_2-z_3)}
\eeq
Combining both results we arrive at the final expression for the correlator over fermionic and ghost fields in a partially twisted sector:
\begin{align}
\label{fgGY}
\nn&- 2 \vartheta_1^{\prime}(0) \vartheta_1^{\prime}(0) \vartheta_1(\theta_1) \vartheta_1(\theta_2) \\
&=-8 \pi^2 \eta^6 \vartheta_1(\theta_1) \vartheta_1(\theta_2)
\end{align}
where we have used $\vartheta_1^{\prime}(0)= 2 \pi \eta^3$. We find that the result does not depend on the worldsheet positions.

\subsection*{D3-D7 models}
For D3-D7 models the presence of Neumann-Dirichlet boundary conditions for the directions along the worldvolume of the D7 brane
implies the theta functions appearing in correlation functions have to be modified as
\beq
\tha{\alpha}{\beta/2}{z} \quad \rightarrow \quad \tha{ 1/2-\alpha/2}{ \beta/2}{z} = \tha{ (\alpha+1)/2}{\beta/2 }{z}
\eeq
where we used a symmetry of the theta function to rewrite it in a more useful form.

Using the above result we can revisit the fermionic and ghost correlators for D3-D7 models. The ghost correlators are unaffected while we have to modify the fermionic correlation functions on the first two sub-tori as described above. By repeating the analysis of the previous section we thus find for the spin-structure-dependent terms:
\begin{align}
\sum_{\alpha \beta} \eta_{\alpha \beta} \ & \vartheta_{\alpha \beta} \left(\frac{z_1-z_2+z_3-z_4}{2} \right) \vartheta_{\alpha \beta} \left(\frac{z_1-z_2+z_3-z_4}{2} \right) \\
\nn  & \times \ \vartheta_{(\alpha+1) \beta} \left(\frac{-z_1-z_2+z_3+z_4}{2} + \theta_1 \right) \vartheta_{(\alpha+1) \beta} \left(\frac{-z_1-z_2+z_3+z_4}{2} + \theta_2 \right) \ .
\end{align}
Using the generalised Riemann identity \eqref{Riemann} with $c_3=c_4=\frac{1}{2}$ this becomes
\beq
=- 2 \vartheta_1(z_1-z_4) \vartheta_1(z_2-z_3) \tha{0 }{ 1/2}{\theta_1} \ \tha{0 }{ 1/2}{\theta_2} \ .
\eeq
The spin-structure independent terms are unaffected and using the result from the D3-D3 case \eqref{spinindep} we find the following result for the spinorial and ghost correlator:
\be
\label{fgGY2}
-8 \pi^2 \eta^6 \tha{ 0 }{ 1/2}{\theta_1} \tha{ 0 }{ 1/2 }{\theta_2}  .
\ee
Again, we find the expression to be independent of the worldsheet positions.

As in section 4, we also need to introduce orbifold images of D7-branes. However just as in section 4 (around eq. (\ref{joeade}) the images give identical results and do not affect the final answer.

\subsection{Completing the calculation - partition functions}

To complete the results of the previous sections we need to combine them with the correlator over momentum exponentials and the appropriate partition functions. The fermionic partition function and the vacuum amplitude over ghosts have already been included implicitly in the result for the fermionic and ghost correlator.

\subsection*{D3-D3 models}
For this setup we combine results \eqref{fgGY} and \eqref{clsolD3D3} with the bosonic partition function. After having included the correlator over momentum exponential we find
\begin{align}
\nn \mathcal{A} \propto & \int \frac{\textrm{d}t}{t} \int \textrm{d}z_1 \textrm{d}z_2 \textrm{d}z_3 \textrm{d}z_4 \textrm{d}z_5 \frac{ \prod_{k=1}^2 (-2 \sin \pi \theta_k)}{(2 \pi^2 t)^2} \left(\prod_{i < j} e^{-k_i \cdot k_j \mathcal{G}(z_i-z_j)} \right) {\left( -{\sqrt{\alpha'} \over \pi ct }\frac{\partial}{\partial r_c} \right)}^3 \mathcal{Z}(t) \ ,
\end{align}
which is identical with the expression obtained earlier for corrections to the gauge kinetic function \eqref{result1D3D3}.

\subsection*{D3-D7 models}
Bringing all results together for this setup we obtain:
\begin{align}
\nn \mathcal{A} \propto & \int \frac{\textrm{d}t}{t} \int \textrm{d}z_1 \textrm{d}z_2 \textrm{d}z_3 \textrm{d}z_4 \textrm{d}z_5 \frac{1}{(2 \pi^2 t)^2} \left(\prod_{i < j} e^{-k_i \cdot k_j \mathcal{G}(z_i-z_j)} \right) {\left( -{\sqrt{\alpha'} \over \pi ct }\frac{\partial}{\partial r_c} \right)}^3 \mathcal{Z}(t) \ .
\end{align}
Again, the resulting expression coincides with the result for the calculation of corrections to the gauge kinetic function \eqref{result2D3D7}.

\subsection{Discussion}

As the results for both the D3-D3 and the D3-D7 models match with the calculations for corrections to the gauge kinetic function we refer readers to the discussion of these previous results in \S\ref{disc1}. The fact that the resulting expressions coincide is a welcome check of our previous calculation. The correlator evaluated here is related by supersymmetry to the amplitude computed in \S\ref{sec:gaugekincorr1} and we expected arrive at the same result. Once we integrate over the worldsheet modulus $t$ we thus arrive at the result:
\be
\mathcal{A} = - \mathcal{C} {\left(\frac{U_2}{\tilde{T}_2}\right)}^{3/2} \frac{\partial^3}{{\partial r_c^3}} \ln \vartheta_1(r_c,\bar{U}) \sim \tilde{T}_2^{-3/2} \sim R^{-3} \ ,
\ee
where as before the scaling is for $R_1 \sim R_2 \sim R$.

\section{Calculation of corrections to Yukawa couplings due to two scalars}
\label{2scalar}

In this section we will evaluate the cylinder amplitude $\langle \tr (\tilde{\phi} \tilde{\phi}) \tr (\psi \psi \phi) \rangle$ of two hidden scalars $\tilde{\phi}$ and a visible sector Yukawa coupling. The calculation follows a similar structure to the earlier ones.
The relevant orbifolds turn out to be
\be
\mathbb{Z}_4: \ \theta=\frac{1}{4}(1,1,-2) \qquad \qquad \mathbb{Z}_6^{\prime}: \ \theta=\frac{1}{6}(1,2,-3) \ .
\ee
The interactions of the form $\langle \tr(\tilde{\phi} \tilde{\phi}) \tr(\psi \psi \phi) \rangle$ consistent with
 gauge-invariance are
\begin{align}
\nn \mathbb{Z}_4: \quad & \langle \tr(\tilde{C}^3 \tilde{C}^3) \tr(C^1 C^2 C^3) \rangle \\
\nn & \langle \tr(\tilde{C}^3 \tilde{C}^3) \tr(C^1 C^1 C^3) \rangle \\
\nn & \langle \tr(\tilde{C}^3 \tilde{C}^3) \tr(C^2 C^2 C^3) \rangle \\
 & \ \\
\nn \mathbb{Z}_6^{\prime}: \quad & \langle \tr(\tilde{C}^3 \tilde{C}^3) \tr(C^1 C^2 C^3) \rangle \\
\nn & \langle \tr(\tilde{C}^3 \tilde{C}^3) \tr(C^2 C^2 C^2) \rangle \ .
\end{align}
 We perform the calculation explicitly for Yukawa couplings with tree-level flavour structure $C^1C^2C^3$ (correlators with anarchic Yukawa structure
 turn out to vanish).

\subsection{Corrections from two scalars: setting up the calculation}
\label{sec-strategy}
The strategy for the  computation of orbifold invariant amplitudes is:
\begin{enumerate}
\item  The canonical vertex operators are taken from the state-operator correspondence,
with orbifold charge assignments imposing $\theta_1+\theta_2+\theta_3=0$.
\item all possible conversions to total $H$-charge zero are enumerated \\
(some may immediately vanish by physical state conditions).
\item Demand
 total picture charge zero (as required on the
cylinder), picture change by contracting with PCO, then take coincidence limit.
\item Poles like $1/(z-w)$ are dropped since they cannot contribute
(see e.g.\ \cite{10075145}).
\item Momentum terms are completed into Lorentz invariant combinations.
\end{enumerate}
It may be useful to note that in this strategy, Lorentz invariance is not maintained in intermediate
steps. In particular, intermediate results may be given in terms of  Lorentz-noncovariant scalars, that then are assumed to be completed to Lorentz invariants, even though the remaining contributions vanish in this formalism.

The vertex operators prior to picture changing are:\footnote{Note the cylinder correlator $\langle \tr(\tilde{\phi} \tilde{\phi}) \tr(\psi \psi \phi) \rangle$ arises from the superfield term $\tr(\tilde{C}^3 \tilde{C}^3) \tr(C^1 C^2 C^3)$. When integrating out the Grassmann variables $\textrm{d}^2 \theta$ this superfield term also gives rise to the term $\tr(\tilde{\psi} \tilde{\psi}) \tr(\phi \phi \phi)$, and so we could also have calculated the amplitude $\langle \tr(\tilde{\psi} \tilde{\psi}) \tr(\phi \phi \phi) \rangle$ to arrive at the same result.}
\bal
\nn \tilde{\phi}_1^{-1}(z_1) \ = & \ \frac12 \left(0,0,0,0,--\right) \\
\nn \tilde{\phi}_2^{-1}(z_2) \ = & \ \frac12 \left(0,0,0,0,--\right) \\
 \psi_1^{-\half}(z_3) \ = & \ \frac12 \left(+,-,-,+,+\right) \\
\nn \psi_2^{-\half}(z_4) \ = & \ \frac12 \left(-,+,+,-,+\right) \\
\nn \phi^{-1}(z_5) \ = & \ \frac12 \left(0,0,0,0,--\right) \ .
\end{align}
The scalars $\tilde{\phi}_1$ and $\tilde{\phi}_2$ are inserted on one boundary of the cylinder while $\psi_1$, $\psi_2$ and $\phi$ are located on the other boundary. It is necessary to picture change 4 operators.

\subsection{Purely internal picture-changing}
\label{purelyint}

The first contribution $\mathcal{A}_1$ arises from picture-changing in internal directions only. The vertex operators become:
\bal
\nn \tilde{\phi}_1^{0}(z_1) \ = & \ \frac12 \left(0,0,0,0,\fbox{0}\right) \\
\nn \tilde{\phi}_2^{0}(z_2) \ = & \ \frac12 \left(0,0,0,0,\fbox{0}\right) \\
 \psi_1^{-\half}(z_3) \ = & \ \frac12 \left(+,-,-,+,+\right) \\
\nn \psi_2^{+\half}(z_4) \ = & \ \frac12 \left(-,+,+,-,\fbox{$-$}\right) \\
\nn \phi^{0}(z_5) \ = & \ \frac12 \left(0,0,0,0,\fbox{0}\right) \ ,
\end{align}
where boxes highlight H-charges that were modified in the process of picture-changing.

We begin the computation by evaluating the contribution of the fermionic and ghost correlators as
\begin{align}
\nn & \vartheta_1^{\prime}(0) \vartheta_1^{-1} (z_{34}) \ \eta_{\alpha \beta} \ \vartheta_{\alpha \beta}\left(-\frac{z_{34}}{2}\right) \vartheta_{\alpha \beta}\left(-\frac{z_{34}}{2} +\theta_1\right) \vartheta_{\alpha \beta}\left(\frac{z_{34}}{2} +\theta_2\right) \vartheta_{\alpha \beta}\left(\frac{z_{34}}{2} +\theta_3\right) \\
= \ & 2 \vartheta_1^{\prime}(0) \vartheta_1^{-1} (z_{34}) \ \vartheta_1(\theta_1+\theta_2+\theta_3) \vartheta_1 \left(-\frac{z_{34}}{2} +\theta_1\right) \vartheta_1(\theta_2) \vartheta_1(\theta_3) =0
\end{align}
where we have used the Riemann identity \eqref{Riemann}. As $\theta_1+\theta_2+\theta_3=0$ for supersymmetric orbifolds the fermionic and ghost contribution is hence exactly zero due to the appearance of $\vartheta_1(0)=0$. Correspondingly, the whole amplitude $\mathcal{A}_1$ vanishes.

\subsection{Picture-change scalars internally, picture-change Yukawa coupling externally}
\label{IntExt}

We can also arrive at a configuration with vanishing H-charge by picture-changing the scalars internally while picture-changing is performed in the external directions for the Yukawa coupling. The external picture-changing can be performed in various ways. One choice is shown below:
\bal
\label{IntExtVertex}
\nn \tilde{\phi}_1^{0}(z_1) \ = & \ \frac12 \left(0,0,0,0,\fbox{0}\right) \\
\nn \tilde{\phi}_2^{0}(z_2) \ = & \ \frac12 \left(0,0,0,0,\fbox{0}\right) \\
 \psi_1^{-\half}(z_3) \ = & \ \frac12 \left(+,-,-,+,+\right) \\
\nn \psi_2^{+\half}(z_4) \ = & \ \frac12 \left(\fbox{$---$},+,+,-,+\right) \\
\nn \phi^{0}(z_5) \ = & \ \frac12 \left(\fbox{$++$},0,0,0,--\right) \ .
\end{align}
External picture-changing introduces a momentum prefactor $k_4^{1+} k_5^{1-}$ while picture-changing the scalars internally contributes a factor $\pd \bar{Z}^3 \pd \bar{Z}^3$.

Similarly, we could have picture changed the fermion and the boson as follows: $\psi_2^{-1/2}(z_4) \rightarrow  \ \frac12 \left(\fbox{$+$},+,+,-,+\right)$ and $\phi^{-1}(z_5) \rightarrow  \ \frac12 \left(\fbox{$--$},0,0,0,--\right)$. This correlator will contribute a momentum prefactor $k_4^{1-} k_5^{1+}$, which vanishes in virtue of the physical state conditions \eqref{k4condition}.\footnote{We can nevertheless evaluate the correlator. The result will be the same as for the amplitude with vertex operators as in \eqref{IntExtVertex}. However, there is a complication: in this case picture-changing cannot be performed before the calculation of the correlator: the OPEs of the fields with the stress-tensor lead to the appearance of poles: $e^{\phi(w)} T_F(w) \psi_2^{-1/2}(z_4) \sim {(w-z_4)}^{-1}  + \ldots$ where ellipses denote less divergent terms. When taking the limit $w \rightarrow z_4$ at this stage it is not clear what to do with the pole. Thus we need to insert the picture-changing operator $O_{\textrm{PC}}^{+1}(w)= e^{\phi(w)}e^{-iH(w)} \pd X^2(w)$ explicitly into the amplitude, evaluate it and only take the limit $w \rightarrow z_4$ at the end. We refer readers to \cite{10075145} for an example of such a calculation.}

Last, we also need to consider external picture-changing along the second complex dimension. The corresponding amplitudes will display a momentum term $(k_4^{2-} k_5^{2+}+k_4^{2+} k_5^{2-})$. Again, we do not need to calculate the corresponding CFT amplitude as this momentum prefactor is identically zero due to the physical state conditions \eqref{k4condition}. All in all, we have to calculate one CFT correlation function: we can identify its prefactor $k_4^{1+} k_5^{1-}$ with the Lorentz-invariant quantity $(k_4k_5)$ as the terms necessary for a Lorentz covariant completion are all zero. Even though we are interested in the result for vanishing momenta only, the amplitude does not vanish in spite of this prefactor: it will be cancelled by the appearance of a momentum pole.

The calculation proceeds in a similar way to those already performed and overall we find
\begin{itemize}
\item D3-D3 models:
\begin{align}
\nn \mathcal{A}_2 \propto (k_4k_5) \int_0^{\infty} & \frac{\textrm{d}t}{t} \int \prod_{i=1}^5 \textrm{d}z_i \frac{\eta^3}{{(2 \pi^2 t)}^2}  {\left(\frac{\vartheta_1(z_{34})}{\vartheta_1^{\prime}(0)} \right)}^{-1+k_3k_4} {\left(\frac{\vartheta_1(z_{45})}{\vartheta_1^{\prime}(0)} \right)}^{-1+k_4k_5} \\
\nn & \left(\frac{\vartheta_1(-z_{34}+\theta_1)}{\vartheta_1(\theta_1)} \right) \left(\frac{\vartheta_1(\theta_2)}{\vartheta_1(\theta_2)} \right) \left(\frac{\vartheta_1(z_{45}+\theta_3)}{\vartheta_1(\theta_3)} \right) \\
& \prod_{i=1}^3 (-2 \sin \pi \theta_i) \ \langle \pd \bar{Z}^3 \pd \bar{Z}^3 \rangle \ ,
\end{align}
\item D3-D7 models:
\begin{align}
\nn \mathcal{A}_2 \propto (k_4k_5) \int_0^{\infty} & \frac{\textrm{d}t}{t} \int \prod_{i=1}^5 \textrm{d}z_i \frac{\eta^3}{{(2 \pi^2 t)}^2}  {\left(\frac{\vartheta_1(z_{34})}{\vartheta_1^{\prime}(0)} \right)}^{-1+k_3k_4} {\left(\frac{\vartheta_1(z_{45})}{\vartheta_1^{\prime}(0)} \right)}^{-1+k_4k_5} \\
& \left(\frac{\vartheta_4(-z_{34}+\theta_1)}{\vartheta_4(\theta_1)} \right) \left(\frac{\vartheta_4(\theta_2)}{\vartheta_4(\theta_2)} \right) \left(\frac{\vartheta_1(z_{45}+\theta_3)}{\vartheta_1(\theta_3)} \right) \ \langle \pd \bar{Z}^3 \pd \bar{Z}^3 \rangle \ .
\end{align}
\end{itemize}
There are further factors of the form $(\ldots)^{k_1k_j}$ which we ignored: as we are only interested in the result for vanishing momenta $k_i$ these terms do not affect the result: $(\ldots)^{k_ik_j} \rightarrow 1$ for $k \rightarrow 0$. The exception are terms of the form $(\vartheta_1(z_{ij}))^{-1+k_1k_j}$ as they will lead to momentum poles when integrated over worldsheet positions.

Picture-changing the scalars left us with the internal bosonic fields $\pd \bar{Z}^3 \pd \bar{Z}^3$ whose correlator we left unevaluated so far. Their quantum correlator vanishes, but we need to include their classical correlator in terms of winding modes \eqref{windingderivative}. Correspondingly we have:
\be
\langle \pd \bar{Z}^3 \pd \bar{Z}^3 \rangle = \left( -{\sqrt{\alpha'} \over \pi ct }\frac{\partial}{\partial r_c} \right)^2 \mathcal{Z}(t) \ .
\ee

The resulting D3-D3 amplitude is ( the D3-D7 amplitude can be calculated accordingly)
\begin{align}
\nn \mathcal{A}_2 \propto (k_4k_5) \frac{\pd^2}{{\pd r_c}^2} \int_0^{\infty} & \frac{\textrm{d}t}{t} \int \prod_{i=1}^5 \textrm{d}z_i \frac{1}{{(2 \pi^2 t)}^2} \ \frac{1}{t^2} \ \mathcal{Z}(t)  \prod_{i=1}^2 (-2 \sin \pi \theta_i) \\
& {\left(\frac{\vartheta_1(z_{34})}{\vartheta_1^{\prime}(0)} \right)}^{-1+k_3k_4} {\left(\frac{\vartheta_1(z_{45})}{\vartheta_1^{\prime}(0)} \right)}^{k_4k_5} \left(\frac{\vartheta_1(-z_{34}+\theta_1)}{\vartheta_1(\theta_1)} \right) \left(\frac{\vartheta_1(\theta_2)}{\vartheta_1(\theta_2)} \right)  \ .
\end{align}
Next we need to integrate this over the worldsheet positions $z_i$. This can be easily done given the result above. The only non-trivial dependence on worldsheet positions is in $z_{34}$. The positions $z_3$ and $z_4$ correspond to vertex operators inserted on the same boundary of the cylinder. Thus, when integrating over them there will be instances where $z_3 \rightarrow z_4$. For small arguments $z_{34}$ the theta function becomes
\be{\vartheta(z_{34})}^{-1+k_3k_4} = {(2 \pi \eta^3 z_{34})}^{-1+k_3k_4} + \mathcal{O}(z_{34}^2) \ .
\ee
Correspondingly, for vanishing momenta we will pick up a momentum pole $\int \textrm{d} z_{34} \ {z_{34}}^{-1+k_3k_4} \rightarrow 1/(k_3k_4)$. As the amplitude also has a momentum prefactor $(k_4k_5)$ the momentum pole gives the only finite contribution to the amplitude for $k_i \rightarrow 0$. We can hence perform the integrations as follows: we set $z_3=z_4$ in the amplitude and pick up the momentum pole. The remaining integrations over worldsheet positions are then trivial leading to $\int \textrm{d}z_1 \textrm{d}z_2 \textrm{d}z_4 \textrm{d}z_5 = {(it/2)}^4$. Note, that we can only pick up a momentum pole since $z_3$ and $z_4$ are on the same boundary of the cylinder. For example, a term ${\vartheta(z_{14})}^{-1+k_1k_4}$ would not have lead to a momentum pole. Implementing these findings we arrive at
\begin{itemize}
\item D3-D3 models:
\begin{align}
\mathcal{A}_2 \propto \frac{k_4k_5}{k_3 k_4} \ \prod_{i=1}^2 (-2 \sin \pi \theta_i) \ \frac{\pd^2}{{\pd r_c}^2} \int_0^{\infty} & \frac{\textrm{d}t}{t} \ \mathcal{Z}(t) \ \left(\frac{\vartheta_1(\theta_1)}{\vartheta_1(\theta_1)} \right) \left(\frac{\vartheta_1(\theta_2)}{\vartheta_1(\theta_2)} \right) \ .
\end{align}
At this stage we do not yet cancel theta-functions in the numerator and denominator: we want to keep the dependence on the orbifold twists as we will now examine the amplitude in various orbifold sectors. As before the result is entirely determined by the $\mathcal{N}=2$-sector and we obtain
\begin{align}
\label{resultIntExtD3D3}
\mathcal{A}_2 \propto \frac{k_4k_5}{k_3 k_4} \ \prod_{i=1}^2 (-2 \sin \pi \theta_i) \ \frac{\pd^2}{{\pd r_c}^2} \int_0^{\infty} & \frac{\textrm{d}t}{t} \ \mathcal{Z}(t) \ .
\end{align}

\item D3-D7 models:\\
Here, we do not encounter any complications in the untwisted sector as $\vartheta_4(0)\neq 0$. This is expected as the untwisted sector preserves $\mathcal{N}=2$ supersymmetry in D3-D7 models. However, we need to include image D7-branes to arrive at a setup that is invariant under orbifold twists. Hence we also need to include winding strings stretching between the D3 and the image D7-branes. Whereas the original D7-stack was located at $r_c$ on the third 2-torus, the images are positioned at $e^{2 \pi i j \theta_3}$ where $j=1\ldots N-1$ for a $\mathbb{Z}_N$-orbifold. we note that the relevant orbifolds, $\mathbb{Z}_4$ and $\mathbb{Z}_6^{\prime}$ have $\theta_3=\pm 1/2$. Correspondingly,
\be
e^{-4 \pi i j \theta_3} \left( -{\sqrt{\alpha'} \over \pi ct }\frac{\partial}{\partial r_c} \right)^2 \mathcal{Z}(t) = e^{\pm 2 \pi i j} \left( -{\sqrt{\alpha'} \over \pi ct }\frac{\partial}{\partial r_c} \right)^2 \mathcal{Z}(t)= \left( -{\sqrt{\alpha'} \over \pi ct }\frac{\partial}{\partial r_c} \right)^2 \mathcal{Z}(t) \ , \ \ j=1\ldots N-1  \ .
\ee
All images contribute equally and thus we arrive at the result
\begin{align}
\mathcal{A}_2 \propto \frac{k_4k_5}{k_3 k_4} \ \frac{\pd^2}{{\pd r_c}^2} \int_0^{\infty} & \frac{\textrm{d}t}{t} \ \mathcal{Z}(t) \ .
\end{align}
\end{itemize}
Before commenting on our results we calculate the remaining contributions.

\subsection{Mixed internal-external picture-changing}
\label{mixedpc}

The last consistent picture-changed configuration differs from the previous cases as we picture-change one of the hidden scalars in an external direction. One example of possible H-charge assignments is:
\bal
\label{mixed}
\nn \tilde{\phi}_1^{0}(z_1) \ = & \ \frac12 \left(0,0,0,0,\fbox{0}\right) \\
\nn \tilde{\phi}_2^{0}(z_2) \ = & \ \frac12 \left(\fbox{++},0,0,0,--\right) \\
 \psi_1^{-\half}(z_3) \ = & \ \frac12 \left(+,-,-,+,+\right) \\
\nn \psi_2^{+\half}(z_4) \ = & \ \frac12 \left(\fbox{$---$},+,+,-,+\right) \\
\nn \phi^{0}(z_5) \ = & \ \frac12 \left(0,0,0,0,\fbox{0}\right) \ .
\end{align}
Again, there are four ways of implementing this. We can picture change in either the external $1$- or the $2$-directions. In addition, the scalar can either obtain an H-charge of $(--)$ or $(++)$ in that direction. In principle one should also picture-change a fermion operator $\psi_2^{-1/2}(z_4) \ =  \ \frac12 \left(-,+,+,-,+\right) \rightarrow \ \frac12 \left(+,+,+,-,+\right)$. All the partial expressions will posses different momentum prefactors which, when combined, give the Lorentz-invariant quantity $(k_2\cdot k_4)$. Further, picture-changing also introduces the fields $\pd \bar{Z}^3 \pd \bar{Z}^3$ which will contribute their classical correlator. The calculation we present here acquires a momentum prefactor of $k_2^{1-}k_4^{1+}$. As it turns out, this will be the only contribution from this choice of picture-changing: the physical state conditions \eqref{k4condition} imply that $k_2^{1+}k_4^{1-}=k_2^{2-}k_4^{2+}=k_2^{2+}k_4^{2-}=0$ allowing us to identify $k_2^{1-}k_4^{1+}=(k_2k_4)$.

Equally, we could have chosen $\phi_1$ to be the operator to be picture-changed in the external directions. The calculation will then be the same with $z_1 \leftrightarrow z_2$ and a momentum prefactor $(k_1 k_4)$.

In the following we sketch the calculation of the amplitude $\mathcal{A}_3$ with H-charges as shown above in \eqref{mixed}.

Evaluating the fermionic and ghost correlators and applying a Riemann identity \eqref{Riemann} we find:
\begin{itemize}
\item D3-D3 models
\be
\frac{2 {\vartheta_1^{\prime}}^3(0) \vartheta_1(-z_{34} +\theta_1) \vartheta_1(\theta_2) \vartheta_1(z_{24}-\theta_3)}{\vartheta_1(z_{24}) \vartheta_1(z_{34})} \ ,
\ee
\item D3-D7 models
\be
\frac{2 {\vartheta_1^{\prime}}^3(0) \vartheta_4(-z_{34} +\theta_1) \vartheta_4(\theta_2) \vartheta_1(z_{24}-\theta_3)}{\vartheta_1(z_{24}) \vartheta_1(z_{34})} \ .
\ee
\end{itemize}
The remaining contributions to this amplitude are given by the bosonic partition function at 1-loop, the correlator over momentum exponentials and the classical correlator $\langle \pd \bar{Z}^3 \pd \bar{Z}^3\rangle$. We calculated these terms already for the amplitude $\mathcal{A}_2$, so that we can use some of the results of the previous section. Again, the classical correlator vanishes unless $\theta_3=0$. Thus, fully twisted sectors do not contribute as before. Untwisted sectors also lead to vanishing results in D3-D3 setups. We continue to evaluate the amplitude for vanishing momenta $k_i$. The integration over worldsheet positions will be trivial unless $z_3 = z_4$ due to the presence of the term ${(\vartheta_1(z_{34}))}^{-1+k_3k_4}$. We perform the integration by setting $z_3=z_4$ and picking up the pole $1/(k_3k_4)$. The remaining worldsheet integrals contribute ${(it/2)}^4$. We find the amplitude to match the result from the previous section except for the momentum prefactor. We thus obtain:
\begin{itemize}
\item D3-D3 models:
\begin{align}
\mathcal{A}_3 \propto \frac{k_2k_4}{k_3 k_4} \ \prod_{i=1}^2 (-2 \sin \pi \theta_i) \ \frac{\pd^2}{{\pd r_c}^2} \int_0^{\infty} & \frac{\textrm{d}t}{t} \ \mathcal{Z}(t) \ .
\end{align}
\item D3-D7 models
\begin{align}
\mathcal{A}_3 \propto \frac{k_2k_4}{k_3 k_4} \ \frac{\pd^2}{{\pd r_c}^2} \int_0^{\infty} & \frac{\textrm{d}t}{t} \ \mathcal{Z}(t) \ .
\end{align}
\end{itemize}
There is an almost identical result from picture changing $\tilde{\phi}_1$, with $k_2 \to k_1$.

\subsection{Combining the partial results}
\label{combine}

In this part we collect all the partial results to arrive at the full result for the amplitude $\langle \tr(\tilde{\phi} \tilde{\phi}) \tr(\psi \psi \phi) \rangle$ at vanishing momenta. We observe that all non-zero contributions are the same except for a momentum prefactor. The overall result will then be proportional to
\be
\frac{k_1k_4+k_2k_4+k_5k_4}{k_3k_4}=\frac{(k_1+k_2+k_5)k_4}{k_3k_4}= - \frac{k_3k_4}{k_3k_4} =-1
\ee
where we have used momentum conservation and the fact that $k_i^2=0$. The overall result is hence momentum-independent and gives a finite contribution.

We can go further and evaluate the integral $\int \textrm{d} t/t \ \mathcal{Z}(t)$ that appears in the result for both the D3-D3 and the D3-D7 calculation. We need to introduce a UV-cutoff in this step which will be annihilated by the derivatives w.r.t.~$r_c$. By rewriting the sum over winding modes in terms of a generalised theta function \eqref{Zt} we can again use \cite{0404087} to arrive at:
\begin{align}
\nn \mathcal{A} = \mathcal{C} \frac{\partial^2}{{\partial r_c}^2} \int_{1/ \Lambda^2}^{\infty} \frac{\textrm{d}t}{t} \mathcal{Z}(t) &=  \mathcal{C} \ \frac{U_2}{\tilde{T}_2} \ \frac{\partial^2}{{\partial r_c}^2} \left(\frac{\Lambda^2}{\tilde{T}_2} - \ln {\left|\frac{\vartheta_1(r_c,\bar{U})}{\eta(\bar{U})} \right|}^2 - \frac{2 \pi{\textrm{Im}(r_c)}^2}{U_2} \right) \\
&= - \mathcal{C}  \frac{\partial^2}{{\partial r_c}^2}{\mathcal G}(0,r_c) \ ,
\end{align}
with the boson propagator on the torus from \eqref{bosprop}.

\bibliographystyle{JHEP}
\bibliography{5ptbib}

\end{document}